\documentclass[prb,amsmath,amssymb,twocolumn]{revtex4-1}
\usepackage[normalem]{ulem}
\usepackage{amsmath,amssymb}
\usepackage[bookmarks=true,colorlinks,linkcolor=blue,urlcolor=blue,citecolor=blue]{hyperref}
\usepackage{graphicx,amsmath,amsfonts,dsfont}
\usepackage[usenames]{color}
\usepackage{epstopdf}
\usepackage{verbatim}
\usepackage{amsthm}
\usepackage{enumitem}
\hypersetup{colorlinks,linkcolor=blue,filecolor=green,urlcolor=blue,citecolor=blue}
\usepackage[titletoc,toc]{appendix}
\usepackage[T1]{fontenc}
\usepackage[utf8]{inputenc}
\usepackage[german,english]{babel}
\usepackage{amsmath}
\usepackage{amsfonts}
\usepackage{amssymb}
\usepackage{mathrsfs}
\usepackage{graphicx}
\usepackage{color}
\usepackage[usenames,dvipsnames]{xcolor}
\usepackage{mathtools}
\usepackage{braket}

\newcommand{\revcomment}[1]{\textcolor{red}{#1}}

\newcommand{\beq}{\begin{equation}}
\newcommand{\eeq}{\end{equation}}

\renewcommand{\oe}{ö}
\let\max\relax  % cancella la definizione built-in
\DeclareMathOperator*{\max}{max} % * mette apici/pedici sopra/sotto anziché in alto/basso a destra
\DeclareMathOperator{\tr}{Tr}

\begin{document}

\title{Impact of non-equilibrium fluctuations  \\
on pre-thermal  dynamical phase transitions 
in long-range interacting  spin chains}

\author{Alessio Lerose}
\affiliation{SISSA --- International School for Advanced Studies, via Bonomea 265, I-34136 Trieste, Italy}
\affiliation{INFN --- Istituto Nazionale di Fisica Nucleare, Sezione di Trieste, I-34136 Trieste, Italy}
\author{Bojan \v{Z}unkovi\v{c}}
\affiliation{Department of Physics, Faculty of Mathematics and Physics, University of Ljubljana, Jadranska 19, 1000 Ljubljana, Slovenia}
\author{Jamir Marino}
\affiliation{
Department of Physics, Harvard University, Cambridge MA 02138, United States\\
Department of Quantum Matter Physics, University of Geneva, 1211, Geneve, Switzerland }
\author{Andrea Gambassi}
\affiliation{SISSA --- International School for Advanced Studies, via Bonomea 265, I-34136 Trieste, Italy\\
INFN --- Istituto Nazionale di Fisica Nucleare, Sezione di Trieste, I-34136 Trieste, Italy}
\author{Alessandro Silva}
\affiliation{SISSA --- International School for Advanced Studies, via Bonomea 265, I-34136 Trieste, Italy}

\begin{abstract}
We study the non-equilibrium phase diagram and the dynamical phase transitions occurring during the  pre-thermalization of non-integrable quantum spin chains, subject to either quantum quenches or linear ramps of a relevant control parameter.
We consider  spin systems in which long-range ferromagnetic interactions compete with short-range, integrability-breaking terms. We capture the pre-thermal stages of the non-equilibrium evolution  via a time-dependent spin-wave expansion at leading order in the spin waves density. In order to access regimes with strong integrability breaking, instead, we perform numerical simulations based on the time-dependent variational principle with matrix product states. % (MPS-TDVP).
By investigating a large class of quantum spin models, we demonstrate  that non-equilibrium  fluctuations can significantly affect the dynamics near critical points of the phase diagram, resulting in a chaotic evolution of the collective order parameter, akin to the dynamics of a classical particle in a multiple-well potential subject to quantum friction.
We also elucidate the signature of this novel dynamical phase on the time-dependent correlation functions of the local order parameter. 
%
%Interestingly enough, correlation functions resemble those of periodically driven quantum many-body systems:  the motion of the collective degree of freedom acts as a periodic self-drive for the spin waves generated by short-range interactions, which manifests itself with a periodic modulation in time of the signal within the causal region of the light-cone  of correlations. 
%
We finally establish a connection with the notion of dynamical quantum phase transition associated with a possible non-analytic behavior of the return probability amplitude, or Loschmidt echo, showing that the latter displays cusps whenever the order parameter vanishes during its real-time evolution.
 \end{abstract}

% 64.60.Ht dy critic ph
% 64.60.ae Rg study
% 05.30.Rt q phase trnas

\date{\today}

\maketitle

\section{Introduction} 
%
%%%%%%%

Consider an extended quantum many-body system in an equilibrium, low temperature, ordered phase (e.g. a ferromagnet), and  drive it out of equilibrium by varying in time a control parameter (e.g., a magnetic field). 
This can occur via an abrupt change from an initial to a final value (the so-called \emph{quantum quench}~\cite{PolkovnikovRMP, Lamacraft2012, Eisert2015a}), or a continuous time-dependent ramp. %  or a periodic driving. 
In these cases the initial long-range ordered state is destabilized, and   it is  therefore natural  to investigate the fate of the order parameter out of equilibrium. 
If thermalization occurs quickly, the order parameter will show a behavior consistent with its equilibrium finite-temperature phase diagram. 
On the other hand, if a metastable,  non-equilibrium quasi-steady state is established at intermediate time scales before thermalization, non-trivial time-dependent phenomena may occur.
This scenario, which focuses on pre-thermal states~\cite{Berges2004a, Berges2004b, kollath, Moeckel2008,Rosch2008, Moeckel2009, Gring2012, Marinolong2012, Langen2013, Nessi2014, Langen2016}, typically occurs in systems close to integrability.
Quantum many body integrable systems are known to relax towards a generalized Gibbs ensemble (GGE), a sort of grand-canonical ensemble accounting for all the local (or quasi-local) conserved quantities of the system~\cite{kino, Cazalilla2006, Rigol2007, cramer, barthel, muss, cauxkon, cef, Langen2015, Essler2016}. 
Under weak integrability-breaking perturbations, GGEs can act as metastable attractors of the dynamics, before the system slowly drifts towards its long-time, asymptotic steady state described by a canonical Gibbs ensemble~\cite{Calabrese2007, Kollar2011, Stark2013, Robinson2014, Bertini2015, Bertini2016,Marcuzzi2013,Demler2015, Marcuzzi2016, roscilde}. 
In particular, the system lingers close to a state violating detailed balance in which conventional equilibrium statistical mechanics does not apply, making the onset of novel types of  phases of matter and critical behavior  possible.

An interesting example of non-equilibrium critical phenomena may emerge after a quantum quench of an interacting quantum many-body system which displays symmetry breaking at equilibrium.
These dynamical phase transitions (DPTs)~\cite{Barankov2004, Barankov2006, Yuzbashyan2006, Gurarie2009, Eckstein2009, Schiro2010, Sciolla2010, SciollaBiroliMF, GambassiCalabrese, Schiro2011, Sandri12, Sciolla2013, Foster2013, Foster2014, Smacchia2014, Peronaci2015, Yuzbashyan2015, Halimeh2016, Halimeh2017} 
are characterized by a non-equilibrium order parameter exhibiting a finite or vanishing long-time temporal average, depending on whether the quantum model under consideration is quenched  below %%(dynamical ordered  phase) 
or above
%(dynamical disordered  phase) 
an associated  dynamical critical point separating the dynamical ordered phase from the dynamical disordered one which depends, inter alia, on the initial conditions. %, corresponding to a specific set of values of its post-quench parameters.
In addition, in systems with local interactions, the scaling of dynamical correlation and response functions can distinguish the different phases%
%when quenching the system close to the dynamical critical point from above, unequal time correlation functions manifest a scaling behaviour reminiscent of aging in classical critical dynamics
~\cite{Janssen1989, Calabrese2005,%
%}, while quenching across  dynamical critical points imprints a scaling behaviour reminiscent of dynamical coarsening \cite{
Chandran2013,Maraga2015, Chiocchetta2015, Chiocchetta2017}.
%%
%Despite dynamical phases are drastically different from their static counterparts, the
%this is highly-non-ground-state physics, and the analogy with quantum phase transitions at finite temperature serves to understand the concept: even if a system is at finite T, it can support different phases separated by a line of critical points provided it is in high enough dimensions (there are exceptions). Temperature is just a convenient way to know how eigenstates are populated (namely via Boltzmann weights, detailed balance), while a NEQ state has more freedom in this respect, but brutally speaking very similar. That?s the reason why you can have NEQ critical behaviour.
%
%
A recent experiment%performed by the experimental group of C. Monroe
~\cite{zhang} has shown that these dynamical phase transitions can be realized with ultracold trapped ions which simulate long-range interacting Ising ferromagnets. 
%\revcomment{CUT? Signatures of non-equilibrium dynamical scaling  have been in turn reported in an experiment monitoring the dynamics of a two-component Bose gas~\cite{Oberthaler2015}.}
%
A second notion of DPT, proposed in Ref.~\onlinecite{Heyl2013}, has been recently studied experimentally in Ref.~\onlinecite{jurc}. 
This notion, however, is not directly related to the existence of a local order parameter characterizing the various dynamical phases, but rather to a non-analytic behavior in the time dependence of the return probability amplitude.
%
%Considering, for example,  global quenches of the transverse field in the one-dimensional quantum Ising chain, a quench across the equilibrium  critical point, starting from the ground state of the ferromagnetic phase and ending into the paramagnetic one, will induce a sequence of cusps in the time-resolved return probability amplitude~\cite{Heyl2013, karr}.
%
%\alessio{The origin of these non-analyticities is rooted in the Fisher zeros of the finite-temperature partition function of the classical Ising model, which is the higher dimensional classical counterpart of the quantum Ising chain in one dimension. DO WE REALLY NEED THIS?}
%
%Indeed, a mapping (involving a rotation into imaginary time) relating the return probabilty amplitude of the quantum model to the partition function of the classical one, directly connects the Fisher zeros to the cusps of the Loschmidt echo~\cite{Heyl2013}.
%
These two notions of dynamical phase transition are in general distinct, and  therefore
 they may even not occur concomitantly in the same model. However, a connection has been pointed out whenever both DPTs are present~\cite{Zunkovic2016} (see also Refs.~\onlinecite{WeidingerO(N),HalimehGeometric}).
The study of the two instances of DPTs is typically restricted to either integrable or mean-field models, or to numerical works, while Gaussian fluctuations have been accounted for in a limited number of cases % and in particular for a vanishing mean-field order parameter 
\cite{Sandri12, Sciolla2013,Chandran2013,Smacchia2014, Maraga2015, Chiocchetta2015, Chiocchetta2017}. The purpose of this paper is to thoroughly study DPTs in non-integrable models that, despite possessing non-trivial fluctuations and a non-vanishing order parameter, are amenable to an analytical approach. %study.

As we anticipated in Ref.~\onlinecite{LeroseShort}, the analysis of the fully-connected quantum Ising model with an additional short-range integrability-breaking perturbation reveals that %Gaussian 
fluctuations may induce unexpected ``chaotic'' behavior in the non-equilibrium dynamics: in the presence of $\mathbb{Z}_2$ symmetry breaking, the asymptotic sign of the order parameter turns out to depend sensitively, and to a large extent unpredictably, on the initial conditions and on the specific values of the parameters of %that characterize 
the post-quench Hamiltonian. We thoroughly show here that this phenomenon % is expected to occur whenever 
turns out to rely on few essential physical ingredients, namely \textit{(i)} the existence of multiple, macroscopically distinct equilibrium configurations for a collective order parameter, and \textit{(ii)} the possibility of dissipating the energy of the collective motion into microscopic non-equilibrium  fluctuations. 
In this work, these ingredients are provided by %mean-field (i.e., infinite-range) models with spontaneous breaking of a discrete symmetry and by general weak variable-range perturbations, respectively. The physics that we unveil is therefore the result of a 
the competition between long- and short-range interactions in quantum many-body systems when quenched near a dynamical critical point:
a mean-field collective degree of freedom moves in a multiple-well landscape and is weakly coupled to an extensive set of microscopic degrees of freedom which represent quantum fluctuations at all length scales. The latter provide a sort of quantum friction on the classical collective motion that in turn makes the eventual ``choice'' of the asymptotic well highly sensitive to the parameters. This behavior is actually reminiscent of what is observed in a coin toss\cite{coin}, with the coin playing the role of the macroscopic collective degree of freedom with two possible stable equilibrium configurations (``heads'' and ``tails''), which dissipates its energy into the microscopic degrees of freedom (phonons of the floor, air molecules,\dots) and finally undergoes a pseudo-random choice of the asymptotic state.
The peculiar behavior described above is based on general properties and is therefore expected to occur in a variety of systems. Here we will first illustrate it in the case of the fully-connected quantum Ising model in a transverse field, i.e., the Lipkin-Meshkov-Glick (LMG) model, and later on we will discuss a much more general class of models where the phenomenon is observed. Moreover, we investigate the signatures of this phenomenon in the non-equilibrium spreading of correlations, as well as its occurrence with more general non-equilibrium protocols such as linear ramps.
%
%
%
%
%%%%%%%
%%%%%%%

The paper is organized as follows. In Sec.~\ref{sec:MF} we review the properties of both the equilibrium and the dynamical criticality of the LMG model  which represents the basis for the analysis presented in the following Sections. %discussing also the corresponding equilibrium and non-equilibrium phase diagrams.
%

%Sec.~\ref{sec:methodweakfluct} has a methodological purpose: we introduce a generic approach, based on the spin-wave expansion, applicable both to static and dynamical settings, in order to treat integrability breaking perturbations of the LMG model (or more general mean-field models) valid for xfmal stages of dynamics.
%

Sec.~\ref{sec:methodweakfluct} illustrates in a pedagogical fashion the methods used in this study, based on a spin wave (Holstein-Primakoff) expansion around the instantaneous average direction of the spins, whose evolution is self-consistently determined by taking into account the feedback from the quantum spin wave fluctuations. This approach is suitable for studying both equilibrium and non-equilibrium problems in a wide range of systems close to mean-field integrability (i.e., long-range or high-dimensional systems), and is therefore of interest by itself.

In Secs.~\ref{sec:quenchchaos} and~\ref{sec:generalchaos} we discuss in detail the impact  of integrability-breaking perturbations on the dynamical phase diagram of mean-field models, showing explicitly, through an extensive  analysis encompassing several different types of perturbations and generalizations of the LMG model, that the \emph{chaotic dynamical phase} found in Ref.~\onlinecite{LeroseShort} has to be expected in general.
%

%Indeed, the latter occurs whenever a collective spin model presenting a phase transition associated to  the spontaneous breaking of a discrete symmetry, is perturbed by a term generating non-equilibrium quantum fluctuations; the latter, manifesting in the form of spin waves,  reshapes the mean-field dynamical critical point into a chaotic dynamical phase, where the order parameter can hop among the available ferromagnetic minima before selecting one of them. This selection occurs with extreme sensitivity  to the values of the parameters of the specific non-equilibrium protocol adopted.

%

Sec.~\ref{sec:corr} is devoted to the calculation of the equal-time correlation function of the order parameter at different space points across the dynamical phase diagram of the LMG model perturbed by nearest-neighbor transverse spin interactions.
These correlation functions exhibit a periodic modulation in time, illustrating  that the dynamics of the spin waves is \emph{periodically self-driven} as a result of the  precession of the collective magnetization of the LMG model induced by the transverse field.%, which in turn drives  the former at a given frequency.

Finally, in Sec.~\ref{sec:ramp} results are presented for a linear ramp of the transverse field as a function of time in the LMG model, generalizing the sudden quench  considered in Sec.~\ref{sec:results}.
As the duration of the ramp increases, the chaotic phase shrinks in the adiabatic limit; on the contrary, the faster the ramp is, the closer the dynamical phase diagram is to the one generated by a sudden quench (Sec.~\ref{sec:results}).

In Sec.~\ref{sec:strongcouple} we confirm the onset of the chaotic phase for strong integrability-breaking perturbations by employing a numerical method based
on matrix product states, %corroborating and 
extending the findings of Ref.~\onlinecite{LeroseShort}.

In Sec.~\ref{sec:loschmidt}  we  discuss the connection between the dynamical phase transition discussed in this paper and the notion of dynamical phase transition associated with cusps of the Loschmidt echo~\cite{Heyl2013}, confirming also in the present  case the prediction of Ref.~\onlinecite{Zunkovic2016}:
whenever the order parameter vanishes during its evolution,  cusps  are concomitantly formed in the real-time dynamics of the return probability amplitude. 

In Sec.~\ref{sec:cat} we discuss the important issue of finite-size effects, relevant to possible experimental realizations of the phenomena hereby discussed.

In Sec.~\ref{sec:conclusions} we %present our conclusions and 
discuss further perspectives.

\section{Dynamical phase transition in the infinite-range Ising model}

%We will consider the dynamical phase diagram of the Lipkin-Meshkov-Glick (LMG) model
%\begin{equation}
%\label{uno}
%H_0 = - \frac{\lambda}{N} \sum_{i,j} \sigma_i^x \sigma_j^x - g \sum_i \sigma_i^z,
%\end{equation}
%%
%(a collective quantum Ising ferromagnet in transverse field) 
%perturbed by a short-range spin interaction \bzcom{Perhaps $U=-J\sum_i \sigma_i^x \sigma_{i+1}^x$?}
%\begin{equation}
%U=-J\sum_i \sigma_i^z \sigma_{i+1}^z.
%\end{equation}
%%which breaks the integrability of the model~\eqref{uno}.
%%
%Preparing the system in  the ferromagnetic ground state of the Hamiltonian~\eqref{uno} with $g=0$, and  quenching of the transverse field from $g=0$ to $g\neq0$, the dynamics of the model $H=H_0+U$,
%hosts three dynamical phases: \textit{(i)} a ferromagnetic one with a non-zero average of the order parameter ($S^x=\sum_i\sigma^x_i$, the collective magnetisation of \eqref{uno}), \textit{(ii)} a paramagnetic one where the order parameter averages to zero at long times, and \textit{(iii)} a critical region  characterised by strong sensitivity of the sign of the asymptotic magnetisation to parameters and initial conditions as discussed in  Ref.~\onlinecite{LeroseShort}. %chaotic jumps of the magnetization between positive and negative values. 
%%
%Below we will uncover further properties of this phase and study its stability by changing the quench protocol.

\label{sec:MF}

%Let us start our brief review on the dynamical properties of the LMG model, considering   quantum spins on a lattice, interacting via ferromagnetic couplings and subject to a transverse magnetic field
In this work we firstly focus on the non-equilibrium dynamics of a general class of Ising-type systems with quantum $s$-spins on a lattice, interacting via ferromagnetic coupling and subject to a transverse magnetic field
\beq\label{eq:model}
H = - \sum_{\mathbf{r},\mathbf{r'}} J_{\lvert\mathbf{r}-\mathbf{r'}\rvert} \, \sigma_{\mathbf{r}}^x \sigma_{\mathbf{r'}}^x - g \sum_{\mathbf{r}} \sigma_{\mathbf{r}}^z,
\eeq
where the sums run over the sites of a lattice, %, labelled by $i,j=1,\dots,N$, 
while $\sigma^\alpha_\mathbf{r} = S^\alpha_\mathbf{r} /s$ are the operators corresponding to the normalized spin components in the $\alpha=x,y,z$ direction, acting on site $\mathbf{r}$. This represents a generalization of the case of spins one-half, where $s=1/2$ and the $\sigma^\alpha_\mathbf{r}$'s reduce to the standard Pauli matrices. Controlling $s$ allows us to keep track of the impact of quantum fluctuations, which is suppressed in the classical limit $s\to\infty$. The ferromagnetic couplings $J_r$ depend on the  distance $r=\lvert\mathbf{r}-\mathbf{r'}\rvert$ between two sites.

For general ferromagnetic interactions $J_r$ (short- or long-range), the system is expected to have an equilibrium zero-temperature phase transition from a unique paramagnetic ground state with $\braket{\sigma^x}=0$ for $g>g_{\text{cr}}$ to a pair of ferromagnetic ground states with $\braket{\sigma^x}_{\pm}=\pm m\ne0$ for $g<g_{\text{cr}}$, characterized by the breaking of the $\mathbb{Z}_2$-symmetry $\sigma^x \mapsto - \sigma^x$. 
The emergence of a non-vanishing order parameter at a finite energy density above the ground state (e.g., in an equilibrium finite-temperature state, or in a non-equilibrium state attained after a quench) depends on the dimensionality and on the range of the interactions. While one-dimensional systems with short-range interactions cannot support order in excited states~\cite{Sachdevbook, CEF1}, this is possible in models with either higher spatial dimensionality  or with long-range interactions.
In these cases, a non-vanishing order parameter may persist in thermal as well as in pre-thermal phases. %, in which case the transition observed by changing the system parameters would be truly dynamical.

The simplest instance of the generic Hamiltonian~\eqref{eq:model} is that %of the  infinite-dimensional model, which is equivalent 
with $J_r = \lambda/N$, corresponding to the infinite-range or fully-connected model~\cite{LMGref,das06} %of the hamiltonian~\eqref{eq:model}, 
\beq
\label{eq:MFH}
H = - \frac{\lambda}{N} \sum_{i,j=1}^N \sigma_i^x \sigma_j^x - g \sum_{i=1}^N \sigma_i^z,
\eeq
where each of the $N$ spins interacts with all the others  with the same ferromagnetic coupling strength, $\lambda/N$.
This is equivalent to the Lipkin-Meshkov-Glick model~\cite{LMGref}.
The rest of Sec.~\ref{sec:MF} is devoted to reviewing the equilibrium and non-equilibrium behavior of this paradigmatic model, focusing on dynamical phase transitions after a quench. The readers familiar with this may skip to Sec.~\ref{sec:methodweakfluct}, in which we discuss fluctuations in the presence of perturbations.

The $1/N$ scaling of %the infinite-range interaction coupling 
$J_r$ in Eq.~\eqref{eq:MFH} is necessary in order to make the energy extensive in the  thermodynamic limit.
As $N\to\infty$ the mean-field approximation  becomes exact for the Hamiltonian~\eqref{eq:MFH}, and therefore the model is exactly solvable in the thermodynamic limit. Indeed, $H$ is solely a function of the total spin components
\beq
\label{eq:quantumH}
H = - \frac{\lambda}{N} ( \tilde{\sigma}_{k=0}^x )^2  - g \, \tilde{\sigma}_{k=0}^z ,
\eeq
where $\tilde{\sigma}_{k=0}^{\alpha}=\sum_i \sigma_i^{\alpha}$ is the Fourier mode with zero momentum $k=0$ of the spins on the lattice. 
All the other degrees of freedom $\tilde{\sigma}_{k\neq0}^{\alpha}$, corresponding to the spatial fluctuations with $k\ne0$ in Fourier space of the spins,  do not contribute to the dynamical properties of the model~\eqref{eq:quantumH}.

The Hamiltonian $H$ is diagonalizable separately in each sector of fixed total spin magnitude $%\big\lvert \tilde{\vec{\sigma}}_{k=0}  \big\rvert^2 = 
(Ns - m)(Ns-m+1)$, with $m=0,1,\dots,Ns$ or $Ns-1/2$ (depending on $Ns$ being integer or half integer respectively). When $N\to\infty$, these sectors can be labelled by a continuous variable,
\beq
 \big\lvert \tilde{\vec{\sigma}}_{k=0}  \big\rvert / N \to \rho,
\eeq
with $0\le\rho\le1$. 
The  ground state always belongs to the maximal total spin sector, $\big\lvert \tilde{\vec{\sigma}}_{k=0} \big\rvert = N$ or $\rho=1$. Accordingly, this state has extensive quantum numbers, and  the thermodynamic limit $N\to\infty$ is equivalent to the semiclassical limit, or, in loose terms, to a classical, continuous spin $\vec{\sigma} \equiv \langle \tilde{\vec{\sigma}}_{k=0} \rangle / N$ of (conserved) length $\rho$. The behavior of the system in that limit is then completely determined by the classical Hamiltonian
\beq
\label{eq:classicalH}
%\frac{H}{N} \quad \leadsto \quad 
\mathcal{H}_{\text{cl}}(\vec{\sigma}) = - \lambda ( \sigma^x )^2  - g \sigma^z ,
\eeq
corresponding to the quantity $H/N$, where $\vec{\sigma}$ is now a classical spin, its phase space being the surface of a sphere of radius $0<\rho\le1$. %; the case $\rho=1$ corresponds to the sector of maximal total spin length. 
%
%
%
%
%As long as we are concerned with equilibrium physics, the
The rigorous version of this statement is the following: When $N\to\infty$, the ground state expectation value of the spins $\Braket{\vec{\sigma}_i}$ is given by the minimum point of $\mathcal{H}_{\text{cl}}$ on the sphere, while its non-equilibrium evolution $\Braket{\vec{\sigma}_i(t)}$ with a possibly time-dependent field $g(t)$, starting from a fully polarized state, is given by the corresponding classical trajectory on the sphere governed by $\mathcal{H}_{\text{cl}}$ via the equations of motion $\dot{\sigma}^{\alpha} = \{ \sigma^{\alpha}, \mathcal{H}_{\text{cl}} \}$, with $\{ \sigma^{\alpha}, \sigma^{\beta} \} = \varepsilon^{\alpha\beta\gamma} \sigma^{\gamma}$ and with time rescaled by $s$. 

\begin{figure}[tb]
\centering
\includegraphics[width=0.6\columnwidth]{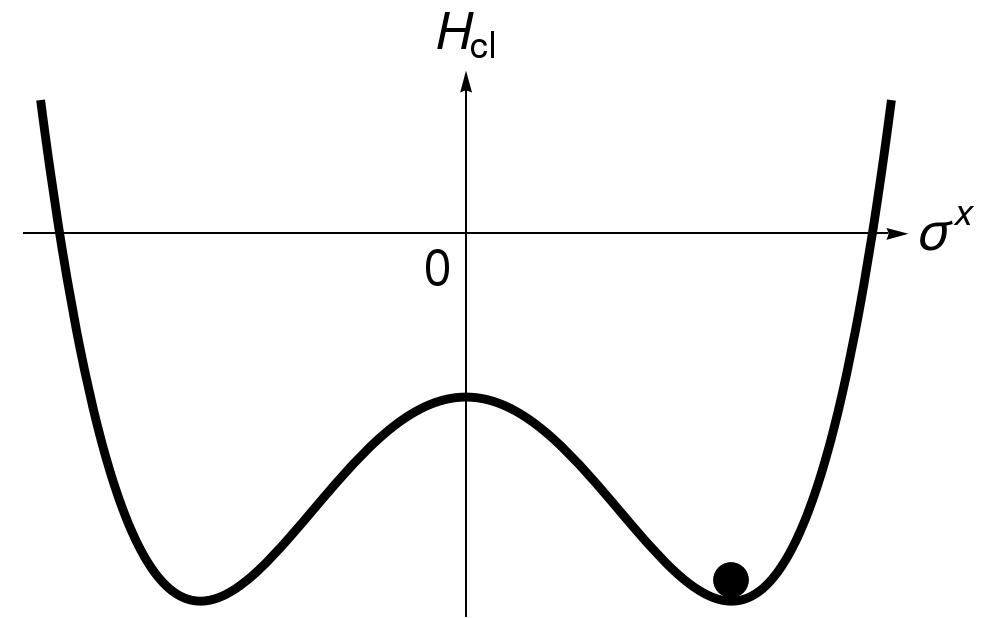}
\caption{
Classical energy landscape \eqref{eq:classicalH} of the collective spin $\vec{\sigma}$ of the LMG model along the plane $\sigma^y\equiv0$ as a function of the magnetization $\sigma^x$, %with $\sigma^z = \sqrt{\rho^2 - (\sigma^x)^2}$.
in the ferromagnetic phase $0<g<g_{\text{cr}}\equiv 2\lambda\rho$. The location of the two symmetric minima is determined by Eq.~\eqref{eq:minima}. In the thermodynamic limit, the degenerate ground state wavefunctions of the collective spin are localized at the two classical minima respectively, and $\vec{\sigma}$ behaves like a classical particle at rest at the bottom of one of the two wells (e.g., black dot in the figure). At finite size, however, quantum tunneling induced by the presence of the other well occurs over an exponentially long time scale, see Sec.~\ref{sec:finitesizeLMG}.
%If the system is prepared in one of the two ferromagnetic ground states corresponding to a certain value $0<g_0<g_{\text{cr}}$ of the magnetic field, and the latter is then suddenly quenched to a different value $0<g<g_{\text{cr}}$, depending on the strength $g-g_0$ of the quench, the resulting non-equilibrium trajectory may display dynamical ferromagnetic or paramagnetic behavior, exemplified by the blue and green line, respectively, separated by a critical trajectory with a diverging period, represented by the red line and corresponding to the \textit{dynamical critical point}. For further details, see Fig.~\ref{fig:MFDPT}.
}
\label{fig:dw}
\end{figure}

\subsection{Equilibrium behavior}

%\revcomment{WE COULD CUT ALL THIS EQUILIBRIUM SECTION. I SUGGEST NO, THOUGH. (IT'S 1.5 PAGES)}

\begin{figure}[tb]
\centering
\begin{tabular}{cc}
\includegraphics[scale=0.06]{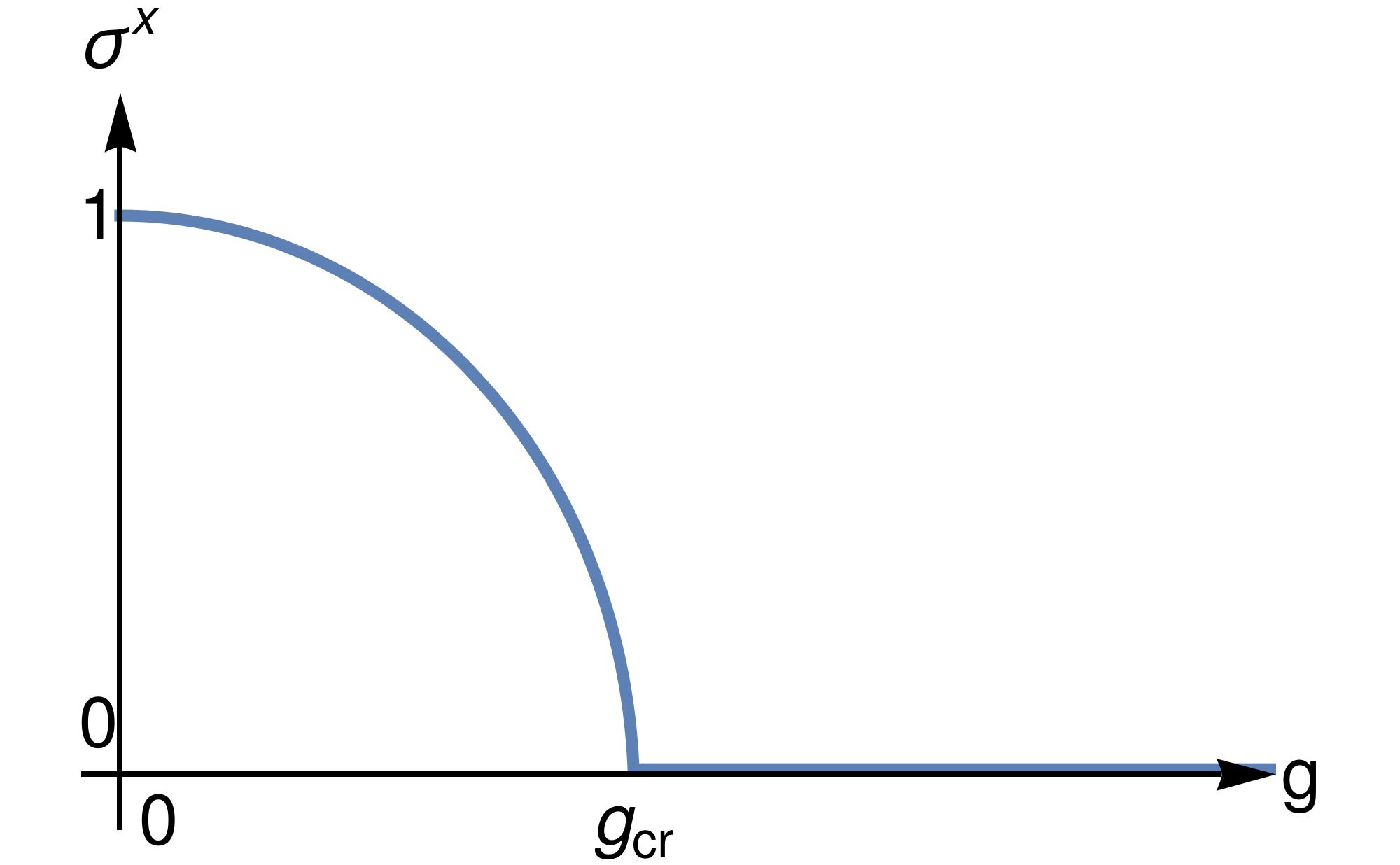} & 
\includegraphics[scale=0.06]{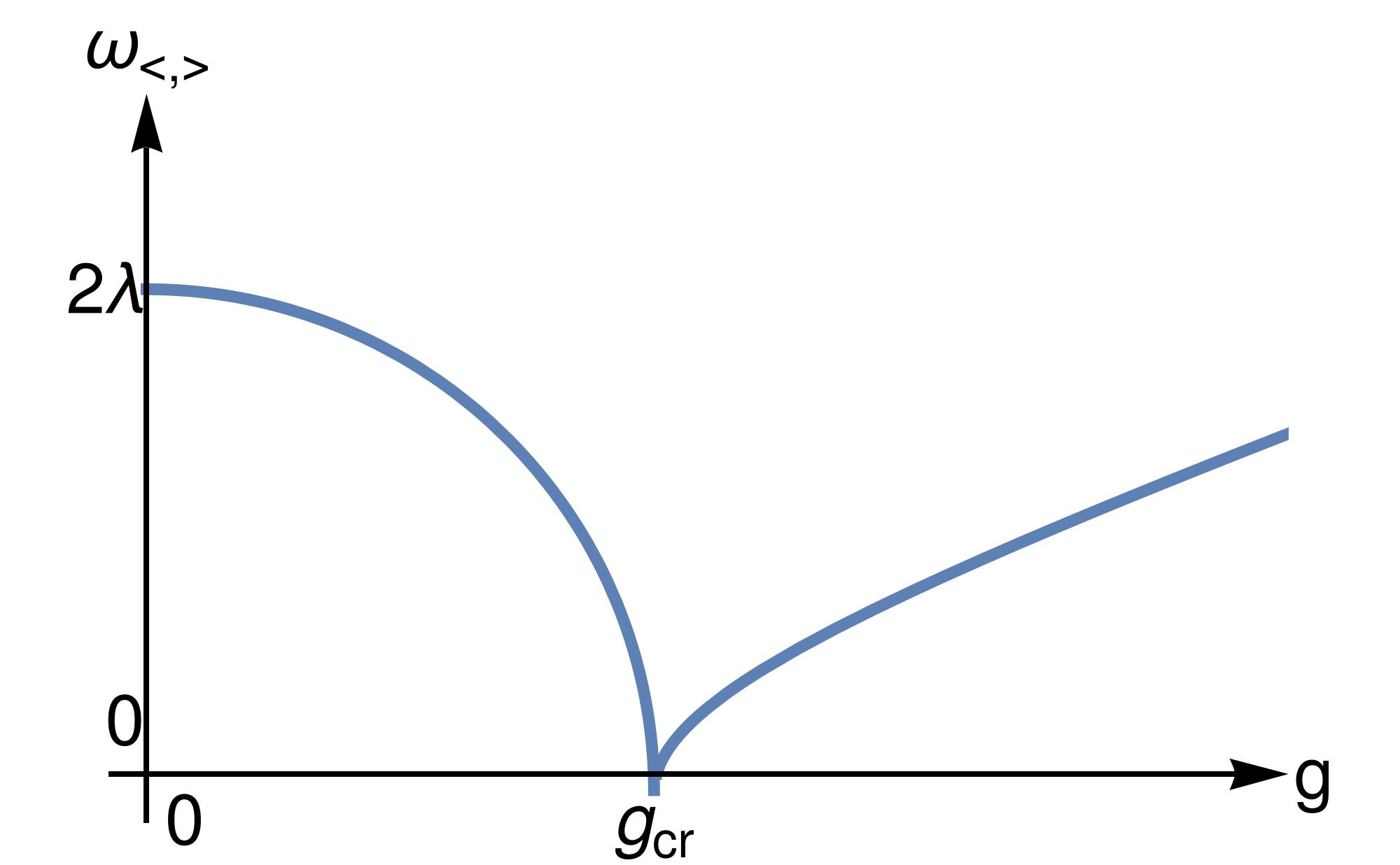}
\end{tabular}
\caption{Left panel: Equilibrium order parameter $\sigma^x$ of the infinite-range Ising model at zero-temperature as a function of the external field $g$, determined by Eq.~\eqref{eq:minima}. %: $\sigma^x = \pm\sin\theta^* \propto \sqrt{2\lambda\rho - g} = \sqrt{g_{cr}-g}$. 
Right panel: Frequency $\omega_{<,>}$ of small oscillations of the collective spin around the minimum, see Eqs.~\eqref{eq:MFsmallosc>} and \eqref{eq:MFsmallosc<}, equal to the energy gap above the ground state. %(or mass gap of the zero-mode); 
In both cases, the critical behavior is characterized by a square root singularity. }
\label{fig:MFequilibrium}
\end{figure}

%From a microscopic standpoint, the state of the system in the maximal total spin sector is well-described at all times by a spin-coherent state $\Ket{\nearrow \nearrow \dots \nearrow}_{\vec{\sigma}(t)}$, where all the spins are constantly aligned with each other, and pointing  in the direction of the instantaneous classical spin $\vec{\sigma}(t)$. Systematic finite-size corrections in powers of $1/N$ can be worked out as semiclassical corrections to the classical limit of the solution to the time-dependent Schr{\oe}dinger.

%\subsection{Ferromagnetic transition at equilibrium}
%\label{sec:ferro}

For a given sphere radius $\rho$, the classical Hamiltonian~\eqref{eq:classicalH} has a single minimum for large $g$ with $\sigma^x=\sigma^y=0$, $\sigma^z=\rho$, corresponding to a paramagnetic phase. As the strength of the field decreases below the critical value \beq g_{\text{cr}}\equiv 2\lambda\rho,\eeq that minimum \emph{bifurcates} into a pair of minima characterized by non-vanishing, opposite magnetizations $\sigma^x$ along the $x$-direction, located on the $xz$-plane symmetrically with respect to the inversion of the $x$-axis [i.e., connected by the  $\mathbb{Z}_2$ symmetry of the Hamiltonian~\eqref{eq:MFH}]. The corresponding double-well energy landscape is represented in Fig.~\ref{fig:dw}. Parameterizing $\vec{\sigma}$ with spherical angles $(\theta,\phi)$, i.e., as $\vec{\sigma} = \rho (\sin\theta\cos\phi,\sin\theta\sin\phi,\cos\theta)$, the two ferromagnetic minima are given by $(\theta^*,0)$ and $(\theta^*,\pi)$, with
\beq
\label{eq:minima}
\cos \theta^* =  \frac{g}{g_{\text{cr}}}. %, \qquad \phi^*_1=0, \;\phi^*_2=\pi .
\eeq
Accordingly, the value of order parameter is \beq \sigma^x = \pm \rho \sin\theta^* = \pm \rho\sqrt{1-(g/g_{\text{cr}})^2}\eeq  (see the left panel of Fig.~\ref{fig:MFequilibrium}). %The energy landscape of the collective spin is therefore shaped as a double well, see Fig.~\ref{fig:dw}.
%This zero-temperature transition realizes the standard mean-field theory of ferromagnetism as surmised by Landau and Ginzburg. Even though it is a ground state transition, it is fully encoded by the classical limit~\eqref{eq:classicalH} of the Hamiltonian.

Let us now determine  the spectrum of the lowest excitations above the ground states discussed above. 
Within each sector with fixed value of the total spin magnitude, labelled by $\rho$, the quantum mechanics of the collective spin is equivalent to that of a quantum particle in a potential well whose depth grows proportionally to $N$. (The absolute ground state sector corresponds to $\rho=1$.) In the thermodynamic limit, the lowest excitations of this particle are harmonic, and are determined by the quadratic expansion of the Hamiltonian around its energy minimum(a).
This can be seen by a simple Holstein-Primakoff transformation, as we discuss below.

For $g>g_{\text{cr}}$ the minimum occurs at $\theta=0$,  and in terms of tangent canonical coordinates $q,p$,  with $[q,p]=i$, the quantum fluctuations around that minimum take the form
\beq\label{eq:qflc}
\begin{split}
 s \, \tilde{\sigma}_{k=0}^z &= Ns \rho - n = Ns\rho - \frac{q^2+p^2-1}{2}, \\
 s \,\tilde{\sigma}_{k=0}^x &\approx \sqrt{Ns\rho} \; q, \\
 s \,\tilde{\sigma}_{k=0}^y &\approx \sqrt{Ns\rho} \; p.
 \end{split}
\eeq
The quantum number $n=0,1,2,\dots$ labels the quantized spin projection along the direction of the minimum. 
The Hamiltonian~\eqref{eq:quantumH} becomes, using Eqs.~\eqref{eq:qflc},
\beq
\label{eq:expansionH>}
\begin{split}
H_> &= - N g \rho + \frac{g}{s} \frac{q^2+p^2-1}{2} -  \frac{\lambda \rho}{s} q^2 = \\
   &=  - N g \rho + \frac{1}{s} \bigg( \frac{\omega_>-\omega_{>}^{(0)}}{2}\bigg) +  \frac{1}{s} \omega_> \; n ,
\end{split}
\eeq
 where
\beq
\label{eq:MFsmallosc>}
\omega_{>}=\sqrt{g(g-g_{\text{cr}})}, \qquad \omega_{>}^{(0)} = g.
\eeq
The first term in the last line of Eq.~\eqref{eq:expansionH>} represents the classical energy [compare with Eq.~\eqref{eq:classicalH}], the second one is the quantum zero-point energy contribution, i.e., the energy increase due to the quantum fluctuations of the spin around the classical minimum configuration, while the last one is the energy of the elementary (harmonic) excitations, with $n=0,1,2,\dots$.

For $g<g_{\text{cr}}$, the two minima of the classical Hamiltonian are determined by Eq.~\eqref{eq:minima}. Introducing the canonical coordinates given by the total spin projection $P$ along $z$ and the conjugated angle $Q$,
\beq
\label{eq:globalPQ}
\begin{split}
P&=s \tilde{\sigma}_{k=0}^z=Ns \rho \cos\theta, \\
Q&=\phi,
 \end{split}
\eeq
respectively, 
and expanding the Hamiltonian~\eqref{eq:quantumH} around one of the two classical minima $(\theta^*,0)$ or $(\theta^*,\pi)$ with $\theta^*$ given by Eq.~\eqref{eq:minima} (by symmetry the excitations spectra  are identical) \[P = Ns\rho\cos\theta^* + \delta P, \qquad Q = \phi^*+ \delta Q ,\] we get %the harmonic excitation tower
\beq
\label{eq:expansionH<}
\begin{split}
H_<  = & - N \Big( g \rho \cos\theta^* + \lambda \rho^2 \sin^2\theta^*\Big)  \\
       &
- \frac{1}{2s} \Big(
 g \cos\theta^* + 2 \lambda \rho  \sin^2\theta^*
 \Big)
 \\
     & + \frac{2\lambda}{s} \bigg[ \frac{1}{Ns} \frac{(\delta P)^2}{2} + N s \rho^2 \sin^2 \theta^* \frac{(\delta Q)^2}{2}\bigg] 
  \\
    = &  - N \bigg(\frac{g^2}{4\lambda} + \lambda \rho^2 \bigg) 
   + \frac{1}{s} \bigg( \frac{\omega_<-\omega_{<}^{(0)}}{2}\bigg)
   + \frac{1}{s}   \omega_< \; n ,
\end{split}
\eeq
where
\beq
\label{eq:MFsmallosc<}
\omega_{<}=\sqrt{g_{\text{cr}}^2 - g^2}, \qquad \omega_{<}^{(0)} = g_{\text{cr}} .
\eeq
Analogously to Eq.~\eqref{eq:expansionH>}, the first term in the last line of Eq.~\eqref{eq:expansionH<} represents the classical energy [compare with Eq.~\eqref{eq:classicalH}], the second one is the quantum zero-point energy contribution, i.e., the energy increase due to the quantum fluctuations of the spin around the classical minimum configuration, while the last one is the energy of the elementary (harmonic) excitations, with $n=0,1,2,\dots$.
%The first term in \eqref{eq:expansionH<} is the purely classical energy (compare with Eq.\eqref{eq:classicalH}), while the last one is the energy of the elementary (harmonic) excitations, with $n=0,1,2,\dots$.
We observe that the energy gap above the ground state closes at the equilibrium critical point $g=g_{\text{cr}}$, with a mean-field critical exponent $1/2$ (see Fig.~\ref{fig:MFequilibrium}).

%The above discussion captures the leading order in $1/N$ in the thermodynamic limit, in which the behavior of the system is classical. A thorough discussion of the finite-size effects, i.e., of the quantum corrections to the classical behavior for finite $N$, can be found in Section \ref{sec:cat}.

%The expansion above captures the leading order in $1/N$. 
In principle, one could think of including modes at finite $k\neq0$ (spin waves) which would however, in this limit, be decoupled from the dynamics of the zero-mode. If $N_{\text{sw}}=0,1,2,\dots$ is the total occupation number of the spin wave modes with $k\ne0$, the collective spin magnitude is $ %s \big\lvert \tilde{\vec{\sigma}}_{k=0} \big\rvert = 
(Ns - N_{\text{sw}}) (Ns - N_{\text{sw}}+ 1) $, i.e.,
\beq
\label{eq:epsilonNsw}
\rho = \frac{\big\lvert \tilde{\vec{\sigma}}_{k=0} \big\rvert}{N} = %1-\epsilon = 
1- \frac{N_{\text{sw}}}{Ns}.
\eeq
Hence one finally obtains from Eqs.~\eqref{eq:MFsmallosc>} and \eqref{eq:MFsmallosc<} the complete spectrum of excitations above the ground state,  in the thermodynamic limit and to leading order in $N_{\text{sw}}$, % $\epsilon$,
\beq
\label{eq:spectrum}
\begin{split}
H_> &= - N g +   \frac{\omega_>-\omega_{>}^{(0)}}{2s} + \frac{1}{s} \big( g  N_{\text{sw}} + \omega_> \, n\big), \\
H_< &= - N \bigg(\frac{g^2}{4\lambda}   + \lambda \bigg) 
+  \frac{\omega_<-\omega_{<}^{(0)}}{2s}
+ \frac{1}{s} \big( 2\lambda  N_{\text{sw}} + \omega_< \, n \big),
\end{split}
\eeq
valid for $g>2\lambda$ and $g<2\lambda$, respectively.
All the spin wave excitations introduced above have finite gap $g/s$ or $2\lambda/s$ and a flat dispersion relation independent of the wavevector $k\ne0$, because the fully-connected interactions carry no information on spatial scales, hence cannot resolve finite wavelengths. As a consequence, the presence of a finite low temperature $T$ leads to exponentially small corrections to the order parameter, with a shift of the critical point that can be computed~\cite{das06} by minimizing the mean-field classical Hamiltonian~\eqref{eq:classicalH} with
\beq
\rho(T) =1 - \frac{1}{s} \; \frac{1}{e^{\frac{2\lambda/s}{T}}-1}.
\eeq

\subsection{Dynamical criticality}

\begin{figure}[tb]
\centering
\includegraphics[width=0.9\columnwidth]{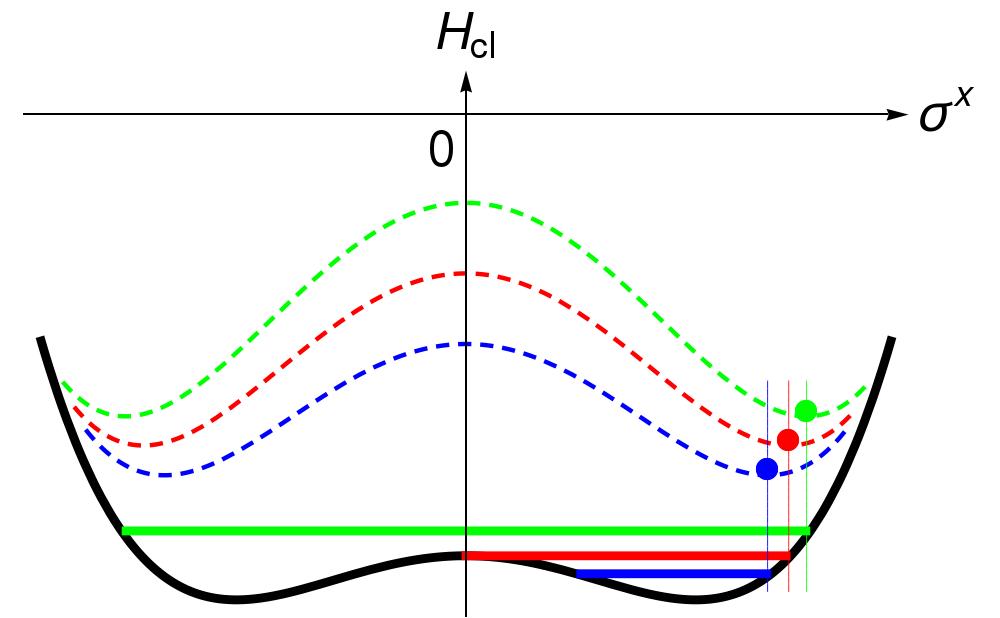}
\caption{
[Color online] Classical energy landscapes \eqref{eq:classicalH} of the collective spin $\vec{\sigma}$ of the LMG model in the plane $\sigma^y\equiv0$ as a function of the magnetization $\sigma^x$ %with $\sigma^z = \sqrt{\rho^2 - (\sigma^x)^2}$.
in the ferromagnetic phase, with a post-quench value $g$ such that $0<g<g_{\text{cr}}$ (black solid  line) and several possible pre-quench values $g_0$ such that $0<g_0<g$ (blue, red and green dashed lines) of the transverse magnetic field. 
%The location of the two symmetric minima is determined by Eq.~\eqref{eq:minima}. In the thermodynamic limit, the degenerate ground state wavefunctions of the collective spin are localized at the two classical minima respectively, and $\vec{\sigma}$ behaves like a classical particle at rest at the bottom of either well, as illustrated e.g. by the black disk. At finite size, however, quantum tunneling to the opposite well occurs over an exponentially long time scale, see Sec.~\ref{sec:cat}.
If the system is prepared in a %pre-quench ferromagnetic 
ground state, e.g., with positive magnetization as illustrated by the blue, red, and green dots for decreasing values of $g_0$, and the magnetic field is suddenly quenched to a larger value $g_0<g<g_{\text{cr}}$, then depending on the strength $g-g_0$ of the quench, the resulting non-equilibrium evolution may display dynamical ferromagnetic or paramagnetic behavior, exemplified by the blue and green line, respectively, separated by a critical trajectory with a diverging period, corresponding to the red line and associated with the \textit{dynamical critical point} $g=g_{\text{dyn}}$. %The vertical distance between a dot and the relative line of the same color represents the work done on the system through the corresponding quench. %Note that here $g$ is fixed and $g_0$ varies. 
In contrast to Fig.~\ref{fig:MFDPT}, here the various resulting evolutions correspond to varying the pre-quench parameter $g_0$, with a fixed post-quench value $g$. 
}
\label{fig:dyndw}
\end{figure}

%\centering
\begin{figure*}[t]
\begin{tabular}{ccc}
\includegraphics[width=0.27\textwidth]{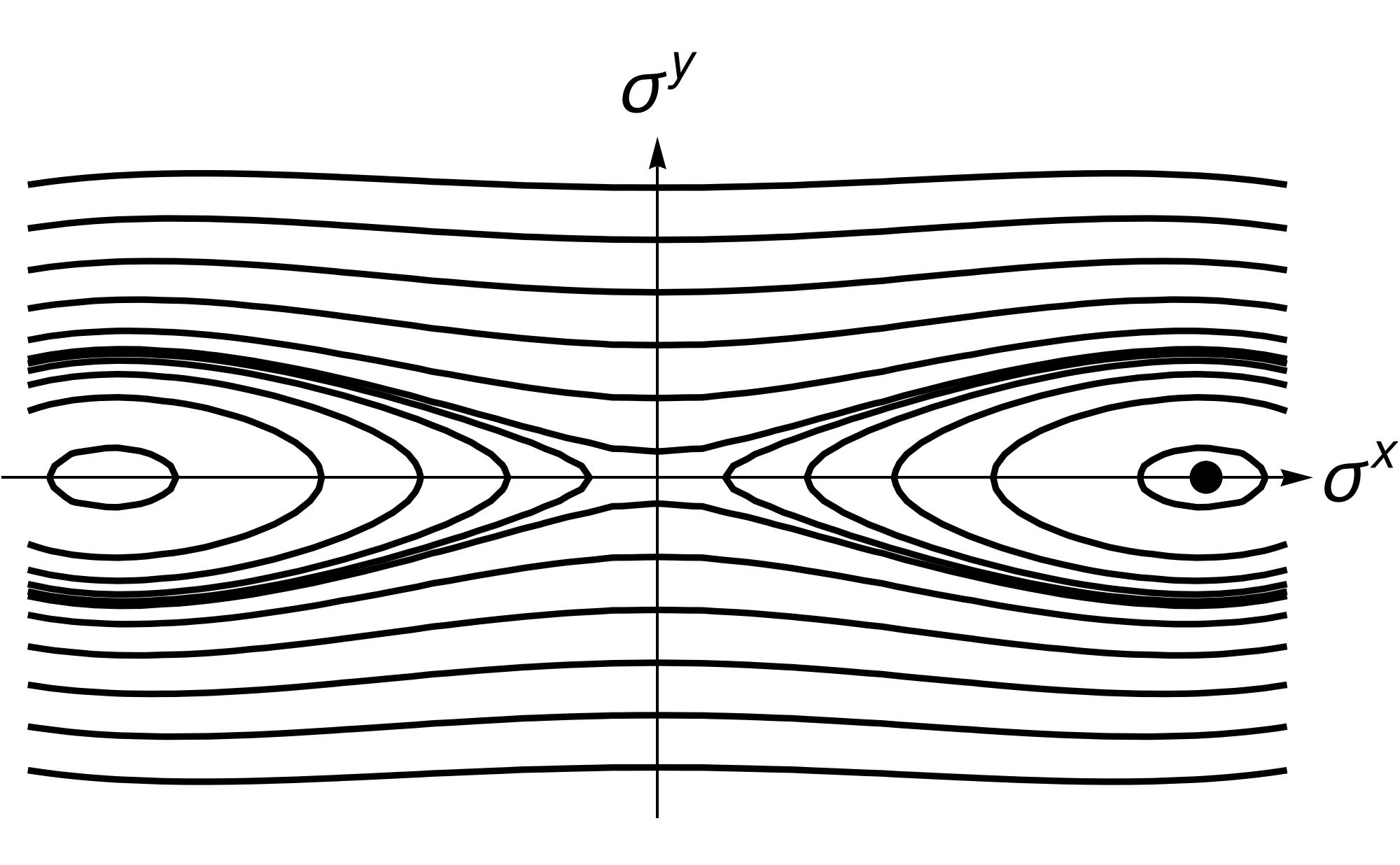} & \includegraphics[width=0.27\textwidth]{below.jpg}  & \includegraphics[width=0.27\textwidth]{below.jpg}  \\
\includegraphics[width=0.27\textwidth]{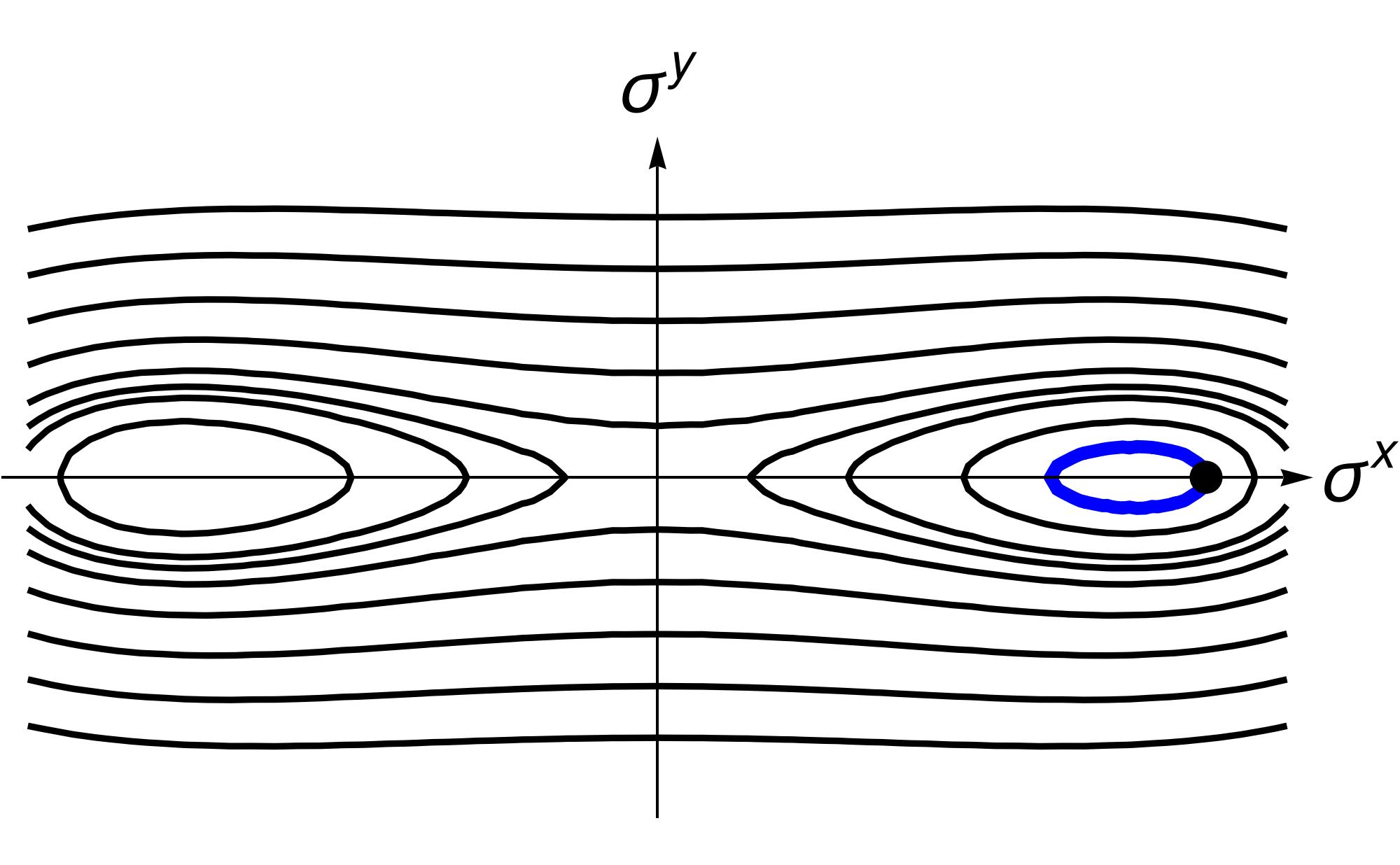} & \includegraphics[width=0.27\textwidth]{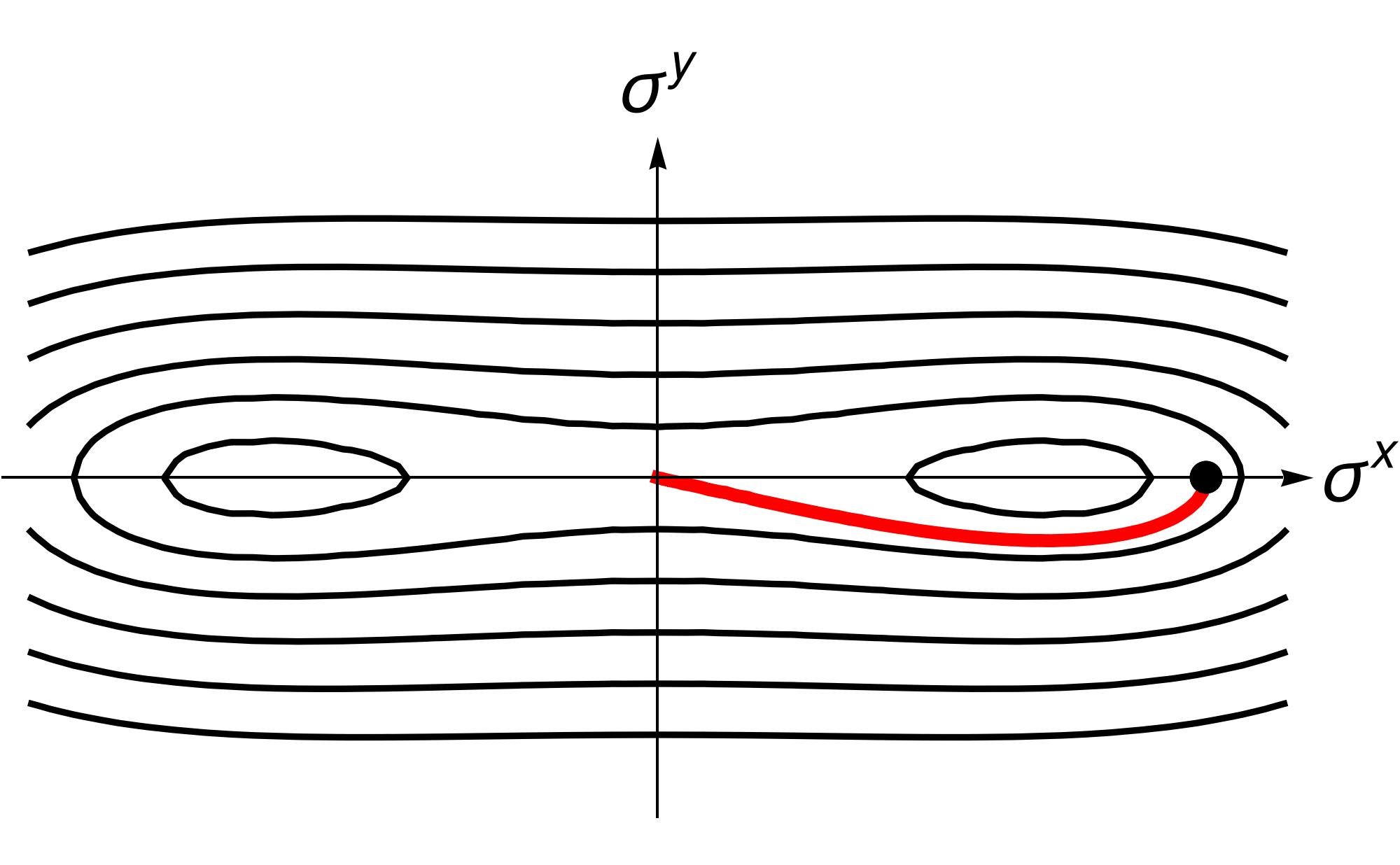}  & \includegraphics[width=0.27\textwidth]{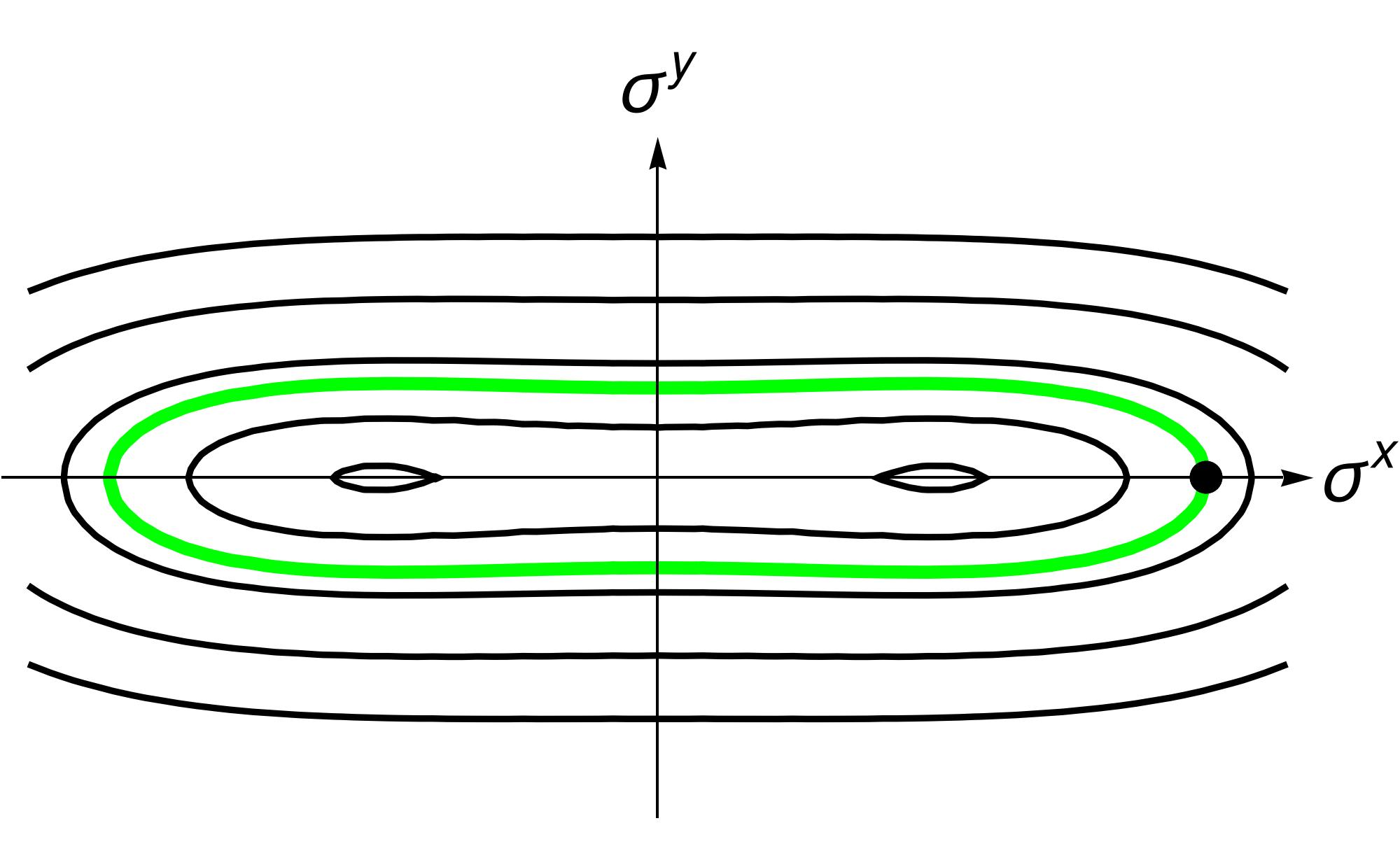}  \\
\includegraphics[width=0.25\textwidth]{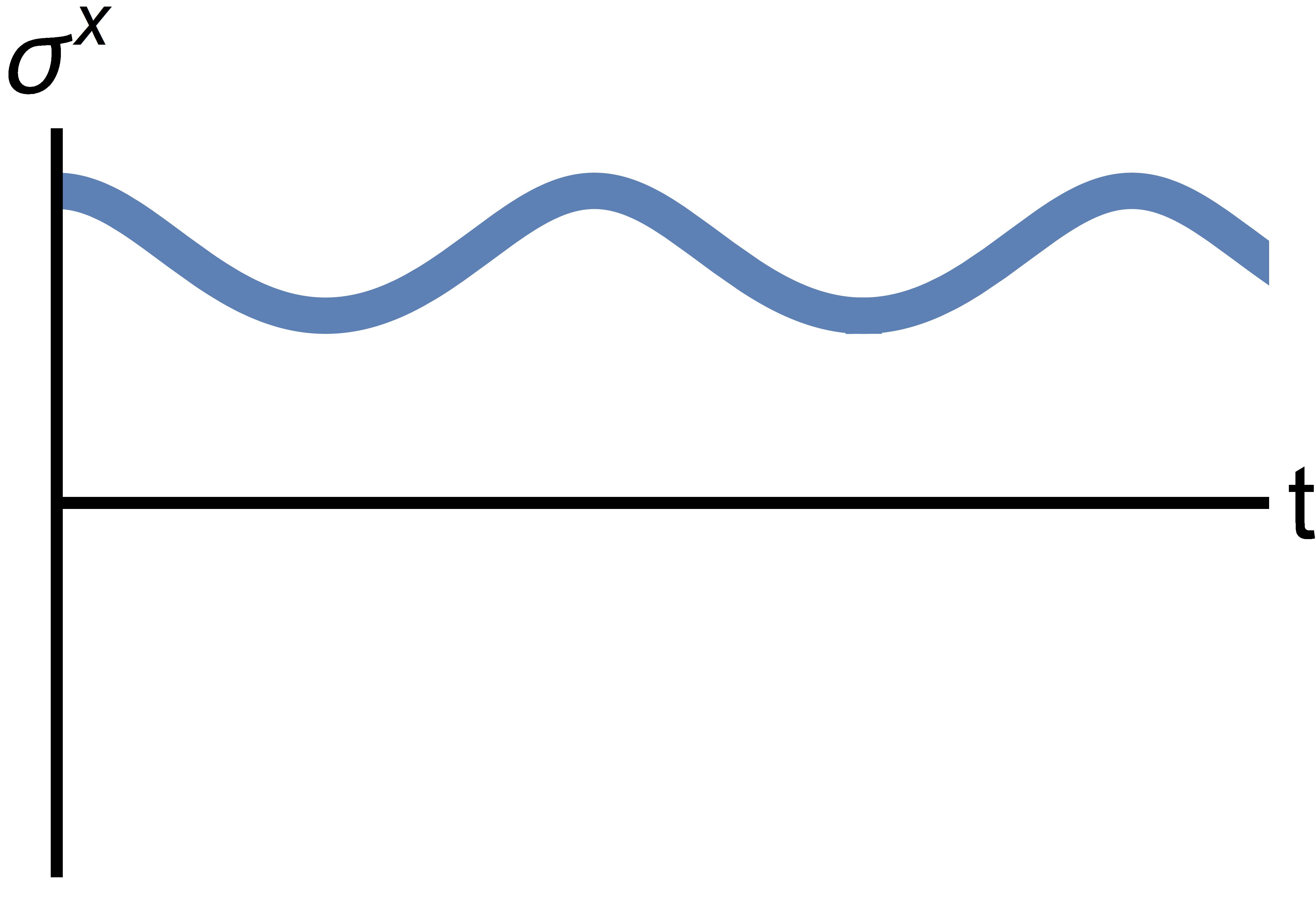} & \includegraphics[width=0.25\textwidth]{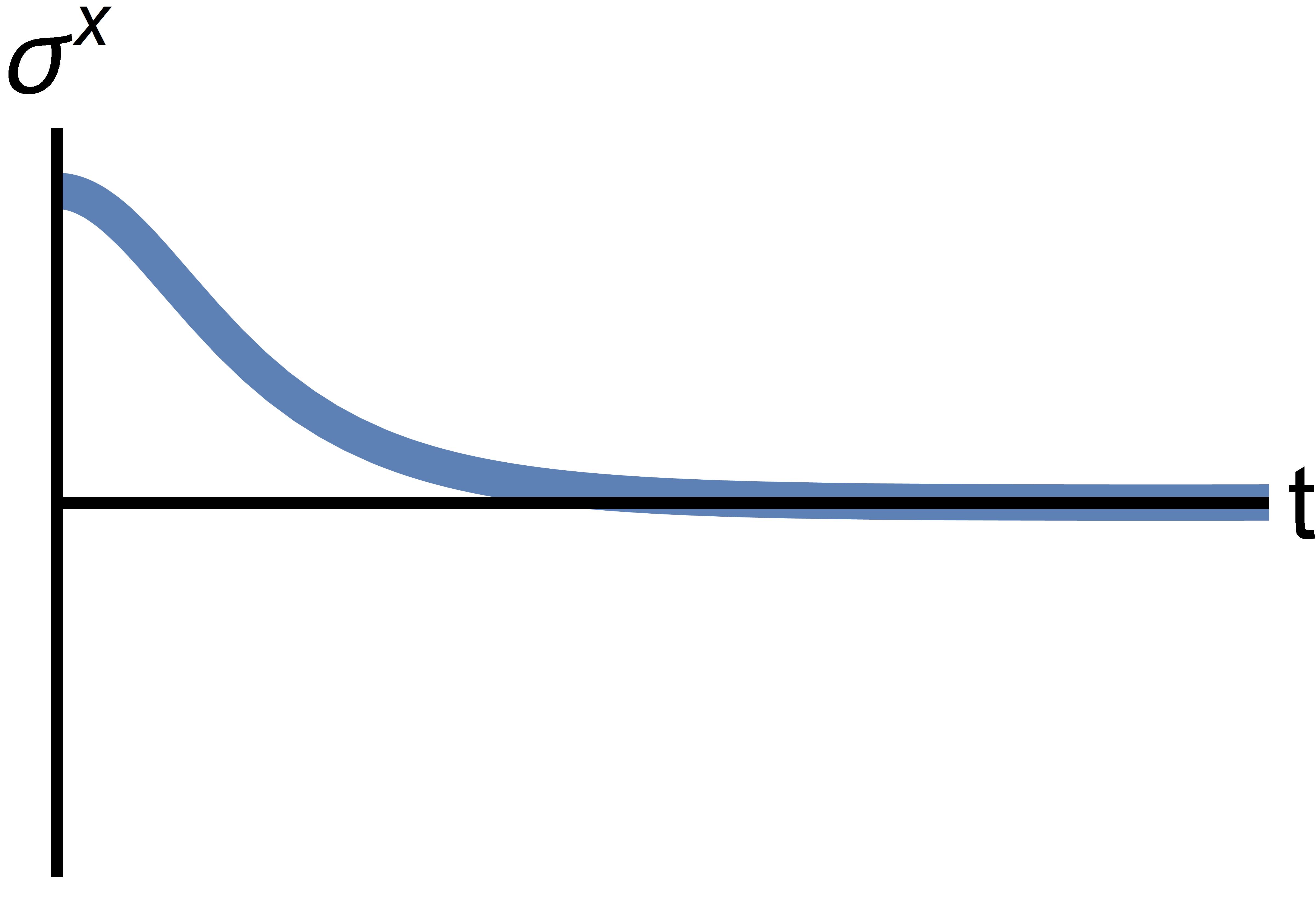}  & \includegraphics[width=0.25\textwidth]{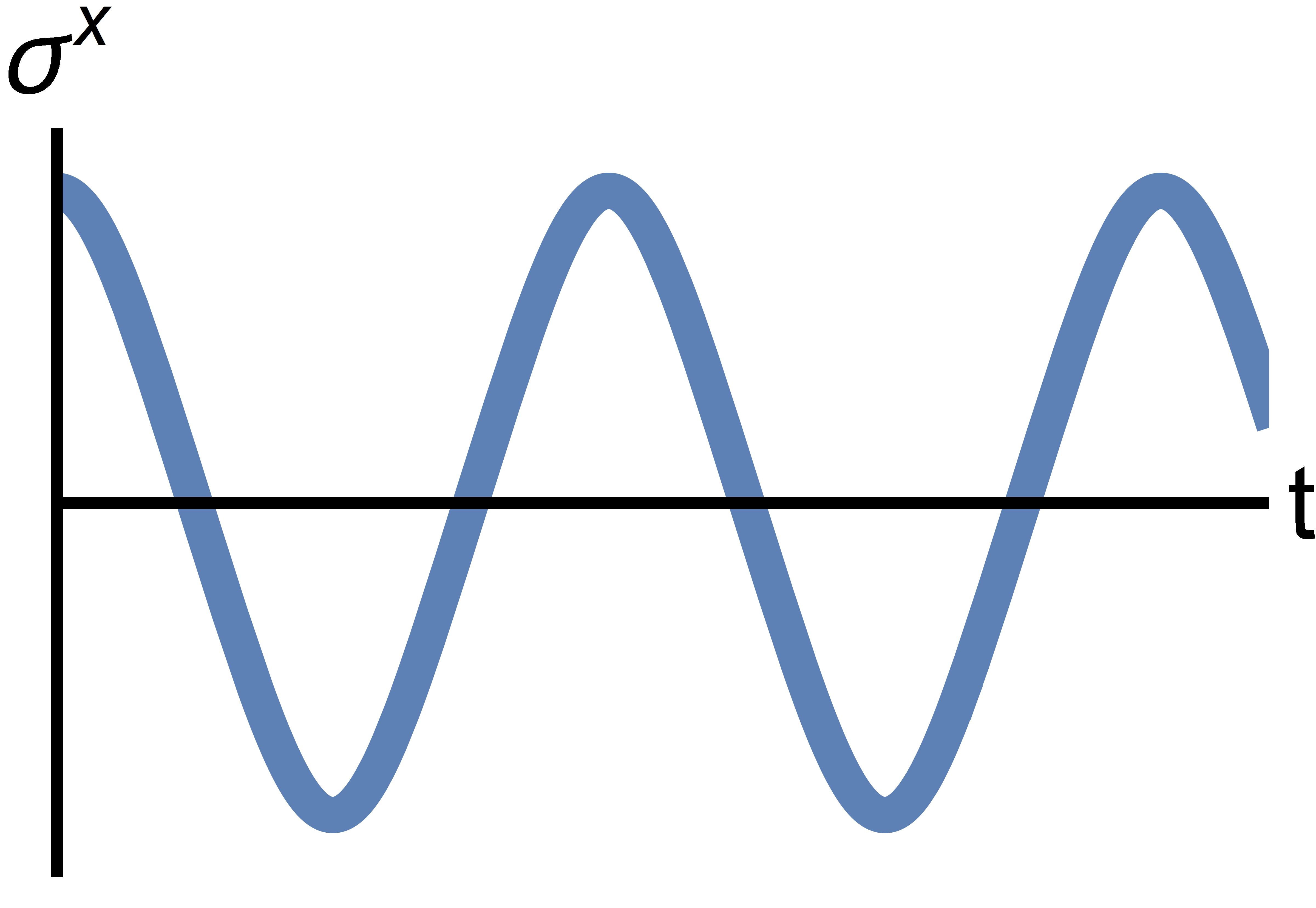}
\end{tabular}
\caption{
[Color online] Non-equilibrium dynamics of the %infinite-range Ising 
LMG model \eqref{eq:MFH} in the thermodynamic limit,
after a sudden quench $g_0 \rightarrow g$ of the transverse magnetic field starting from a ferromagnetic ground state of $H(g_0)$. 
The first row shows the semiclassical phase portrait of the pre-quench Hamiltonian $\mathcal{H}_{\text{cl}}(g_0)$, where the initial state is represented by one of the two minima.
The second row shows the semiclassical phase portrait of the post-quench Hamiltonian $\mathcal{H}_{\text{cl}}(g)$, where the initial state is no longer a stationary point but moves along a non-trivial non-equilibrium trajectory, in the three qualitatively different cases corresponding to $g<g_{\text{dyn}}$, $g=g_{\text{dyn}}$ and $g>g_{\text{dyn}}$ in the first, second and third column, respectively. 
The third row shows the dynamics of the order parameter as a function of time for the three cases.
First column: for a weak quench, the dynamics remain trapped within the starting ferromagnetic sector; second column: for the critical quench, the initial state lies on a separatrix of the post-quench Hamiltonian and its subsequent evolution approaches the unstable equilibrium point at infinite time; third column: for a strong quench, the semiclassical orbit encircles both ferromagnetic minima, hence the symmetry is dynamically restored and the time-averaged order parameter is zero. 
%Note that here $g_0$ is fixed and $g$ varies.
In contrast to Fig.~\ref{fig:dyndw}, here the different trajectories correspond to a varying post-quench parameter $g$, with a fixed pre-quench value $g_0$. 
%See also Fig.~\ref{fig:dyndw}.
} 
\label{fig:MFDPT}
\end{figure*}
%\end{widetext}

\begin{figure}[tb]
\centering
\begin{tabular}{cc}
\includegraphics[scale=0.06]{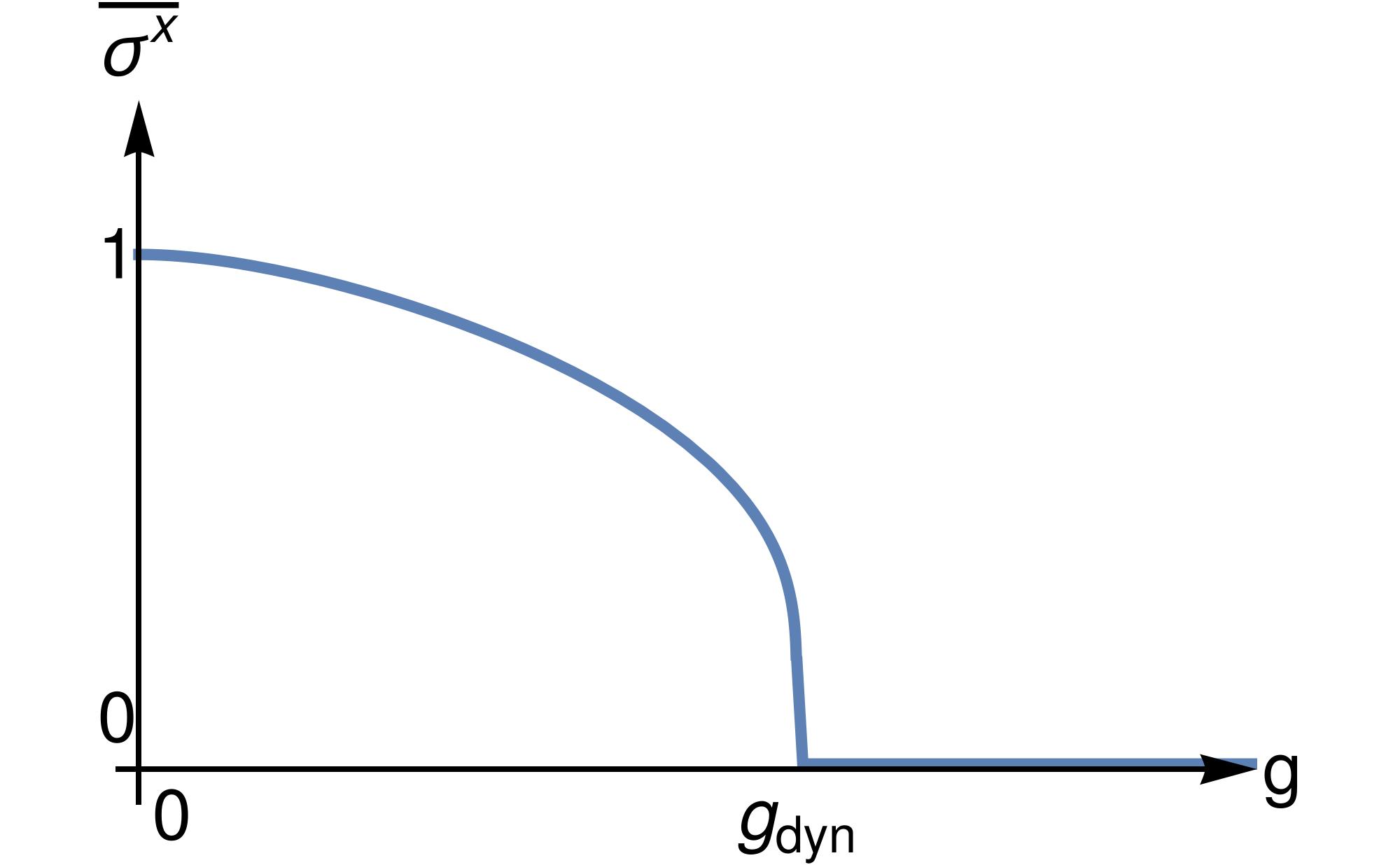} &
 \includegraphics[scale=0.06]{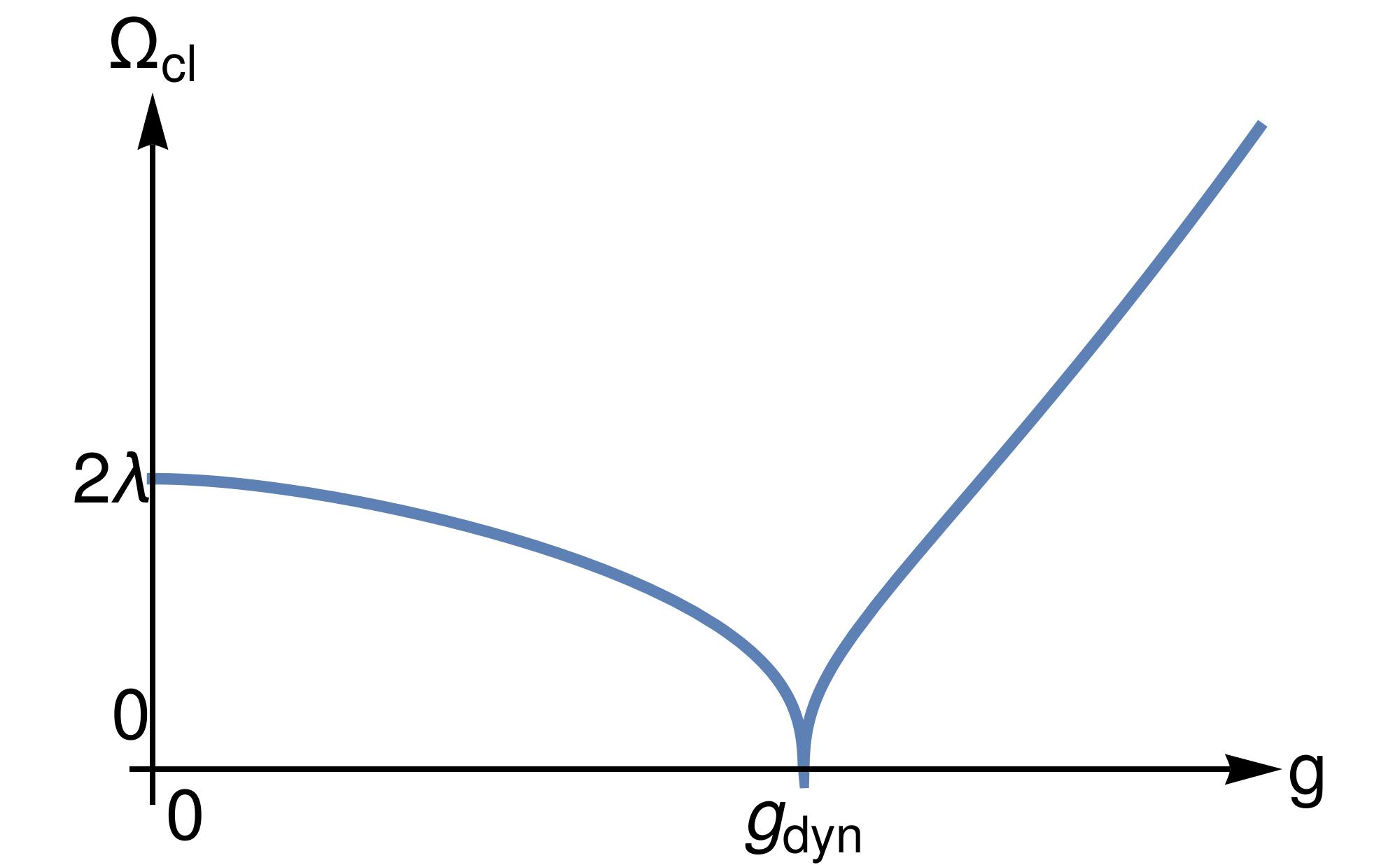}
\end{tabular}
\caption{Left panel: Non-equilibrium order parameter $\overline{\sigma^x}$, defined in Eq.~\eqref{eq:neqop}, of the infinite-range Ising model \eqref{eq:MFH} after a quench of the external magnetic field starting from a ferromagnetic ground state with $g_0=0$ and positive magnetization, as a function of the post-quench field $g$. Right panel: Classical frequency $\Omega_{\text{cl}}$ of the mean-field dynamical trajectory, which represents the characteristic time scale of the non-equilibrium evolution, as a function of the post-quench field $g$. For both quantities, the nature of the singular behavior at the dynamical critical point $g=g_{\text{dyn}}$ is logarithmic, as explained in the text. These plots can be compared with the analogous ones in equilibrium conditions in Fig.~\ref{fig:MFequilibrium}. }
\label{fig:dyncrit}
\end{figure}

After setting the stage, let us now focus on the dynamics of interest in this work.
Suppose that the system is prepared at time $t=0$ in a ferromagnetic ground state of the Hamiltonian~\eqref{eq:MFH} with a transverse field $g_0 < g_{\text{cr}} = 2\lambda$. Microscopically, this state is close to a spin-coherent state with all the spins aligned in the direction $(\theta^*,0)$ or $(\theta^*,\pi)$ with $\cos\theta^* = g/g_{\text{cr}}$, see Eq.~\eqref{eq:minima}, and with subextensive zero-point fluctuations of the collective spin, see Eq.~\eqref{eq:spectrum}. Then, the external field is suddenly increased to $g>g_0$,  faster than the typical timescale of the system's dynamics. As argued above, the spins will %retain almost perfect coherence in the non-equilibrium evolution, and
initiate a collective precession, and the evolution of their %collective spin 
direction on the sphere will be described by the classical trajectory of the post-quench Hamiltonian $\mathcal{H}_{\text{cl}}(g)$, with the initial data corresponding to the minimum of the pre-quench Hamiltonian $\mathcal{H}_{\text{cl}}(g_0)$, see Eq.~\eqref{eq:classicalH}.

%\begin{widetext}

Depending on the strength $g-g_0$ of the quench $g_0 \rightarrow g$ of the transverse field, starting from a ferromagnetic pre-quench Hamiltonian, the resulting dynamics display qualitatively different orbits \cite{SciollaBiroliMF,Zunkovic}, as shown in Figs.~\ref{fig:dyndw} and~\ref{fig:MFDPT}:
\begin{enumerate}
\item For a shallow quench [$g<g_{\text{dyn}}\equiv (g_0 + g_{\text{cr}})/2$], the post-quench energy remains below the top of the barrier that separates the two ferromagnetic sectors. Correspondingly, the spin will precess within the starting ferromagnetic sector (blue lines in Figs.~\ref{fig:dyndw} and \ref{fig:MFDPT}). 
\item As the strength of the quench increases, the precession period $T_{\text{cl}}=2\pi/\Omega_{\text{cl}}$ (which depends on both $g_0$ and $g$) increases, until for $g \nearrow g_{\text{dyn}} $ it takes an infinite time to complete one cycle, and the unstable point at the top of the energy barrier is approached exponentially fast along the classical separatrix (red lines in Figs.~\ref{fig:dyndw} and~\ref{fig:MFDPT}). %\review{which separatrix? comment more about it}. 
\item For deep quenches above this threshold $g>g_{\text{dyn}}$, the corresponding post-quench energy is larger than the barrier and the orbit of the collective spin on the sphere encircles both minima, such that the symmetry is \emph{dynamically} restored after taking time-averages. 
\end{enumerate}
In fact, the time-average 
\beq
\label{eq:neqop}
\overline{\sigma^x} = \lim_{T\to\infty} \frac{1}{T} \int_0^T dt \, \sigma^x(t)
\eeq
of the equilibrium order parameter $\sigma^x$
as a function of the quench strength, vanishes abruptly at the dynamical critical value $g_{\text{dyn}}$ of the transverse field  which depends also on the initial condition. % , where $\Omega_0(g\to g_{cr})\to0$), 
This dynamical critical point separates a \emph{dynamical ferromagnetic phase} with $\overline{\sigma^x}\ne0$ from a \emph{dynamical paramagnetic phase} with $\overline{\sigma^x}=0$.  %The dynamical critical point is determined by the initial condition and is given by $g_{cr} = (g_0 + g_{\text{cr}})/2 = \lambda + g_0/2$.

The vanishing of an order parameter and the divergence of a characteristic time scale such as those reported in Fig.~\ref{fig:dyncrit} are usually associated with critical phenomena.
However, the system under consideration is clearly out of thermal equilibrium, as all microscopic spins perform a coherent, undamped precession. For this reason the above phenomenology can be described as \emph{dynamical criticality}. 
In order to reinforce the idea that this behavior is distinct from the corresponding equilibrium phase transition, we emphasize that the equilibrium singularity of the order parameter upon approaching a critical point has a critical exponent $1/2$, see Fig.~\ref{fig:MFequilibrium}, whereas the non-equilibrium order parameter $\overline{\sigma^x}$ actually displays a \emph{logarithmic} singularity.
Indeed, the divergence of the period of the classical oscillations as $g \nearrow g_{\text{dyn}}$ is of the same form as that of a classical pendulum as the initial position approaches the upper configuration, with vanishing initial velocity \cite{landau}, and therefore the time average $\overline{\sigma^x}$ inherits the same type of singularity.

%
%\alessio{ change here} Derivation in brief:
%\begin{widetext}
%\begin{equation}
%\label{eq:logsing}
%T_{\text{period}}= 2\int_{q_1}^{q_2} \frac{dq}{v(q)} \thicksim \int_0^\Delta \frac{dq}{\sqrt{a(g-g_{cr})+b q^2}} \thicksim \int_0^{\Delta'/\sqrt{g-g_{cr}}} \frac{d\eta}{\sqrt{1+\eta^2}} \thicksim \log \frac{\Delta}{\sqrt{g-g_{cr}}},
%\end{equation}
%\end{widetext}
%
%and hence $\frac{1}{T}\int_0^T dt \, \sigma^x(t) \underset{T\to\infty}{\thicksim} \frac{c}{T_{\text{period}}}\thicksim \frac{c'}{\log (g-g_{cr})}$.

%The review of the non-equilibrium dynamics of the infinite-range Ising model by observing that 
The dynamical criticality thoroughly discussed here is not peculiar of the Ising ferromagnets or of  sudden quenches. Rather, it is a  general feature of mean-field models driven away from equilibrium \cite{SciollaBiroliMF,GambassiCalabrese}. If the driving is chosen to be a slow  ramp of the value of  $g$ instead of a sudden quench, the dynamical critical point retains its nature, although it gets shifted towards the equilibrium critical point, until it merges with the latter in the limit of adiabatic variation.

\subsection{Finite-size (quantum) corrections}
\label{sec:finitesizeLMG}

\begin{figure*}[t]
\centering
\begin{tabular}{cc}
\includegraphics[scale=0.18]{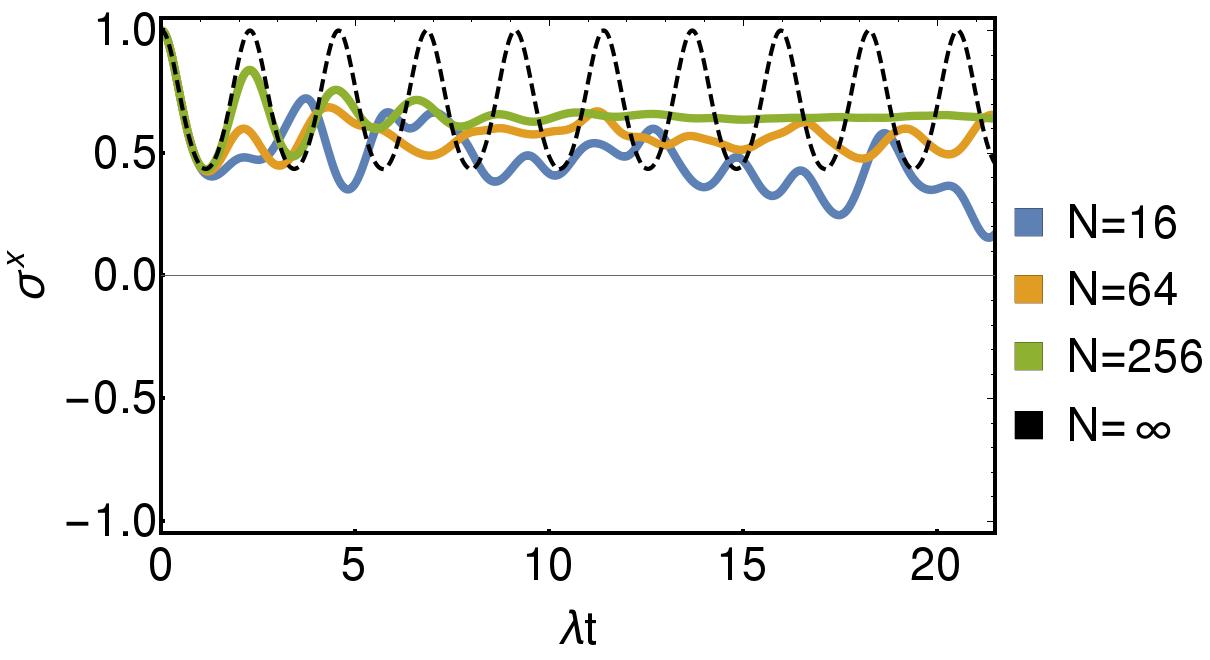} & \includegraphics[scale=0.17]{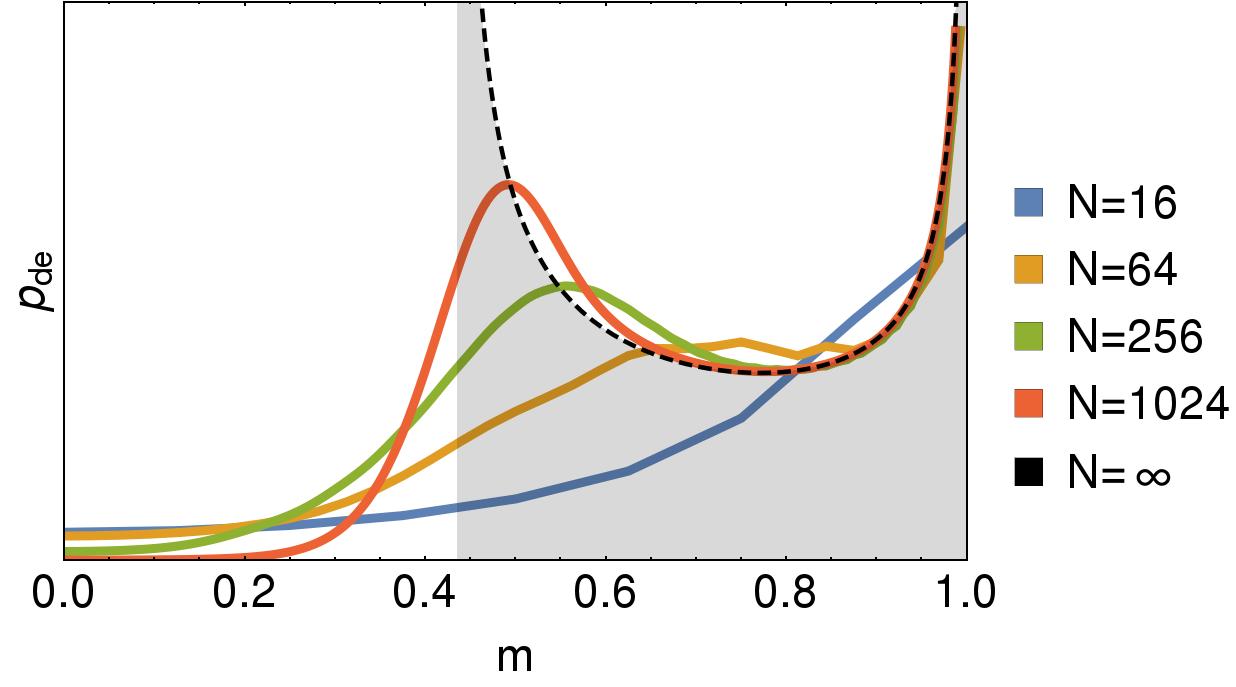} \\
\includegraphics[scale=0.18]{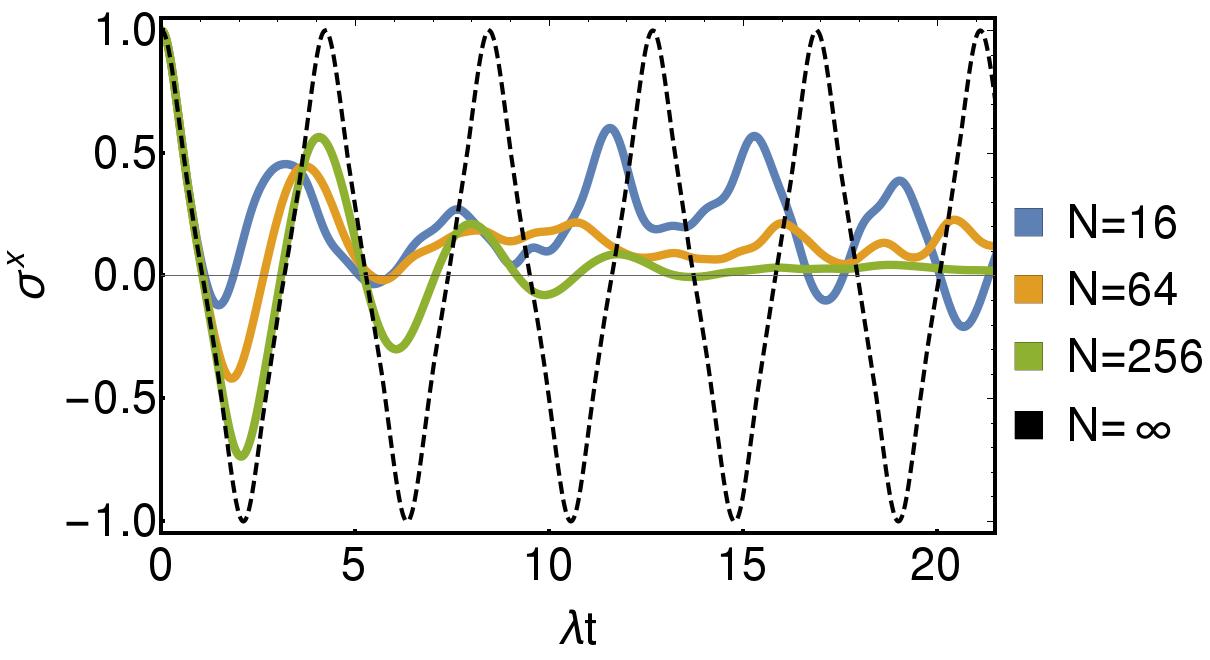} &  \includegraphics[scale=0.17]{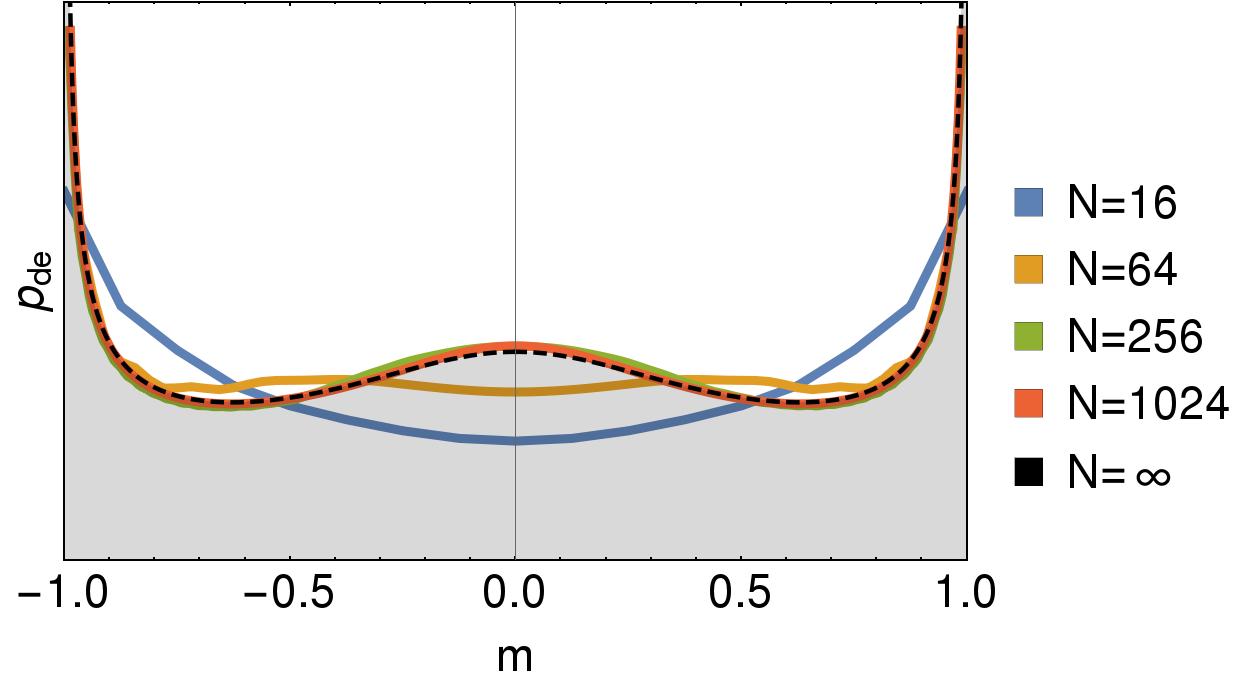} \\
 \includegraphics[scale=0.18]{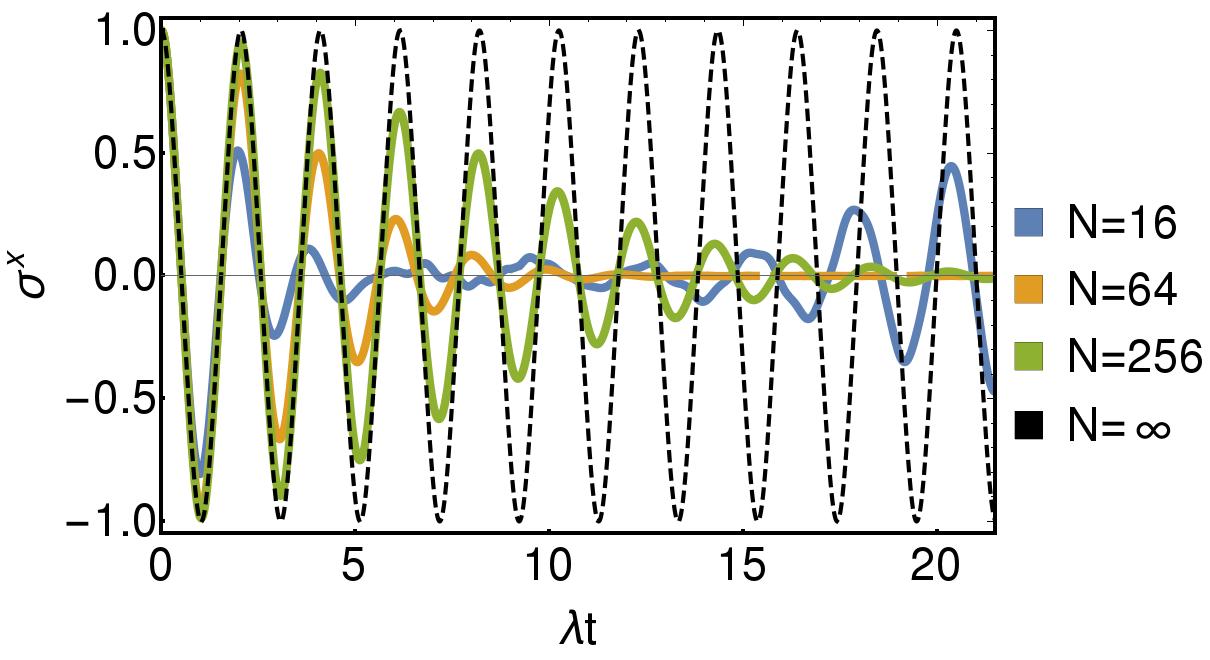} & \includegraphics[scale=0.17]{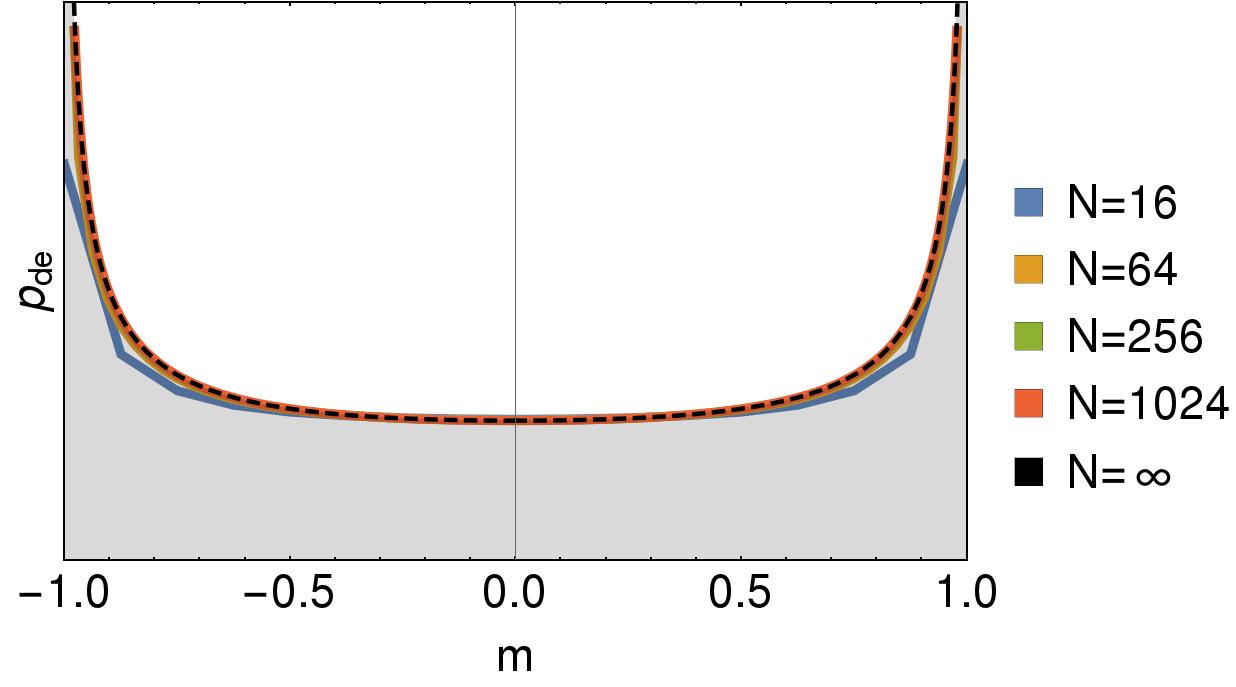}
\end{tabular}
\caption{
[Color online] Convergence to the classical behavior in the thermodynamic limit $N\to\infty$ of the quantum dynamics governed by the Hamiltonian~\eqref{eq:MFH} with finite size $N$ and with $s=1/2$. This is studied via exact diagonalization in the maximal spin sector. 
Left panels: Evolution of the dynamical order parameter  with $g/\lambda = 0.9$ (top), $g/\lambda = 1.1$ (center), $g/\lambda = 1.7$ (bottom), and increasing system size $N=16$, $64$, $256$, starting from a fully polarized state along the $\hat{x}$-direction, i.e., from a ground state with $g_0=0$. The classical limit is shown by the black dashed curve.
Right panels: corresponding infinite-time average distribution of the order parameter, as obtained from the diagonal ensemble $p_{\text{de}}(m) = \overline{\big\lvert\Braket{\Psi(t)|m}\big\rvert^2}$, where $\Ket{m}$ is the state with magnetization $m$ and the overline stands for infinite-time-average. The classical ``microcanonical'' distributions, obtained by averaging over the trajectory of $\mathcal{H}_{\text{cl}}$ with energy $E=\braket{\psi_0 |H|\psi_0}/N=-\lambda$, are shown by the black dashed curve.
Note that the quantum evolution agrees with its classical limit over a time window that increases with $N$. After this time, quantum phenomena emerge. 
In all cases, damping of the classical oscillations takes place as a consequence of the quantum spreading of the wavepacket.  
Furthermore, for system sizes $N$ as small as $16$, additional quantum effects become observable.
In the top left panel,
quantum tunneling to the opposite well can be observed  in the dynamical ferromagnetic phase at relatively small time, which scales as $T_{\text{tun}}=\mathcal{O}(e^{c N})$; note that the corresponding infinite-time distribution of the magnetization is suppressed in the classically forbidden region $m\approx 0$ as $N\to\infty$.
In the center left panel, a remnant of ferromagnetic behavior can be observed in the dynamical paramagnetic phase, %for smaller sizes $N$, 
due to contributions to the wavepacket coming from ferromagnetic initial conditions (in order to visualize this, one should replace the small black dot in Fig.~\ref{fig:MFDPT} with an extended circle of radius $1/\sqrt{N}$).
In the bottom left panel, recurrences in the evolution of the order parameter emerge at relatively small time $T_{\text{rec}}=\mathcal{O}(N)$, due to wavepacket refocusing after spreading.
All these three effects occur at larger times for $N=64$, $256$, and thus do not appear in the relative plots.
 }
\label{fig:ConvergenceToClassical}
\end{figure*}

In order to understand the possible connection with experimental realizations of long-range models, we now discuss the quantum corrections to the above classical behavior, which are relevant when the size $N$ of the system is finite. \footnote{The LMG model is  solvable by Bethe-ansatz for all $N$ (see Ref.~\onlinecite{GGA}), which allows in principle to compute analytically the ground state as well as the non-equilibrium properties. However, this exact solution in quite unpractical for large $N$, and a semiclassical approach turns out to be simpler and more powerful in order to understand the behavior of the system.}

As we have argued above, the infinite-range Hamiltonian~\eqref{eq:MFH} describes the dynamics of the a single collective degree of freedom, namely $\vec{\sigma} \equiv \sum_{i=1}^N \vec{\sigma}_i / N$. 
%Within the sector of the Hilbert space with maximal total spin magnitude, the components of the collective spin have quantum numbers $m=Ns,Ns-1,\dots,-(Ns-1),-Ns$, the range of which grows proportionally to the system size. Hence, the behavior of this degree of freedom approaches its classical limit for $N\to\infty$. 
In fact, %by rescaling the size of the collective spin by its magnitude $Ns$, one gets 
the  operators $\sigma^{\alpha}$, with $\alpha=x,y,z$, have spectrum in $[-1,1]$ and satisfy
\beq
\big[\sigma^{\alpha},\sigma^{\beta}\big]= \frac{1}{Ns}i \epsilon^{\alpha\beta\gamma} \sigma^{\gamma},
\eeq
which implies that an effective Planck's constant $\hbar_{\text{eff}} \equiv 1/(Ns)$ characterizes the quantum dynamics \cite{SciollaBiroliMF}. For this reason, the corrections to the classical motion can be investigated via a semiclassical expansion in inverse powers of $N$ of the solution of the Schr{\oe}dinger equation.

Let us now discuss the non-equilibrium dynamics within the semiclassical approximation\cite{Littlejohn,Bhaduri}. The first quantum correction to the classical evolution starting from a spin-coherent state is equivalent to treating the corresponding Gaussian Wigner function in phase space as a probability distribution and considering its classical (Liouville) evolution. To this level of approximation, known as the truncated Wigner approximation (TWA), the role of quantum mechanics amounts just to providing a degree of uncertainty to the classical phase space point which represents the initial state of the system\cite{TWA}. The amount of uncertainty is quantified by the phase space extension of the wavepacket, which covers an area equal to Planck's constant $h$, corresponding to the maximal phase space resolution allowed by the Heisenberg uncertainty relation.

In the presence of a non-quadratic Hamiltonian, like the one in Eq.~\eqref{eq:classicalH} in which we are interested in this work, closeby points in phase space separate linearly in time, due to their different periods, with the sole exception of the critical trajectory with diverging period, around which two points separate exponentially fast in time\footnote{This fact is crucial for the phenomenology of the chaotic dynamical ferromagnetic phase, see Sec.~\ref{sec:results}.}. Accordingly, since the linear extension of the initial wavepacket in phase space is $\sqrt{\hbar_{\text{eff}}}\sim 1/\sqrt{N}$, after a timescale of order $\mathcal{O}(\sqrt{N})$ [or $\mathcal{O}(\log{\sqrt{N}})$ around the separatrix] (the so-called Ehrenfest time $t_{\text{Eh}}$) the wavepacket spreads over the whole classical trajectory, and the observables relax to their ``microcanonical'' average\cite{Littlejohn,Bhaduri,SciollaBiroliMF,MoriLRETH}.

In the light of the above, the qualitative modifications of the classical dynamics discussed in the previous section due to finite-size effects can be summarized as follows:
\begin{enumerate}
\item a fully polarized spin-coherent initial state actually corresponds to a broad wavepacket of linear extension $ \propto 1/\sqrt{N}$ on the sphere of radius $1$, rather than to a single  point in phase space;
\item in order to observe the classical evolution described in the previous section, the thermodynamic limit must be taken first: at finite $N$, instead, quantum-mechanical effects such as the wavepacket spreading set in after the time scale $T_{\text{Eh}} \sim \mathcal{O}(\sqrt{N})$ [or $\mathcal{O}(\log{N})$ around the dynamical critical point] and the persistent classical oscillations are correspondingly damped;
\item the sharp dynamical phase transition highlighted in the previous section is smoothed out by quantum fluctuations, resulting in a crossover. % which becomes increasingly sharper in the limit $N\to\infty$;
%\item within the dynamical ferromagnetic phase, tunneling to the opposite ferromagnetic well occurs over a time scale $T_{\text{tun}}$ that grows exponentially with $N$.
\end{enumerate}

The four relevant time scales $T_{\text{cl}}=\mathcal{O}(1)$ (classical period), $T_{\text{Eh}}=\mathcal{O}(\sqrt{N})$ (wavepacket-spreading time scale), $T_{\text{rec}}=\mathcal{O}(N)$ (wavepacket recurrence time), $T_{\text{tun}}=\mathcal{O}(e^{c N})$ (tunneling time), are all well separated in the thermodynamic limit. Despite all quantum phenomena set in at increasingly longer time with $N$, in small systems they become important. In order to highlight the relevance and consequences of these finite-size effects, we report in Fig.~\ref{fig:ConvergenceToClassical} the time evolution of the order parameter $\sigma^x$ as well as the infinite-time averaged distribution $p_{\text{de}}$ of the magnetization for increasing system sizes $N$, as obtained from numerical diagonalization of the Hamiltonian~\eqref{eq:MFH} in the maximal spin sector.

\section{Static and dynamical \\ spin-wave expansions}

The  lack of interaction between the collective mode discussed in Sec. \ref{sec:MF} and the spin waves is an artifact of the infinite-range limit. In any  realistic model,  quantum fluctuations contribute  to the dynamics and, as a result of this interaction, the system is expected to eventually thermalize. 
It is thus natural to investigate the possible persistence of instances of dynamical criticality discussed above  in the \emph{pre-thermal} stage of the  dynamics, together with the possible onset of qualitatively new phenomena generated by these additional fluctuations. %, in addition, to the new dynamical phases  generated by the interaction of the collective, large, classical spin and  spin wave fluctuations.
%
%We address this point in Sec.~\ref{sec:results} of this work, while 
For this aim, we present in this Section a method to account systematically for the effect of  fluctuations on the dynamics of general interacting spin models, which was briefly introduced in Ref.~\onlinecite{LeroseShort}.

\label{sec:methodweakfluct}

%In order to discuss these aspects, we consider models that weakly break the trivial, mean-field, integrability, introducing a weak coupling of the collective degree of freedom with  spatial quantum fluctuations. The weak interaction is provided by couplings which break the full permutational symmetry and possess a non-trivial, spatial structure.

\subsection{Perturbative corrections to the equilibrium transition}

\label{sec:equilibrium}

In order to understand the impact of quantum fluctuations on the physics of the fully-connected Ising ferromagnet of Sec.~\ref{sec:MF}, we consider perturbations in the form of additional spatially-decaying interactions,
\beq
\label{eq:startingH}
H = - \frac{\lambda}{N} \sum_{\mathbf{r},\mathbf{r'}} \sigma_{\mathbf{r}}^x \sigma_{\mathbf{r'}}^x - g \sum_{\mathbf{r}} \sigma_{\mathbf{r}}^z -  \sum_{\mathbf{r},\mathbf{r'}} J_{\lvert\mathbf{r}-\mathbf{r'}\rvert}\sigma_{\mathbf{r}}^x \sigma_{\mathbf{r'}}^x,
\eeq
where $\mathbf{r},\mathbf{r'}$ run over a $d$-dimensional lattice with $N$ sites, and the coupling $J_r$ decays to zero upon increasing the geometrical distance $r=\lvert\mathbf{r}-\mathbf{r'}\rvert$. 
For simplicity we will focus on the one-dimensional case $d=1$ with periodic boundary conditions, even though all of the results we find do not rely on this assumption, as will become clear in the following. Accordingly, we denote by $i,j=1,\dots,N$ the lattice sites.

The perturbation makes the Hamiltonian a function  not only of the spin Fourier component at $k=0$ (as occurs for $J_r\equiv 0$), but of all the Fourier components with $k\ne0$. 
When the perturbation is small, %it is reasonable to assume that 
the amplitude of the  modes with $k\neq0$ is expected to be small, so that we can treat them perturbatively at the lowest non-trivial order corresponding to a quadratic approximation.
In order to do so, we introduce canonical coordinates representing small fluctuations around the mean-field spin-coherent states % that completely characterize the dynamics for $J_r=0$ (see Sec.~\ref{}). 
 by using a Holstein-Primakoff transformation relative to the direction of the average collective spin vector $\langle \tilde{\vec{\sigma}}_{k=0}\rangle$. 
Let us start by describing this approach in equilibrium. We first rewrite the Hamiltonian~\eqref{eq:startingH} in terms of Fourier components,
\begin{equation}
\label{eq:Hs_fourierbar}
H = 
- \frac{\bar{\lambda}}{N} \left(\tilde{\sigma}_{k=0}^x\right)^2
- g \, \tilde{\sigma}_{k=0}^z
- \frac{1}{N } \sum_{k\ne0} \tilde{J}_k \; \tilde{\sigma}_{k}^{x} \tilde{\sigma}_{-k}^{x},
\end{equation}
where $\bar{\lambda}\equiv\lambda+\tilde{J}_0$, $\tilde{J}_k=\tilde{J}_{-k}= \sum_{r=0}^{N-1} e^{-ikr} J_r$ and $\tilde{\sigma}_k^{\alpha}= \sum_j e^{-ikj} \sigma_j^{\alpha}$, where  $k$ varies in the Brillouin zone.
Let us now introduce a rotated reference frame  $(\hat{X},\hat{Y},\hat{Z})$, whose components in the original fixed frame $(\hat{x},\hat{y},\hat{z})$ are parameterized by the polar angles $\theta$ and $\phi$ as
\begin{equation}
\label{eq:newbasis}
\hat{X} \equiv 
\left( \begin{matrix}
\cos\theta \cos\phi \\
\cos\theta\sin\phi \\
-\sin\theta
\end{matrix} \right) ,  
\hat{Y} \equiv 
\left( \begin{matrix}
-\sin\phi \\
\cos\phi \\
0 
\end{matrix} \right)
 , 
\hat{Z} \equiv 
\left( \begin{matrix}
\sin\theta \cos\phi \\
\sin\theta\sin\phi \\
\cos\theta
\end{matrix} \right) .
\end{equation}
The spins can  be then decomposed on the basis of $\mathcal R$ as
\begin{equation}
\vec{\sigma}_j =  \hat{X} \, \sigma_j^X \, + \hat{Y} \, \sigma_j^Y   + \hat{Z} \,  \sigma_j^Z .
\end{equation}
Accordingly, the Hamiltonian~\eqref{eq:Hs_fourierbar} can be rewritten as
\begin{widetext}
\beq
\label{eq:Hrotatedframe}
\begin{split}
\frac{H}{N} = & - \bar{\lambda}\left[ \left(\hat{X}\cdot\hat{x}\right) \frac{\tilde{\sigma}_{0}^X}{N} + \left(\hat{Y}\cdot\hat{x}\right) \frac{\tilde{\sigma}_{0}^Y}{N} +\left(\hat{Z}\cdot\hat{x}\right) \frac{\tilde{\sigma}_{0}^Z}{N}  \right]^2 \\
&- g \left[ \left(\hat{X}\cdot\hat{z}\right) \frac{\tilde{\sigma}_{0}^X}{N} + \left(\hat{Y}\cdot\hat{z}\right) \frac{\tilde{\sigma}_{0}^Y}{N} +\left(\hat{Z}\cdot\hat{z}\right) \frac{\tilde{\sigma}_{0}^Z}{N}  \right]
  \\ 
&-  \sum_{k\ne0} \tilde{J}_k \left[ \left(\hat{X}\cdot\hat{x}\right) \frac{\tilde{\sigma}_{k}^X}{N} + \left(\hat{Y}\cdot\hat{x}\right) \frac{\tilde{\sigma}_{k}^Y}{N} +\left(\hat{Z}\cdot\hat{x}\right) \frac{\tilde{\sigma}_{k}^Z}{N}  \right]\cdot \left[ 
\left(\hat{X}\cdot\hat{x}\right) \frac{\tilde{\sigma}_{-k}^X}{N} + \left(\hat{Y}\cdot\hat{x}\right) \frac{\tilde{\sigma}_{-k}^Y}{N} +\left(\hat{Z}\cdot\hat{x}\right) \frac{\tilde{\sigma}_{-k}^Z}{N}
  \right]
\end{split}
\eeq
in terms of the Fourier transforms $\tilde{\sigma}_k^{X,Y,Z}$ of $\sigma_j^{X,Y,Z}$.
%%%
\end{widetext}

In the rotated frame $\mathcal{R}$, we introduce the spin wave canonical variables via the Holstein--Primakoff transformation \cite{wannier}, expanded to lowest order in $1/\sqrt{s}$, i.e.,
\begin{equation}
\label{eq:approxH-P}
\left\{
\begin{split}
\sigma_j^X &=\frac { q_j}{\sqrt{s}}  + \dots \, , \\
\sigma_j^Y  &= \frac { p_j}{\sqrt{s}}+ \dots \, , \\
%\mathcal{O}\bigg(\frac{q_j,p_j}{\sqrt{s}}\bigg)^3  \\
\sigma_j^Z  &= 1-\frac{n_j}{s} \equiv 1 - \frac{q_j^2+p_j^2-1}{2s},
\end{split} 
\right.
\end{equation}
where $q_j$ and  $p_j$ are the conjugate canonical variables representing small deviations of the spin away from the $\hat{Z}$-axis, and along the  directions $\hat{X}$ and $\hat{Y}$, respectively. 
In our notation, the bosonic number operator, $n_j=b^\dag_j b_j$, is defined via   $b_j =  (q_j + i p_j)/\sqrt{2}$. 
Accordingly, after introducing the coordinates
 $\tilde{q}_k= N^{-1/2} \sum_j e^{-ikj}q_j$ and $\tilde{p}_k= N^{-1/2} \sum_j e^{-ikj}p_j$ in Fourier space we get
 \begin{widetext}
%%%
\begin{equation}
\label{eq:approxH-Pfourier}
\left\{
\begin{split}
\frac{\tilde{\sigma}_{k}^X}{N} =&  \frac{\tilde{q}_{k}}{\sqrt{Ns}} +  \dots \, ,  \\
\frac{\tilde{\sigma}_{k}^Y}{N} =&  \frac{\tilde{p}_{k}}{\sqrt{Ns}} + \dots \, , \\
\frac{\tilde{\sigma}_{k}^Z}{N}   =& \delta_{k,0}-\sum_{k'}\frac{\tilde{q}_{k'}\tilde{q}_{k-k'}+\tilde{p}_{k'}\tilde{p}_{k-k'}-\delta_{k,0}}{2Ns}  .
 \end{split} 
 \right.
 \end{equation}

The Hamiltonian~\eqref{eq:Hrotatedframe} can now be written in terms of the canonical spin wave coordinates,%
\beq
\label{eq:Hsw}
\begin{split}
H = & - \bar{\lambda} N \left(\hat{Z}\cdot\hat{x}\right)^2 \left(1-\frac{n_0+N_{\text{sw}}}{Ns}\right)^2
                  - g N \left(\hat{Z}\cdot\hat{z}\right) \left(1-\frac{n_0+N_{\text{sw}}}{Ns}\right) \\
                      & - 2 \bar{\lambda}\frac{\sqrt{N}}{\sqrt{s}} \left(\hat{Z}\cdot\hat{x}\right) \left(1-\frac{n_0+N_{\text{sw}}}{Ns}\right) \left[
                      \left(\hat{X}\cdot\hat{x}\right) \tilde{q}_0 + \left(\hat{Y}\cdot\hat{x}\right) \tilde{p}_0
                      \right] \\
                      &   - g\frac{\sqrt{N}}{\sqrt{s}} \left[
                      \left(\hat{X}\cdot\hat{z}\right) \tilde{q}_0 + \left(\hat{Y}\cdot\hat{z}\right) \tilde{p}_0
                      \right] \\
                  & - \frac{\bar{\lambda}}{s}\left[
                   \left(\hat{X}\cdot\hat{x}\right)^2 \tilde{q}_0^2
                   + \left(\hat{Y}\cdot\hat{x}\right)^2 \tilde{p}_0^2
                   + 2 \left(\hat{X}\cdot\hat{x}\right) \left(\hat{Y}\cdot\hat{x}\right) \frac{\tilde{q}_0\tilde{p}_0+\tilde{p}_0\tilde{q}_0}{2} 
                  \right]
  \\ 
&+ U_2 + U_3 + U_4   ,
\end{split}
\eeq
with the $k\ne0$ contribution of the short-range interaction  split into the three terms:
\beq
\label{eq:U123}
\begin{split}
U_2&=-  \sum_{k\ne0} \frac{\tilde{J}_k}{s} \bigg[ \left(\hat{X}\cdot\hat{x}\right)^2 \tilde{q}_k \tilde{q}_{-k}
+\left(\hat{Y}\cdot\hat{x}\right)^2 \tilde{p}_k \tilde{p}_{-k} \\ & \qquad \qquad \qquad \qquad \qquad
+ 2 \left(\hat{X}\cdot\hat{x}\right) \left(\hat{Y}\cdot\hat{x}\right) \frac{\tilde{q}_k \tilde{p}_{-k} + \tilde{p}_k \tilde{q}_{-k}}{2} \bigg],\\
U_3&= +\frac{1}{\sqrt{Ns}} \sum_{k\ne0}  \frac{\tilde{J}_k}{s} \left(\hat{Z}\cdot\hat{x}\right) \\ 
& \qquad \qquad \times \Bigg\{
 \left(\hat{X}\cdot\hat{x}\right) 
\left[ \tilde{q}_k \sum_{k'}\frac{\tilde{q}_{k'}\tilde{q}_{-k-k'}+\tilde{p}_{k'}\tilde{p}_{-k-k'}}{2} + (k\leftrightarrow -k) \right] + \\
 &   \qquad \qquad
+\left(\hat{Y}\cdot\hat{x}\right) 
\left[ \tilde{p}_k \sum_{k'}\frac{\tilde{q}_{k'}\tilde{q}_{-k-k'}+\tilde{p}_{k'}\tilde{p}_{-k-k'}}{2} + (k\leftrightarrow -k) \right]
 \Bigg\},\\
U_4&= -\frac{1}{Ns}\sum_{k\ne0} \frac{ \tilde{J}_k}{s}  
\left(\hat{Z}\cdot\hat{x}\right)^2 
\sum_{k'}\frac{\tilde{q}_{k'}\tilde{q}_{k-k'}+\tilde{p}_{k'}\tilde{p}_{k-k'}}{2}
\sum_{k''}\frac{\tilde{q}_{k''}\tilde{q}_{-k-k''}+\tilde{p}_{k''}\tilde{p}_{-k-k''}}{2},
\end{split}
\eeq
standing for  the quadratic, %($U_2$), 
cubic, % ($U_3$), 
and quartic %($U_4$) 
terms in the spin waves, respectively. 
%%%
\end{widetext}

In Eq.~\eqref{eq:Hsw}, the quantity $N_{\text{sw}}$ is the total number of spin waves, %divided by the maximal total spin $Ns$, 
i.e.,
\beq
\label{eq:Nsw}
N_{\text{sw}} = \sum_{k\ne0} n_k  = \sum_{k\ne0} \frac{\tilde{q}_k \tilde{q}_{-k}+\tilde{p}_k \tilde{p}_{-k}-1}{2}
\eeq
[cf. Eq.~\eqref{eq:epsilonNsw}]. 
The expansion in Eq.~\eqref{eq:Hsw}  is valid as long as the spin waves have a low density $N_{\text{sw}} \ll Ns$, i.e., the collective spin magnitude is close to its maximal value $Ns$. In this regime,  spin waves  behave as free bosonic excitations which interact with the macroscopic collective spin only, corresponding to the $k=0$ mode.  Higher-order terms, which account for non-linear scattering among the spin waves, can be neglected: they are expected to contribute significantly to the dynamics only at longer times and to drive the system away from the pre-thermal regime relevant for the DPT discussed here.
%\end{widetext}

%It is important to observe that the Hamiltonian starts from the linear order in the zero-mode quantum fluctuations $\tilde{q}_0,\tilde{p}_0$, and from the quadratic order in the spatial fluctuation modes $\tilde{q}_k,\tilde{p}_k$, $k\ne0$. 

Our approach is equivalent to  treating fluctuations within the Gaussian approximation, which is the lowest non-trivial order beyond mean-field. This is expected to be sufficiently accurate  when the interaction $\tilde{J}_{k\ne0}$ introduces a small perturbation to the mean-field dynamics, such that a small spin-wave density  $N_{\text{sw}}/(Ns)$ is generated during the dynamics. 
In this case, similarly to the well-known Bogolyubov theory of weakly-interacting Bose gases \cite{LiebYngvason}, we can treat them as free particles. Accordingly, the only relevant interaction is that between the collective mode $\tilde{q}_0,\tilde{p}_0$ and the spin waves, given by terms in $U_3$, which describe scattering of the zero-momentum mode  into a pair of spin waves with opposite momenta $(k,-k)$, and viceversa.
This approximation amounts to neglecting terms of order $\mathcal{O}(\braket{N_{\text{sw}}}/Ns)^2$. Thereby, we arrive at the following form of the Hamiltonian~\eqref{eq:startingH}, truncated to linear order in the collective $k=0$ mode and to quadratic order in the spin-wave fluctuations with $k\ne0$, 

\begin{widetext}
%%%
\beq
\label{eq:effectiveH}
\begin{split}
H \simeq
% classical energy
& - \bar{\lambda} N ( \hat{Z} \cdot \hat{x} )^2 - g N ( \hat{Z} \cdot \hat{z} ) \\
% quadratic energy of the quantum fluctuations
      & + \frac{1}{s} \sum_{k\ne0} \left[ 2 \bar{\lambda} ( \hat{Z} \cdot \hat{x} )^2 + g ( \hat{Z} \cdot \hat{z} )\right] \frac{\tilde{q}_k \tilde{q}_{-k}+\tilde{p}_k \tilde{p}_{-k}-1}{2}
      \\
      & - \frac{1}{s} \sum_{k\ne0} \tilde{J}_k \bigg[ 
      ( \hat{X} \cdot \hat{x} )^2 \; \tilde{q}_k \tilde{q}_{-k} +
      ( \hat{Y} \cdot \hat{x} )^2 \; \tilde{p}_k \tilde{p}_{-k} \\ & \qquad \qquad \qquad+
      2 ( \hat{X} \cdot \hat{x} )( \hat{Y} \cdot \hat{x} ) \; \frac{\tilde{q}_k \tilde{p}_{-k}+\tilde{p}_k \tilde{q}_{-k} }{2}
       \bigg] \\
% interaction energy of the vacuum and the quantum fluctuations
      & +  \frac{\sqrt{N}}{\sqrt{s}}\tilde{q}_0 \Bigg\{
      -2 \bar{\lambda} \bigg(1-\frac{N_{\text{sw}}}{Ns}\bigg)  ( \hat{Z} \cdot \hat{x} )  ( \hat{X} \cdot \hat{x} )  - g  ( \hat{X} \cdot \hat{z} ) 
      \\
      & \qquad
      + 2   ( \hat{Z} \cdot \hat{x} ) \frac{1}{Ns} \sum_{k\ne0} \tilde{J}_k \bigg[ ( \hat{X} \cdot \hat{x} ) \; \tilde{q}_k \tilde{q}_{-k} +  ( \hat{Y} \cdot \hat{x} ) \; \frac{\tilde{q}_k \tilde{p}_{-k}+\tilde{p}_k \tilde{q}_{-k} }{2} \bigg]
      \Bigg\}
      \\
      & +  \frac{\sqrt{N}}{\sqrt{s}}\tilde{p}_0 \Bigg\{
      -2 \bar{\lambda} \bigg(1-\frac{N_{\text{sw}}}{Ns}\bigg)  ( \hat{Z} \cdot \hat{x} )  ( \hat{Y} \cdot \hat{x} )  - g  ( \hat{Y} \cdot \hat{z} ) 
      \\
      & \qquad
      + 2  ( \hat{Z} \cdot \hat{x} ) \frac{1}{Ns} \sum_{k\ne0} \tilde{J}_k \bigg[  ( \hat{Y} \cdot \hat{x} ) \; \tilde{p}_k \tilde{p}_{-k} +    ( \hat{X} \cdot \hat{x} ) \;\frac{\tilde{q}_k \tilde{p}_{-k}+\tilde{p}_k \tilde{q}_{-k} }{2} \bigg]
      \Bigg\},
\end{split}
\eeq
where the explicit expressions of the various scalar products between versors in terms of the rotation angles $\theta$ and $\phi$ can be inferred from Eq.~\eqref{eq:newbasis}. 
The Hamiltonian~\eqref{eq:effectiveH} is our starting point for  assessing the impact of  fluctuations on the equilibrium and dynamical phase transition occurring in the LMG model.

\end{widetext}

We first study the equilibrium behavior in the presence of fluctuations. %\review{from here: rewrite a bit, it is unclear}
The average total spin in equilibrium can  be determined at the Gaussian level by imposing vanishing expectation values of $\tilde{q}_0$ and $\tilde{p}_0$, i.e., %that the quantum Hamiltonian has no linear terms in the fluctuations,
\beq
\label{eq:self-consistency}
\langle \tilde{q}_0 \rangle = \langle \tilde{p}_0 \rangle = %\overset{!}{=} 
0 .
\eeq
Equation \eqref{eq:approxH-Pfourier} with $k=0$ shows that this is equivalent to requiring that the average total spin $\langle \tilde{\vec{\sigma}}_{k=0}\rangle$ is aligned along the $\hat{Z}$-direction determined by the spherical angles $\theta,\phi$. 
%or, equivalently, by treating the total spin as a classical variable and minimizing the variational energy over the manifold of Gaussian states\review{end of my confusion :)}. 
In the mean-field limit $\tilde{J}_{k\ne0}=0$, the spin waves are frozen in their vacuum state and the problem becomes equivalent to finding the ground state of the single classical spin $\vec{\sigma} = \langle \tilde{\vec{\sigma}}_{k=0} \rangle / N$ on the sphere. As $\tilde{J}_{k\ne0}\ne0$, %the oscillators are squeezed and 
the spin waves are generically excited even in the ground state, analogously to the depletion of the condensate fraction in a dilute Bose gas in the presence of weak interactions.
Eqs.~\eqref{eq:self-consistency} are actually satisfied when the values of $\theta$ and $\phi$ are chosen in such a way  that
the equilibrium expectation values %the averages 
of the two curly brackets in Eq.~\eqref{eq:effectiveH} vanish.  The second one %is automatically satisfied \review{collegato al commento di sopra: e di conseguenza modificare quanto segue..} 
does it if $ \hat{Y} \cdot \hat{x} =  \hat{Y} \cdot \hat{z} = 0$, and %accordingly
\beq
 \sum_{k\ne0} \tilde{J}_k \bigg\langle \frac{\tilde{q}_k \tilde{p}_{-k}+\tilde{p}_k \tilde{q}_{-k} }{2} \bigg\rangle=0,
\eeq
which implies that $\phi^*=0$ or $\pi$,  meaning that the collective spin lies in the $xz$-plane, as could be anticipated based on symmetry arguments. The remaining equation determines the value of $\theta^*$. In particular, $\theta^*=0$ is always a solution: however, it is stable only for $g$ large enough. For small $g$, stable solutions are calculated as follows. First we diagonalize the quadratic part of the Hamiltonian [second and third sums on the r.h.s. of Eq.~\eqref{eq:effectiveH}] obtaining a parametric spin wave dispersion relation  $\omega_k/s$, %\review{say some words how we derive it} 
%(we neglect the renormalization of the dispersion relation due to the fluctuations, which derives from the quartic and higher order terms in the Hamiltonian)
\begin{widetext}
\beq
\label{eq:disprel}
\omega_k = %\frac{1}{s}
\sqrt{\big( 2 \bar{\lambda} \sin^2\theta + g \cos\theta \big)\big(  2 \bar{\lambda} \sin^2\theta + g \cos\theta - 2\tilde{J}_k \cos^2\theta\big)}.
\eeq
Denoting by $\omega^{(0)}_k/s$ the ``unperturbed'' common frequency of the spin wave modes,
\beq
\omega^{(0)}_k \equiv \omega^{(0)} = %\frac{1}{s} 
%\big( 
2 \bar{\lambda} \sin^2\theta + g \cos\theta ,
%\big),
\eeq
the zero-temperature Gaussian expectation values of the relevant observables can then be expressed as %in terms of the parametric and unperturbed frequencies as  
\beq
\label{eq:gaussianparam}
\left\{
\begin{split}
\big\langle
\tilde{q}_k \tilde{q}_{-k}
\big\rangle
&=\frac{1}{2} \frac{\omega^{(0)}_k}{\omega_k} ,
\\
\big\langle
\tilde{p}_k \tilde{p}_{-k}
\big\rangle
&= \frac{1}{2} \frac{\omega_k}{\omega^{(0)}_k},
\\
\bigg\langle
\frac{\tilde{q}_k \tilde{p}_{-k}+\tilde{p}_k \tilde{q}_{-k}}{2}
\bigg\rangle
&= 0.
\end{split}
\right.
\eeq
Accordingly, the equation of state which determines $\theta^*$ reads
\begin{equation}
\label{eq:eqstate}
     \sin\theta^*\Bigg[ -2 \bar{\lambda} (1-\epsilon)   \cos\theta^*  + g   
             +   \cos\theta^*  \frac{1}{Ns} \sum_{k\ne0} \tilde{J}_k   \sqrt{\frac{ 2 \bar{\lambda} \sin^2\theta^* + g \cos\theta^* }{  2 \bar{\lambda} \sin^2\theta^* + g \cos\theta^* - 2\tilde{J}_k \cos^2\theta^*}} \; \Bigg] =0 .%\overset{!}{=}0,
\end{equation}
\end{widetext}
Combining Eqs.~\eqref{eq:Nsw} and~\eqref{eq:gaussianparam}, we get an explicit expression for the total spin depletion $\epsilon$ defined by the equation
\beq
\label{eq:epsilondef}
\frac{\Big\lvert \Braket{\tilde{\vec{\sigma}}_{k=0}} \Big\rvert}{N} = 1 - \frac{\Braket{N_{\text{sw}}}}{Ns} \equiv 1-\epsilon,
\eeq
i.e.,
\beq
\label{eq:epsilongaussian}
\epsilon \equiv \frac{\Braket{N_{\text{sw}}}}{Ns} = \frac{1}{2Ns} \sum_{k\ne0} \bigg( \frac{1}{2} \frac{\omega^{(0)}_k}{\omega_k} + \frac{1}{2} \frac{\omega_k}{\omega^{(0)}_k} - 1\bigg)_{\theta=\theta^*}. %= \\ =  \frac{1}{2Ns} \sum_{k\ne0} \Bigg( \frac{1}{2} \sqrt{\frac{ 2 \bar{\lambda} \sin^2\theta^* + g \cos\theta^* }{  2 \bar{\lambda} \sin^2\theta^* + g \cos\theta^* - 2\tilde{J}_k \cos^2\theta^*}} \\ + \frac{1}{2}\sqrt{\frac{  2 \bar{\lambda} \sin^2\theta^* + g \cos\theta^* - 2\tilde{J}_k \cos^2\theta^*}{ 2 \bar{\lambda} \sin^2\theta^* + g \cos\theta^* }} - 1 \Bigg).
\eeq
%which has to be determined consistently with the Eq \eqref{eq:eqstate}.
Note that $\epsilon\ge0$ and $\epsilon=\mathcal{O}(\tilde{J}_{k\ne0}^2)$. Moreover, in the limits $g\to0$ and $g\to\infty$, the depletion $\epsilon$ at equilibrium vanishes, %($\epsilon\to0$), 
whereas it is arguably maximal at the critical point $g_{\text{cr}}=2\bar{\lambda}-\mathcal{O}(\tilde{J}_{k\ne0}^2)$ (see below). 

As a check, in the mean-field case $\tilde{J}_{k\ne0} \equiv 0$, Eqs.~\eqref{eq:disprel} and~\eqref{eq:eqstate} imply $\omega^{(0)}_k = \omega_k$, $\epsilon=0$, and $\cos\theta^* = g/2\lambda$ in the ferromagnetic phase $g<2\lambda$,  retrieving the mean-field equilibrium properties. As soon as a spatially-decaying interaction $\tilde{J}_{k\ne0} \ne 0$ is turned on, quantum fluctuations modify the equilibrium state.

In the equilibrium paramagnetic phase $g>g_{\text{cr}}$, the ground state has $\theta^*=0$, and  from Eq.~\eqref{eq:disprel} we find
\beq
\label{eq:dispreltheta0}
\omega_{k,>} =%\frac{1}{s} 
\sqrt{g(g-2 \tilde{J}_k)}.
\eeq
Deep in the equilibrium ferromagnetic phase,  with $g\to0$, the system approaches instead  a full ferromagnetic ordering with $\theta^*\to\pi/2$, and 
therefore the corresponding dispersion relation derived from Eq.~\eqref{eq:disprel} becomes independent of $k$, i.e. the  band becomes flat,
\beq
\label{eq:dispreldeepferro}
\omega_{k,<} \quad\underset{g\to0}{\longrightarrow}\quad 2\bar{\lambda} .
\eeq
%as it can be easily seen a priori 
This could have been anticipated by observing that  in the limit $g\to0$ the spin Hamiltonian~\eqref{eq:startingH} becomes diagonal in the $\sigma^x$-basis.

Let us determine now the perturbative corrections to the critical point, employing an equivalent variational approach. The critical point corresponds to the value of $g$ at which the paramagnetic configuration $\theta=0$ becomes an unstable saddle-point of $H$. We compute the variational energy $\mathcal{E}(\theta) = \langle H \rangle_\theta$ as a function of $\theta$ (with fixed $\phi=0$), by taking the average of $H$ in Eq.~\eqref{eq:effectiveH} with $\langle \tilde{q}_0 \rangle =\langle \tilde{p}_0 \rangle =0$ and $\langle \tilde{q}_k  \tilde{q}_{-k}\rangle$ given by Eq.~\eqref{eq:gaussianparam}, thereby obtaining
\beq
\frac{\mathcal{E}(\theta)}{N} =  -  \bar{\lambda} \sin^2 \theta - g \cos\theta
 + \frac{1}{Ns} \sum_{k\ne0} \frac{\omega_k - \omega_k^{(0)}}{2} .
\eeq
In order to determine the stability of the solution $\theta^*=0$, we expand $\sim\mathcal{E}(\theta)$ at small $\theta$, finding
\begin{widetext}
\begin{equation}
\begin{split}
\frac{\mathcal{E}(\theta)}{N} \underset{\theta\to0}{\thicksim}  & - g + \frac{1}{Ns} \sum_{k\ne0}\frac{1}{2} \Big( \sqrt{g(g-2\tilde{J}_k)}-g\Big)  
          \\ & 
          + \Bigg\{
      g - 2\bar{\lambda} + \frac{1}{Ns} \sum_{k\ne0} \bigg[ 
      \sqrt{g(g-2\tilde{J}_k)} \frac{1}{2}\bigg( \frac{2\bar{\lambda}-g/2 + 2 \tilde{J}_k}{g-2\tilde{J}_k} + \frac{2\bar{\lambda}-g/2}{g}\bigg)
      %\qquad \qquad \qquad \qquad \qquad \qquad \qquad \qquad \qquad \qquad 
      -\bigg(2\bar{\lambda}-g/2\bigg)
      \bigg]
      \Bigg\} \frac{\theta^2}{2} 
       + \mathcal{O}(\theta^4).
\end{split}
\end{equation}
\end{widetext}
The critical point is then determined by the vanishing of the coefficient of the quadratic term in curly bracket, %proportional to the ``massive'' term $\propto \theta^2$, 
which yields an equation for $g_{\text{cr}}$.
The corrections for small $\tilde{J}_{k\ne0}$ may be found perturbatively by expanding the solution $g_{\text{cr}}(\tilde{J}_{k\ne0})$ with respect to $\tilde{J}_{k\ne0}$ and by equating both sides order by order. The explicit calculation yields a quadratic correction:
\beq
\label{eq:quantumcorrectioncrit}
g_{\text{cr}} = 2\bar{\lambda} \Bigg\{ 
1 - \frac{5}{16} \frac{1}{Ns} \sum_{k\ne0} \bigg( \frac{\tilde{J}_k}{\bar{\lambda}}\bigg)^2
\Bigg\} + \mathcal{O}(\tilde{J}_{k\ne0}^3).
\eeq
As expected on physical grounds, the spin waves destabilize the ferromagnetic ordering and thereby lower the critical value $g_{\text{cr}}$. % (as expected on physical grounds). %Notice that the correction is of order $\epsilon$: compare with equation \eqref{eq:epsilongaussian} to lowest order in $\tilde{J}_k$.%\footnote{In the thermodynamic limit, $\frac{1}{N} \sum_{k\ne0} \tilde{J}_k^2 \to \int_{-\pi}^{\pi}\frac{dk}{2\pi} \tilde{J}_k^2 $.}

%WRITE FORMULAS AT FINITE TEMPERATURE

The ground state equations discussed above can immediately be generalized to the case with a finite temperature $T>0$. In fact, it is sufficient to substitute in Eq.~\eqref{eq:gaussianparam} the pre-factor $1/2$ with
\beq
%\frac{1}{2} \quad \mapsto \quad 
\frac{1}{2} + \langle n_k \rangle_T =  \frac{1}{2} + \frac{1}{e^{\omega_k/T}-1},
\eeq
where $\langle n_k \rangle_T$ is the Bose-Einstein distribution of the excited spin waves.
The expression of $\epsilon$ in Eq.~\eqref{eq:epsilongaussian}  and the equation of state \eqref{eq:eqstate} are modified accordingly. As in the mean-field case,  thermal corrections are exponentially suppressed at low temperature as long as the gap in the dispersion relation \eqref{eq:disprel} is non-vanishing.

\subsection{Dynamics: Time-dependent spin wave theory}
\label{sec:tdswt}

The non-equilibrium dynamics in the presence of weak fluctuations can be analyzed by generalizing the approach developed in the previous section to a time-evolving state. The spin wave expansion will be performed with respect to a time-dependent  rotated frame $\mathcal R$, with the angles $\theta(t),\phi(t)$ co-moving with  %the evolution of 
the average collective spin \cite{LeroseShort}.
%
%The time-dependent change of frame 
This is implemented by the unitary operator 
\beq V(\theta(t),\phi(t))= e^{-i\phi\, s\sum_j\sigma_j^z} \, e^{-i\theta\, s\sum_j\sigma_j^y}, \eeq
acting on the spins as  a time-dependent global rotation: 
%\begin{widetext}
\begin{equation}
\left\{
\begin{split}
V \sigma_j^x V^{\dagger} &= \hat{X} \cdot \vec{\sigma}_j \equiv \sigma_j^X, \\
V \sigma_j^y V^{\dagger} &= \hat{Y} \cdot \vec{\sigma}_j \equiv \sigma_j^Y, \\
V \sigma_j^z V^{\dagger} &= \hat{Z} \cdot \vec{\sigma}_j \equiv \sigma_j^Z.
\end{split}
\right.
\end{equation}
%\end{widetext}
The Heisenberg  equations of motion for $\sigma_j^{\alpha}$ ($\alpha \in \{ X,Y,Z\}$), in the mobile frame $\mathcal R$, read then
\begin{equation}
\label{eq:inertial}
\frac{d}{dt} \sigma_j^{\alpha} =  \frac{1}{i}\Big[\sigma_j^{\alpha},\widetilde{H}\Big],
\quad \text{with } \; \widetilde{H} \equiv  H + i V \dot{V}^{\dagger},
\end{equation}
where %$H$ is still given by Eq.~\eqref{eq:effectiveH} and 
the last term is the inertial force contribution equal to
\beq
\label{eq:inertialterm}
i V \dot{V}^{\dagger} = - s \, \vec{\omega}(t) \cdot %\left( 
\sum_j \vec{\sigma}_j = - s \, \vec{\omega}(t) \cdot \tilde{\vec{\sigma}}_{0},% \right),
\eeq
where we introduced the vector $\vec{\omega}=(\omega^X,\omega^Y,\omega^Z)$, with $\omega^X =
%-\sin\theta \; \dot{\phi}$
\dot{\hat{Y}} \cdot \hat{Z}$,
$\omega^Y  =  %\dot{\theta}
\dot{\hat{Z}} \cdot \hat{X}$,  and
$\omega^Z = %\cos\theta \; \dot{\phi}
\dot{\hat{X}} \cdot \hat{Y}$.
%
%\beq
%\omega^x \equiv \hat{x}\cdot\vec{\omega} = -\sin\theta \; \dot{\phi}, \qquad 
%\omega^y \equiv \hat{y}\cdot\vec{\omega} =  \dot{\theta}, \qquad 
%\omega^z \equiv \hat{z}\cdot\vec{\omega} = \cos\theta \; \dot{\phi}.
%\eeq
%As a generalization of Larmor's theorem in electrodynamics, the effect a time-dependent rotation of the reference frame is equivalent to the presence of a time-dependent external magnetic field.

The resulting Hamiltonian $\widetilde{H}(t)$ is thus given by the expression in Eq.~\eqref{eq:effectiveH} for $H$ with time-dependent $\hat{X}(t),\hat{Y}(t),\hat{Z}(t)$ [i.e., with time-dependent $\theta(t),\phi(t)$], and with the additional terms
\beq
\begin{split}
   -s  \big( \dot{\hat{X}} \cdot \hat{Y} \big)  \, \tilde{\sigma}_0^Z  =&  -s  \big( \dot{\hat{X}} \cdot \hat{Y} \big)  \\ & +\frac{1}{s}\sum_{k\ne0} \Big[ s \big(\dot{\hat{X}} \cdot \hat{Y}\big) \Big] \frac{\tilde{q}_k \tilde{q}_{-k}+\tilde{p}_k \tilde{p}_{-k} -1 }{2}, \\
  -s  \big( \dot{\hat{Y}} \cdot \hat{Z} \big)  \, \tilde{\sigma}_0^X  =&  \frac{\sqrt{N}}{\sqrt{s}}\tilde{q}_0 \bigg\{  -  s  \big(\dot{\hat{Y}} \cdot \hat{Z}\big) \bigg\}, \\
  -s  \big( \dot{\hat{Z}} \cdot \hat{X} \big)  \, \tilde{\sigma}_0^Y  =&  \frac{\sqrt{N}}{\sqrt{s}}\tilde{p}_0 \bigg\{ -   s  \big(\dot{\hat{Z}} \cdot \hat{X}\big) \bigg\},
    \end{split}
\eeq
to be added to the second, fourth and fifth line of Eq.~\eqref{eq:effectiveH}, respectively.
This time-dependent Hamiltonian governs the self-consistent coupled evolution equations of the angles $\theta,\phi$ and of the excitations $\tilde{q}_k,\tilde{p}_k$. %writing down the equations of motion for the dynamical variables and by
In particular, the motion of the angles is obtained by imposing the condition \eqref{eq:self-consistency} to hold at all times, which corresponds to setting the 
%two terms contained in curly brackets in Eq.~\eqref{eq:effectiveH} equal  to $s \, \omega^X = - s ( \sin\theta ) \dot{\phi}$ and $s \, \omega^Y = s\dot{\theta}$, respectively, where the quantities on the r.h.s. of these equations are due to %(these latter two come from the inertial force contribution in Eq.~\eqref{eq:inertialterm}.
coefficients of $\tilde{q}_0$ and $\tilde{p}_0$ in $\widetilde{H}(t)$  equal to zero.
This procedure yields a pair of classical evolution equations for $\theta(t),\phi(t)$, which depend also on
the spin-wave correlation functions. The excitation of spin waves  thereby affects the mean-field trajectory of $\theta(t),\phi(t)$ at order $\tilde{J}_{k\ne0}$ for weak integrability breaking. Concurrently, the motion of the vacuum $\theta(t),\phi(t)$ drives the non-equilibrium evolution of the spin excitations $\tilde{q}_k, \tilde{p}_k$. %The final system of coupled evolution equations (at Gaussian level) is
Explicitly, we obtain
%\begin{widetext}
\begin{equation}
\label{eq:vacuummotion}
\left\{
\begin{split}
 s \frac{d}{dt}\theta =& + 2 \bar{\lambda} [1-\epsilon(t)]  \sin\theta \cos\phi \sin\phi \\
                         & - 2  \Big( \frac{1}{Ns} \sum_{k\ne0} \tilde{J}_k  \Delta^{pp}_k (t) %\left\langle \tilde{p}_k \tilde{p}_{-k} \right\rangle 
                         \Big)  
                         \sin\theta \cos\phi \sin\phi \\ &
                          +2 \Big( \frac{1}{Ns} \sum_{k\ne0}  \tilde{J}_k  \Delta^{qp}_k (t) % \frac{\left \langle \tilde{q}_k \tilde{p}_{-k} + \tilde{p}_k \tilde{q}_{-k}\right\rangle}{2} 
                         \Big) 
                         \cos\theta  \sin\theta \cos^2\phi, \\
 s \frac{d}{dt}\phi =& -g + 2 \bar{\lambda} [1-\epsilon(t)]   \cos\theta \cos^2\phi  \\
                                          & -2 \Big( \frac{1}{Ns} \sum_{k\ne0}  \tilde{J}_k \Delta^{qq}_k (t) %\left\langle  \tilde{q}_k \tilde{q}_{-k} \right\rangle 
                                          \Big) 
                                          \cos\theta \cos^2\phi  \\ &
                                           +2 \Big( \frac{1}{Ns} \sum_{k\ne0}  \tilde{J}_k   \Delta^{qp}_k (t) % \frac{\left\langle\tilde{q}_k \tilde{p}_{-k} + \tilde{p}_k \tilde{q}_{-k}\right\rangle}{2} 
                                          \Big) 
                                          \sin\phi \cos\phi  , 
\end{split}
\right.
\end{equation}
%\end{widetext}
where $\Delta^{qq}_k (t)$, $\Delta^{qp}_k (t)$, $\Delta^{qq}_k (t)$ are the equal-time correlation functions
\begin{equation}
\label{eq:deltadef}
\left\{
\begin{split}
\Delta^{qq}_k (t) &\equiv  \big\langle \tilde{q}_k(t) \tilde{q}_{-k}(t) \big\rangle ,\\
\Delta^{pp}_k (t) &\equiv  \big\langle \tilde{p}_k(t) \tilde{p}_{-k}(t) \big\rangle, \\
\Delta^{qp}_k (t) &\equiv \frac{1}{2}\big\langle \tilde{q}_k(t) \tilde{p}_{-k}(t) + \tilde{p}_k(t) \tilde{q}_{-k}(t) \big\rangle,
\end{split}
\right.
\end{equation}
and, as in Eq.~\eqref{eq:epsilondef}, the non-equilibrium density $\epsilon(t)$ of spin waves reads 
\begin{equation}
\label{eq:epsilondef}
\epsilon(t) \equiv  \frac{1}{Ns} \sum_{k\ne0}\big\langle n_k (t) \big\rangle =% \frac{1}{Ns} \sum_{k\ne0}\bigg\langle\frac{\tilde{q}_{k}(t)\tilde{q}_{-k}(t)+\tilde{p}_{k}(t)\tilde{p}_{-k}(t)-1}{2}\bigg\rangle =
 \frac{1}{Ns} \sum_{k\ne0} \frac{\Delta^{qq}_k (t) + \Delta^{pp}_k (t) - 1}{2}.
\end{equation}
Using now the equations of motion for the spin wave coordinates,
\begin{widetext}
\begin{equation}
\label{eq:swmotion}
\left\{
\begin{split}
 s\frac{d}{dt} \tilde{q}_k 
                                   =& +2\bar{\lambda}  \cos^2 \phi \; \tilde{p}_k 
                                  % \\ & 
                                   -2\tilde{J}_k \sin^2\phi \; \tilde{p}_k +2\tilde{J}_k \cos\theta\cos\phi\sin\phi  \; \tilde{q}_k , \\  
 s\frac{d}{dt} \tilde{p}_k
                                   =& -2\bar{\lambda}  \cos^2 \phi \; \tilde{q}_k %\\& 
                                   +2\tilde{J}_k \cos^2\theta\cos^2\phi \; \tilde{q}_k -2\tilde{J}_k \cos\theta\cos\phi\sin\phi  \; \tilde{p}_k ,
\end{split}
\right.
\end{equation}
one obtains the evolution of the  parameters $\Delta^{qq}_k,\Delta^{qp}_k,\Delta^{pp}_k$, which describe the dynamics of the Gaussian wavefunction of the spin waves:
\begin{equation}
\label{eq:swmotiondelta}
\left\{
\begin{split}
s\frac{d}{dt}\Delta^{qq}_k  = \, & 4\tilde{J}_k  \cos\theta \cos\phi\sin\phi \,  \Delta^{qq}_k %\\& 
+4\left(\bar{\lambda} \cos^2\phi -\tilde{J}_k  \sin^2\phi \right)\, \Delta^{qp}_k ,\\
s\frac{d}{dt}\Delta^{qp}_k  = & -2\left(\bar{\lambda} \cos^2\phi - \tilde{J}_k   \cos^2\theta \cos^2\phi \right) \Delta^{qq}_k %\\& 
+ 2 \left(\bar{\lambda} \cos^2\phi -\tilde{J}_k \sin^2\phi \right) \Delta^{pp}_k ,\\
s\frac{d}{dt}\Delta^{pp}_k  = & -4\left( \bar{\lambda} \cos^2\phi - \tilde{J}_k \cos^2\theta \cos^2\phi \right) \Delta^{qp}_k %\\& 
-4\tilde{J}_k  \cos\theta \cos\phi\sin\phi \,  \Delta^{pp}_k.
\end{split}
\right.
\end{equation}
\end{widetext}
Note that the evolution does not conserve the occupation numbers $\{n_k\}$, except in the mean-field limit $\tilde{J}_{k\ne0}= 0$.
The equations of motion in Eq.~\eqref{eq:swmotiondelta} are actually not independent, as the quantities $\Delta^{qq},\Delta^{qp},\Delta^{pp}$ are  related by the condition  
\beq
\label{eq:gaussianconstraint}
4 \left(\Delta^{qp}\right)^2 = 4 \Delta^{qq} \Delta^{pp} - 1,
\eeq 
which is an exact property of Gaussian quantum states with minimal uncertainty, and which is then  satisfied at all times and for all values of $k$. 

The dynamical problem is now fully specified by the system of $2N$  coupled evolution equations~\eqref{eq:vacuummotion} and~\eqref{eq:swmotiondelta}, taking into account the constraints~\eqref{eq:gaussianconstraint}, together with suitable initial conditions, which can correspond to the ground state or to a thermal state of the pre-quench Hamiltonian. These equilibrium states, already determined in Sec.~\ref{sec:equilibrium} via the equation of state~\eqref{eq:eqstate} (and its generalization to thermal states), may be  retrieved by looking for stationary solutions of the dynamical equations with the initial parameters $g_0,\lambda_0,\tilde{J}_{k,0}$. A variation in time of $g=g(t)$, corresponding to the driving under consideration, will then yield the non-equilibrium evolution at Gaussian level according to the  dynamical equations of motion derived above.

\subsection{Dynamics: Time-independent approach}
\label{sec:timeindependent}

The system of coupled evolution equations for the collective spin and the spin waves discussed above can also be derived by using a time-independent approach. Indeed, the original Hamiltonian $H$ in Eq.~\eqref{eq:startingH} can be written in terms of two global canonical variables, the total spin projection $P$ along the $\hat{z}$-direction
\beq
\label{eq:defP}
P = s \, \tilde{\sigma}^z_{k=0} = (Ns- N_{\text{sw}})\cos\theta,
\eeq
and its conjugated angle
\beq
Q = \phi
\eeq
[see Eq.~\eqref{eq:globalPQ}], in addition to the canonical spin wave variables $\tilde{q}_k,\tilde{p}_k$ analogous to the ones introduced in the previous section [cf. Eq.~\eqref{eq:approxH-Pfourier}]. 
An explicit calculation shows that these observables provide a complete set of $2N$ canonical variables for the spin system, i.e.,
%\[
%\begin{split}
%\tilde{q}'_k & = \cos \alpha(Q,P) \; \tilde{q}_k - \sin \alpha(Q,P) \; \tilde{p}_k, \\
%\tilde{p}'_k & = \sin \alpha(Q,P) \; \tilde{q}_k + \cos \alpha(Q,P) \; \tilde{p}_k.
%\end{split}
%\]
%However, for the commutation relation \eqref{eq:timeindepcommutation} to hold, the condition $\alpha(Q,P)\equiv 0$ must hold, i.e., the directions of the two spin-wave deformations $\tilde{q}_k, \tilde{p}_k$ have to coincide with the two tangent directions to the Bloch's sphere at the point $(Q,P)$, determined by the separate variation of the two coordinates $Q,P$, respectively. This is guaranteed by the choice of frame as in equation \eqref{eq:newbasis}, as enforced by the commutation relations
\beq
\label{eq:timeindepcommutation}
[ Q, \tilde{q}_k ] = [ Q, \tilde{p}_k ] = [ P, \tilde{q}_k ] = [ P, \tilde{p}_k ] = 0.
\eeq
Expanding $H$ up to the quadratic order in the modes $\tilde{q}_k$ and $\tilde{p}_k $, while retaining the full non-linearity %and spherical nature  
in the collective spin coordinates $Q$ and $P$, one has\footnote{The ordering of the operators $Q,P$ is actually immaterial, as differently ordered expressions differ by terms of higher order in $1/N$, suppressed in the thermodynamic limit. Indeed, as explained in Sec.~\ref{sec:MF}, when $N\to\infty$ the behavior of the collective mode is classical.}
\begin{widetext}
\beq
\label{eq:timeindepH}
\begin{split}
H \simeq & - N g \frac{P}{Ns}  - N \bar{\lambda} \bigg[  \bigg(1-2\frac{N_{\text{sw}}}{Ns}\bigg) - \frac{P^2}{N^2s^2} \bigg] \cos^2 Q \\
       & -  \sum_{k\ne0} \frac{\tilde{J}_k}{s} \bigg[
       \frac{P^2}{N^2 s^2} \cos^2 Q \; \tilde{q}_k \tilde{q}_{-k} + \sin^2 Q \; \tilde{p}_k \tilde{p}_{-k} %\\
       %& 
       - 2 \frac{P}{Ns} \cos Q \sin Q \; \frac{\tilde{q}_k \tilde{p}_{-k} + \tilde{p}_k \tilde{q}_{-k}}{2}
       \bigg],
       \end{split}
\eeq
\end{widetext}
where $N_{\text{sw}}$ is defined as in Eq.~\eqref{eq:Nsw}.
%\beq
%\epsilon = \frac{1}{Ns} \sum_{k\ne0} \frac{\tilde{q}_k \tilde{q}_{-k} + \tilde{p}_k \tilde{p}_{-k} - 1}{2}.
%\eeq
Conceptually, this corresponds to promoting $\theta$ and $\phi$ in the Hamiltonian to proper dynamical variables rather than treating them as external parameters to be self-consistently adjusted, as was  the case in the time-dependent approach discussed in the previous Section. Accordingly, in this derivation, there is no need to introduce the variables $\tilde{q}_0, \tilde{p}_0$.

The equations of motion derived from the time-independent Hamiltonian~\eqref{eq:timeindepH} are
\begin{widetext}
\beq
\label{eq:eomtimeindep}
\left\{
\begin{split}
s \, \dot{Q} = & - g + 2 \bar{\lambda} \frac{P}{Ns} \cos^2 Q 
                - \frac{1}{Ns}\sum_{k\ne0}   2 \tilde{J}_k \bigg[
       \frac{P}{N s}  \cos^2 Q  \, \Delta^{qq}_k 
       -    \cos Q \sin Q  \, \Delta^{qp}_k
       \bigg] , \\
s\frac{\dot{P}}{Ns} = & - 2 \bar{\lambda} \bigg[ \bigg(1-2\frac{N_{\text{sw}}}{Ns}\bigg) - \frac{P^2}{N^2s^2} \bigg] \cos Q \sin Q 
              \\&  - \frac{1}{Ns}\sum_{k\ne0}   2 \tilde{J}_k \bigg[
       \frac{P^2}{N^2 s^2}  \cos Q \sin Q \, \Delta^{qq}_k -  \cos Q \sin Q \, \Delta^{pp}_k
       +  \frac{P}{Ns} \big( \cos^2 Q- \sin^2 Q \big) \, \Delta^{qp}_k
       \bigg] ,\\
s \, \dot{\tilde{q}}_k = & + 2\bar{\lambda} \cos^2 Q \; \tilde{p}_k - 2\tilde{J}_k \sin^2 Q \; \tilde{p}_k + 2\tilde{J}_k \frac{P}{Ns} \cos Q \sin Q \; \tilde{q}_k , \\
s \, \dot{\tilde{p}}_k = &  - 2\bar{\lambda} \cos^2 Q \; \tilde{q}_k + 2\tilde{J}_k \frac{P^2}{N^2s^2} \cos^2 Q \; \tilde{q}_k - 2\tilde{J}_k \frac{P}{Ns} \cos Q \sin Q \; \tilde{p}_k ,
\end{split}
\right.
\eeq
\end{widetext}
where the $\Delta_k$'s are defined as in Eq.~\eqref{eq:deltadef}. 

Crucially, the quantum fluctuations of the collective operators $P/N$ and $Q$ in the initial state are of order $1/\sqrt{N}$, and hence  they behave like uncertainty-free classical variables in the thermodynamic limit (see Sec. \ref{sec:finitesizeLMG} for details on the convergence to the classical behavior). By identifying $Q=\phi$ and by changing variable from $P$ to $\theta$ via Eq.~\eqref{eq:defP}, after taking quantum averages with $\Braket{N_{\text{sw}}}/(Ns)\equiv \epsilon$ [cf. Eq.~\eqref{eq:epsilondef}], one finds
\[
%P = Ns (1-\epsilon) \cos\theta \qquad \Leftrightarrow \qquad 
\dot{\theta}= \frac{\frac{\dot{P}}{Ns}+\cos\theta \; \dot{\epsilon}}{-(1-\epsilon)\sin\theta},
\]
and one easily verifies that the equations of motion~\eqref{eq:eomtimeindep} obtained here are equivalent to Eqs.~\eqref{eq:vacuummotion} and~\eqref{eq:swmotion} obtained within the time-dependent spin wave approach, to the quadratic order in the quantum fluctuations.

The quadratic spin-wave expansion discussed in this Section indicates that  the system with Hamiltonian~\eqref{eq:startingH} can be alternatively regarded  as being composed by a macroscopic classical degree of freedom $(Q,P)$, corresponding to the collective spin, interacting with an extensive ensemble of quantum oscillators $\{(\tilde{q}_k,\tilde{p}_k) \}_{k\ne0}$, see Eq.~\eqref{eq:timeindepH}. %This is reminiscent of Caldeira-Leggett models, although the bath of microscopic degrees of freedom is in our case self-generated by the many-body Hamiltonian dynamics of the  spin system.

\section{Impact of the spin waves \\ on the mean-field dynamical phase transition}
\label{sec:results}

Let us now apply the methods described in the previous section in order to study the impact of  fluctuations on the mean-field  dynamical criticality discussed in Sec. \ref{sec:MF}.
%of the LMG model and generalizations thereof after a quench or a ramp of the transverse field. %, and the order/chaos transition and the dynamical stabilization in the presence of periodic driving. 
For the sake of definiteness, we will first depart from the exactly solvable mean-field limit by considering a model in one dimension where a nearest-neighbor interaction is added to the infinite-range interaction of the LMG model \cite{LeroseShort}. % and which determine its mean-field character in the thermodynamic limit.% a short-range interaction on top of the mean-field Hamiltonian~\eqref{eq:MFH}, that brings into the model a non-trivial spatial structure. 
%As remarked in the introduction and further detailed in Sec.~\ref{sec:generalchaos} below, however, 
%the time-dependent spin wave theory applies equally well to any spatially-decaying couplings in any spatial dimensionality, and similar results can be obtained.) 
A similar analysis is then carried out and similar results are obtained in Sec.~\ref{sec:generalchaos} for a much wider class of models.

We thus consider the Hamiltonian
\beq
\label{eq:startingH1dnn}
H = - \frac{\lambda}{N} \sum_{i,j=1}^N \sigma_i^x \sigma_j^x - g \sum_{i=1}^N \sigma_i^z -  J \sum_{i=1}^N \sigma_i^x \sigma_{i+1}^x,
\eeq
where the strength of the nearest-neighbor perturbation is controlled by the parameter $J$ and periodic boundary conditions are understood. In the opposite limit $\lambda \to 0$ with finite $J$, the model reduces to the well-known quantum Ising chain in a transverse field, which is  exactly solvable in terms of free Bogolubov fermions \cite{Sachdevbook}. In this case, however, dynamical criticality  disappears, as discussed in the Introduction.

In order to study the resulting dynamics, we will use Eqs.~\eqref{eq:vacuummotion},~\eqref{eq:deltadef},~\eqref{eq:epsilondef},~\eqref{eq:swmotiondelta}, where $\tilde{J}_k=J \cos k$ with $k=(2\pi/N)j$, $j=-(N/2)+1,\dots,-1,0,1,\dots,(N/2)-1,N/2$ for this periodic one-dimensional chain.

\subsection{Quench: Chaotic dynamical %ferromagnetic 
phase}
\label{sec:quenchchaos}

\begin{figure}[t]
\centering
\includegraphics[width=0.9\columnwidth]{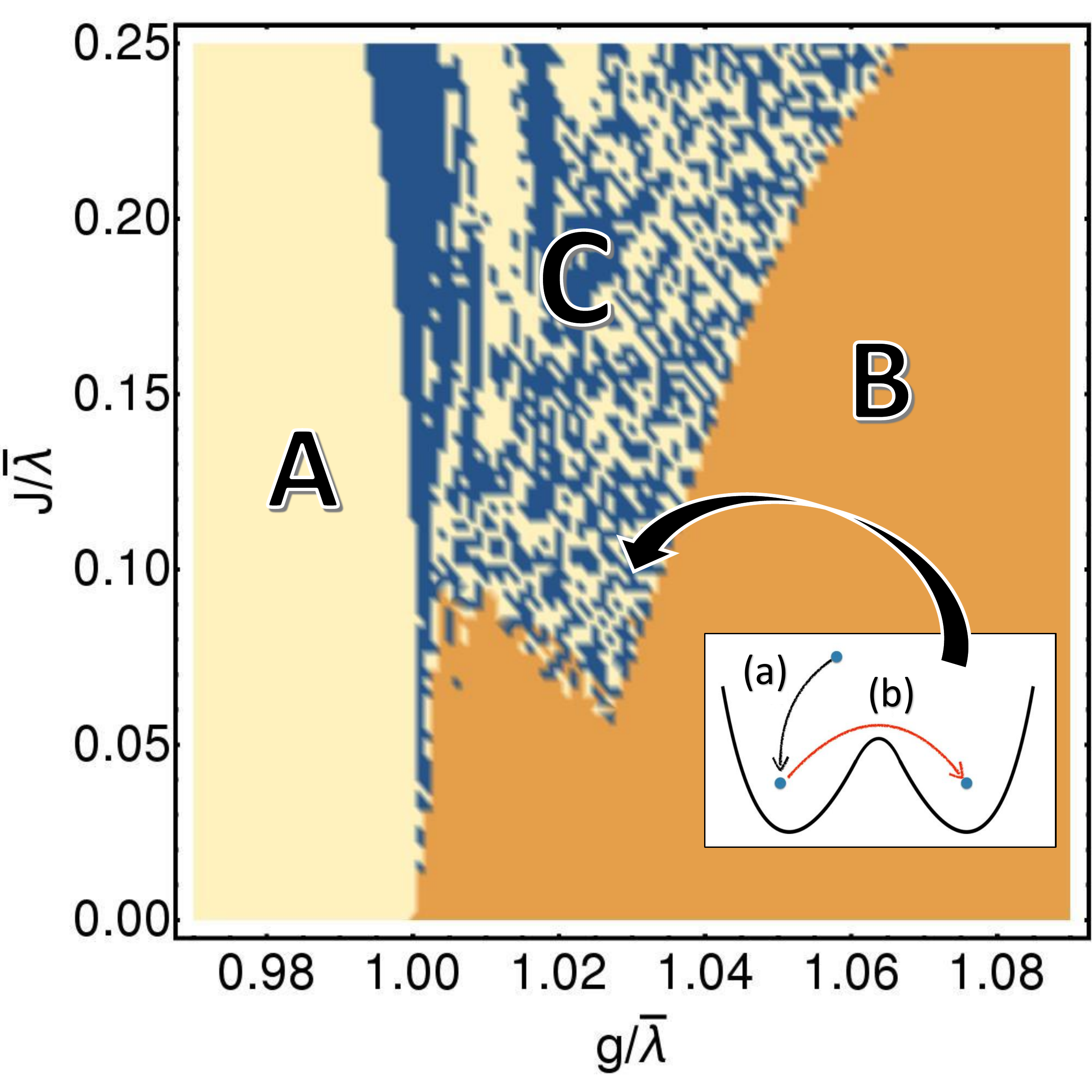}
\caption{
%\bzcom{Perhaps it would be interesting to show next to this phase diagram also the phase diagram which shows the absolute value of the final magnetization. Also the labels need to be corrected here.}
[Color online]
Dynamical phase diagram of the model in Eq.~\eqref{eq:startingH}  after a quench of the magnetic field $g_0=0 \to g$ starting from the fully polarized ground state with positive magnetization, as a function of $g$ and $J$. Here $N=100$. As energy scale we choose $\bar{\lambda}\equiv \lambda+J$.
The color of each point of the diagram is determined by the sign of the long-time average $\overline{ \sigma^x}$ of $\sigma^x(t)$: light yellow for $\overline{\sigma^x}>0$, orange for $\overline{\sigma^x}=0$, and blue for $\overline{ \sigma^x}<0$. 
Regions A and B are perturbative extensions of the dynamical ferromagnetic and paramagnetic phases of the LMG model with $J=0$, corresponding to the horizontal axis (see Fig.~\ref{fig:robustness} for an illustration of the dynamics within A and B). 
Upon increasing $J$ at fixed $g$, in a neighborhood   of the mean-field critical point $g=\bar{\lambda}$,  a new \emph{chaotic dynamical ferromagnetic phase} C emerges, within which the magnetization $\sigma^x(t)$, after an initial dynamical paramagnetic behavior, gets trapped in one of the two  symmetry-broken sectors with opposite signs of the collective magnetization  [process (a) in the inset], in some cases followed by  hopping between them  [process (b) in the inset] (see Fig.~\ref{fig2} for an illustration of the dynamics within C). 
The extent  and features of the three phases A, B, C are stable as $N$ is increased. 
%Note the emergence of more regular ``stripes'' on the left of region C.
%See Fig.~\ref{fig2} for an illustration of the dynamics in region C.% \jm{The boundaries separating the various phases in the $(g,J)$-diagram become sharper as $N$ increases.}
}
\label{fig1}
\end{figure}

We numerically integrated the evolution equations~\eqref{eq:vacuummotion} and~\eqref{eq:swmotiondelta} for a range of post-quench values of $g$ and $J$ and starting from a fully polarized ferromagnetic initial state with $\Braket{\sigma^x_j(t=0)}=1$ (i.e., the pre-quench Hamiltonian is chosen with $g_0=0$, and the value of $J_0$ is thus actually immaterial as long as $\lvert J_0 \rvert < \lambda$). 
At each  integration time, we compute the time-dependent components of the average collective spin $\vec \sigma$,
\beq
\vec{\sigma}(t)\equiv \frac{1}{N}\Big\langle \tilde{\vec{\sigma}}_{k=0}(t)\Big\rangle= [1-\epsilon(t)]
\begin{pmatrix}
\sin\theta(t) \cos\phi(t) \\
\sin\theta(t) \sin\phi(t) \\
\cos\theta(t)
 \end{pmatrix},
\eeq
verifying that the non-equilibrium density $\epsilon(t)$ of spin waves [see Eq.~\eqref{eq:epsilondef}] approaches asymptotically a small value at long times within the range of parameters considered.
From this $\vec{\sigma}(t)$, we compute the long-time average of the magnetization along the ferromagnetic direction $\hat{x}$, i.e., the dynamical order parameter $\overline{\sigma^x}$,
and plot it for different values of $J$ and $g$, coloring the corresponding point in light yellow if $\overline{\sigma^x}>0$ (dynamical ferromagnetic ordering in the initial sector), in orange if $\overline{\sigma^x} = 0$ (dynamical paramagnetic behavior), and in blue if  $\overline{\sigma^x}<0$ (reversed dynamical ferromagnetic ordering).
The result is the dynamical phase diagram reported in Fig.~\ref{fig1}.
%
%

%In Fig.~\ref{fig1}, we portray  the two dynamical ferromagnetic and paramagnetic phases  present for $J=0$, surviving %even for
%at finite coupling $J>0$: their perturbative robustness against the  presence of spin wave fluctuations is discussed in   Fig.~\ref{fig:robustness}.

\begin{figure*}
\centering
\begin{tabular}{cc}
\includegraphics[width=7cm]{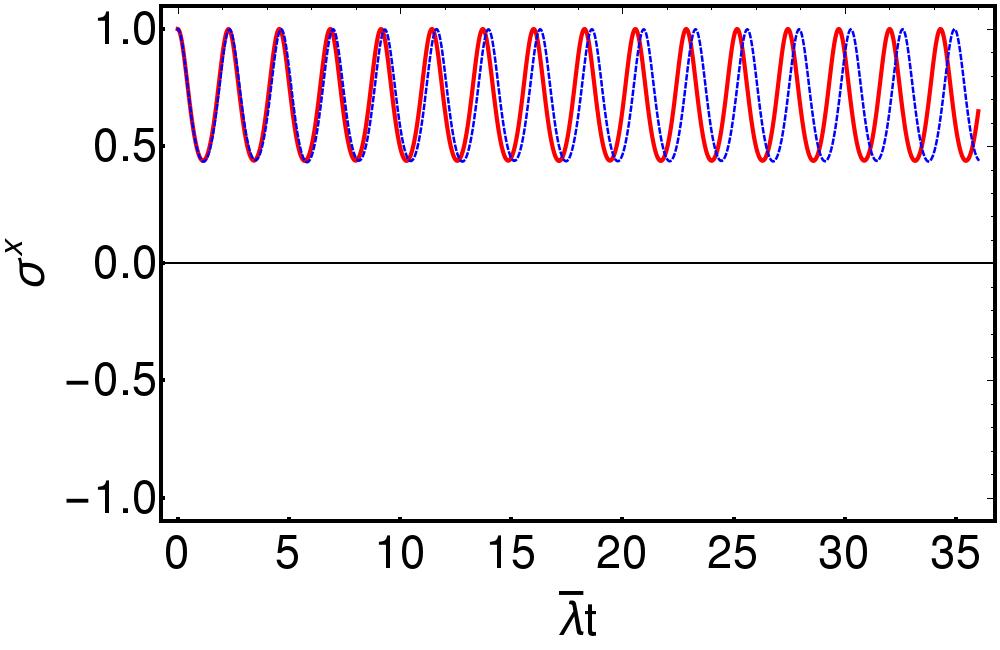} & \includegraphics[width=7cm]{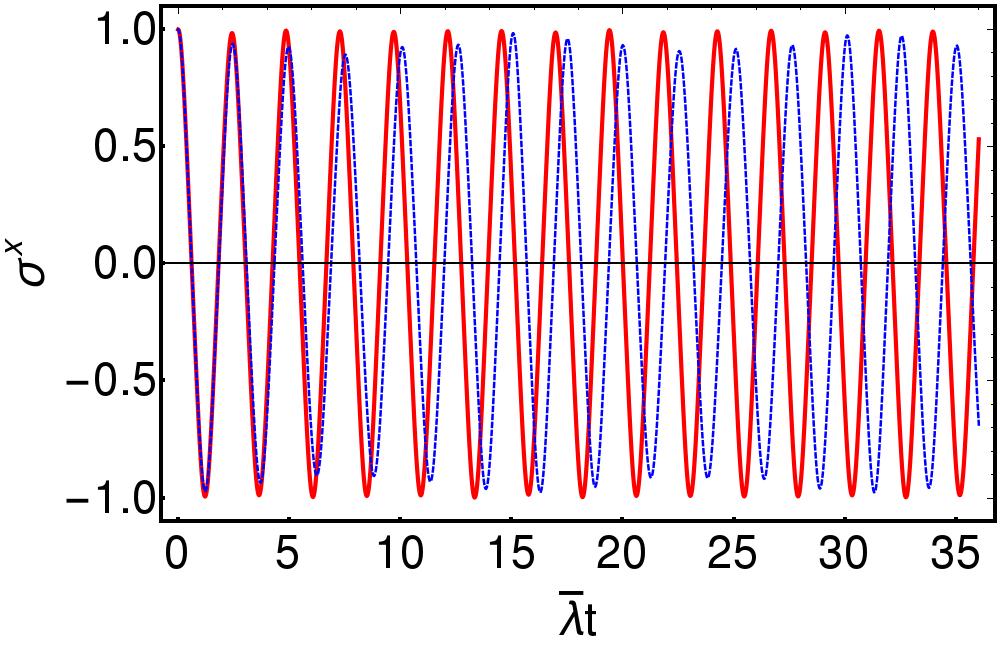} \\
\includegraphics[width=7cm]{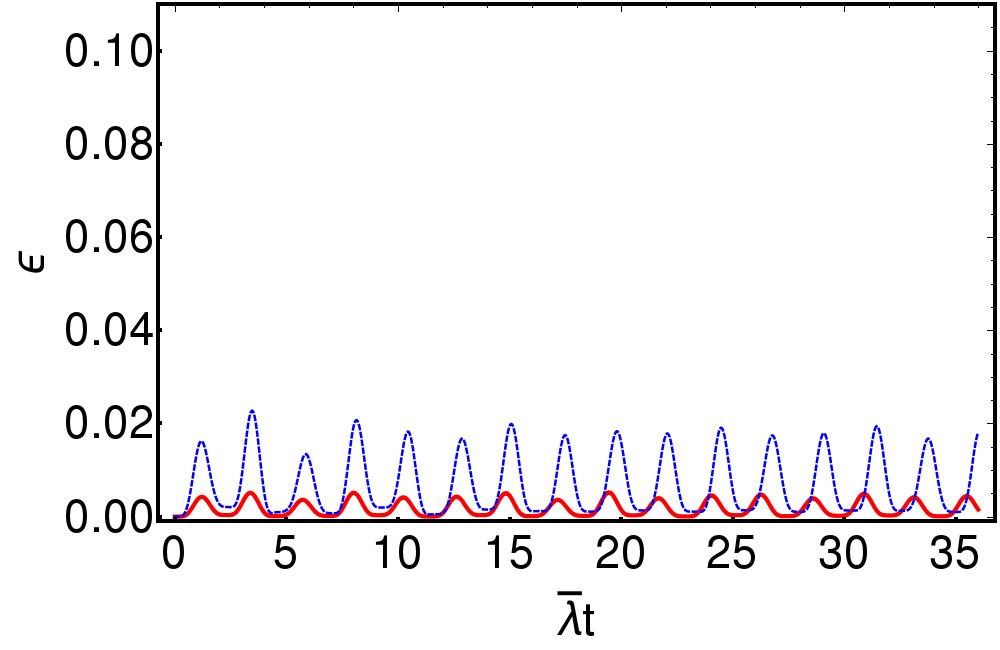} & \includegraphics[width=7cm]{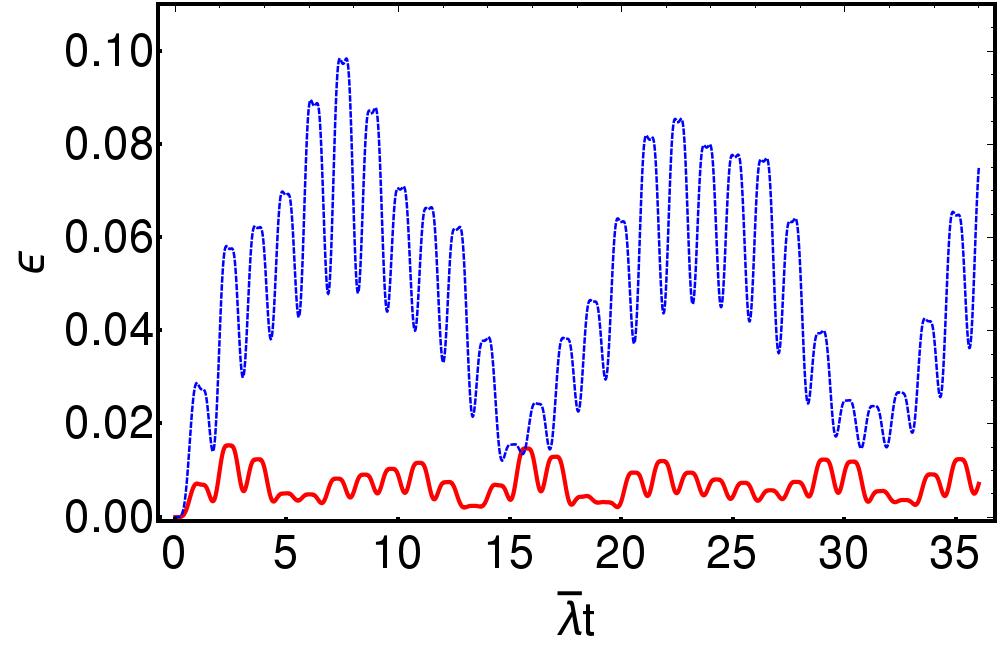}
\end{tabular}
\caption{
[Color online]
Dynamical behavior of the order parameter $\sigma^x(t)$ (first row) and of the spin wave density $\epsilon(t)$ (second row) in the presence of a short-range interaction $J/\bar{\lambda}=0.1$ (solid red line) and $0.2$ (dashed blue), after a quench from a fully polarized ferromagnetic state ($g_0=0$). 
Left panels: dynamical ferromagnetic phase with $g/\bar{\lambda}=0.9$. 
Right panels: dynamical paramagnetic phase with $g/\bar{\lambda}=1.5$. 
These dynamical phases are characterized by the sign of the time-average of $\sigma^x(t)$, shown in the top panels. 
The quantity $\epsilon(t)$ shown in the bottom panels represents the total amount of spin wave excitations generated during the non-equilibrium evolution. This is the control parameter for the validity of the low-density expansion, which is consistent if $\epsilon\ll 1$, i.e., if the length of the total spin $\lvert \vec{\sigma}(t) \rvert = 1 - \epsilon(t)$ remains close to its maximal value. The presence of a short-range interaction, even of sizable strength $J/\bar{\lambda}=0.2$, produces a perturbative modification of the mean-field evolution and,   correspondingly, a small amount of spin waves. In particular, the mean-field persistent oscillations are not damped by the self-generated ``bath''. In the plots, $N=100$ and the mean-field dynamical critical point is $g_{\text{dyn}}/ \bar{\lambda}=1$.
}
\label{fig:robustness}
\end{figure*}

This figure shows that the dynamical ferromagnetic and paramagnetic phases A and B respectively, which touch each other at the dynamical critical point for $J=0$, withstand the effects of the quantum fluctuations introduced by having $J\ne0$, apart from %a possible magnetization reversal occurring rather 
getting separated by a new phase C
close to $g\simeq \bar{\lambda}$ (note the horizontal scale of Fig.~\ref{fig1}). %, corresponding to phase C. 
The robustness of phases A and B is further demonstrated in Fig.~\ref{fig:robustness}, which shows the time-evolution of the order parameter $\sigma^x(t)$ (first row) and of the spin wave density $\epsilon(t)$ (second row) within the dynamical ferromagnetic (first column) and paramagnetic (second column) phases, %away from the dynamical transition point, i.e.,
 with $g/\bar{\lambda}=0.9$ and $g/\bar{\lambda}=1.5$ respectively. (Note that these values are well outside the range covered by Fig.~\ref{fig1}.) The red solid and blue dashed lines correspond to increasing values of the coupling $J$ with spin waves, which, as anticipated, do not alter significantly the qualitative features of the dynamics. %The only visible effect is that, in 
Note that in both the dynamical phases A and B, $\epsilon(t)$ remains sufficiently small and therefore we expect the spin wave treatment developed in Sec.~\ref{sec:methodweakfluct} to be accurate and these two phases to be robust.

%\begin{widetext}

\begin{figure}
\centering
\includegraphics[width=7cm]{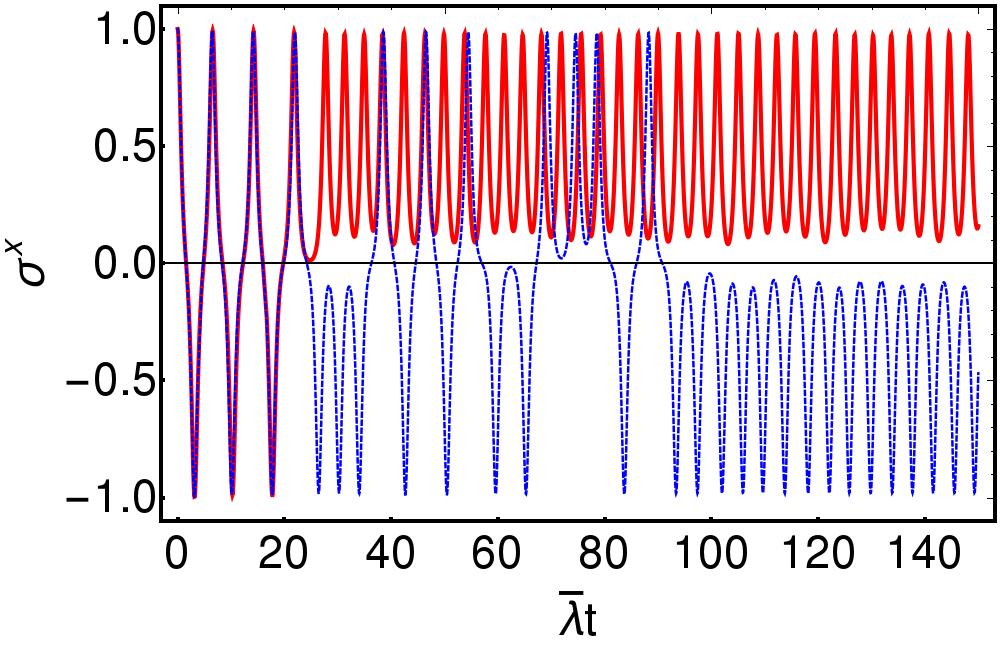}
\caption{
[Color online]
Evolution of the order parameter $\sigma^x(t)$ within the chaotic dynamical ferromagnetic phase 
C in Fig.~\ref{fig1}, for $g/\bar{\lambda}=1.03$ (solid red line) and $1.031$ (dashed blue), and with $J/\bar{\lambda}=0.1$. Here $N=100$. %The inset refers to the case $J=0.1$ and shows the 
The two lines are practically indistinguishable during the initial  paramagnetic transient, but they have markedly distinct fates at the onset of the critical process denoted by  (a) in the inset of Fig.~\ref{fig1} and they eventually end up into 
{distinct wells}.
%opposite ferromagnetic sectors. 
In both cases, $\epsilon(t)$ grows from $\epsilon(t=0)=0$ to values around $0.04$. % -- see also \cite{SM}).
This extreme sensitivity on the value of $g$ (and of $J$) is at the origin of the ``mosaic'' structure of region C in Fig.~\ref{fig1}.}
\label{fig2}
\end{figure}

Close to the mean-field dynamical transition point $g = \bar{\lambda}$, however, the system becomes extremely sensitive to non-equilibrium  fluctuations, resulting in the peculiar  phase C. In a typical point of this region, 
%\ag{Within C,} 
the dynamics of $\sigma^x(t)$ is driven by two  processes, illustrated in the inset of Fig.~\ref{fig1}: (a) the decay from a transient paramagnetic behavior to one of the two possible ferromagnetic sectors, 
and (b) the possible hopping between them.  
%
%\end{widetext}
%
Heuristically, these phenomena occur when the energy of the macroscopic collective spin $\vec{\sigma}(t)$ is  slightly above the barrier  separating the two ferromagnetic minima. In this case the dynamical production of spin waves reduces the energy carried by $\vec{\sigma}$ and hence causes the dynamical trapping into one of the two ferromagnetic wells, accompanied by an increase of the spin wave density $\epsilon(t)$. 
The system is dynamically ferromagnetic, although it can occasionally hop to the opposite well, %sector, 
with a process assisted by the absorption of energy from the spin wave bath. The asymptotic sign of $\sigma^x(t)$, %magnetization, 
and therefore the sign of $\overline{\sigma^x}$,  sensitively depends  on the specific values of the parameters in a large portion of this novel dynamical ferromagnetic region (C in Fig.~\ref{fig1}), with a collective pseudo-aleatory character of the dynamics, which is illustrated in Fig.~\ref{fig2}. Due to this sensitive dependence on the post-quench values of the parameters, which actually implies the same for the choice of the initial state, this phase C is referred to as ``chaotic''.

This  dynamical behavior, obtained on the basis of the time-dependent spin wave theory, persists up to values $J/\bar{\lambda} \simeq 0.67$ of the coupling $J$, i.e., $J=2\lambda$; at such strong coupling $\epsilon(t)$   grows significantly, invalidating the low-density spin-wave expansion.  %Accordingly, 
In order to explore this strong coupling regime, we relied on 
a
time-dependent variational principle developed on matrix product states, see Sec.~\ref{sec:bojan}.

The dynamics in the chaotic dynamical ferromagnetic region C may be understood qualitatively %in terms of 
%rationalized 
via a simple analogy:~a coin toss. The toss corresponds to the sudden quench of the external field, where a macroscopic amount of energy is injected into the system in the form of regular macroscopic motion. The coin repeatedly hitting the ground and exciting its phonons  corresponds to the loss of energy in favor of the microscopic degrees of freedom, i.e., the spin waves. Finally, the coin settling into one of the two macroscopically distinct stable configurations (heads or tails) corresponds to the trapping into one of the two ferromagnetic sectors ($\overline{\sigma^x}>0$ or $\overline{\sigma^x}<0$). A diagram of the outcomes of the coin (heads or tails) as a function of the variables which  parameterize the toss, would result in a picture very similar to region C of Fig.~\ref{fig1}, as indeed shown in Ref.~\onlinecite{coin}. 
{Although the equations of motion in both cases are completely deterministic, the final outcome is extremely sensitive to the details of the dynamics and it can be considered as an \emph{effectively random} process}.
We emphasize that we checked that the %picture is stable 
numerical results reported above in Fig.~\ref{fig1} and in the figures which follow are not affected as $N$ is increased (up to $N=400$).

\subsection{Generality of the chaotic dynamical phase}
\label{sec:generalchaos}

\begin{figure*}
\centering
\begin{tabular}{cc}
\includegraphics[width=7cm]{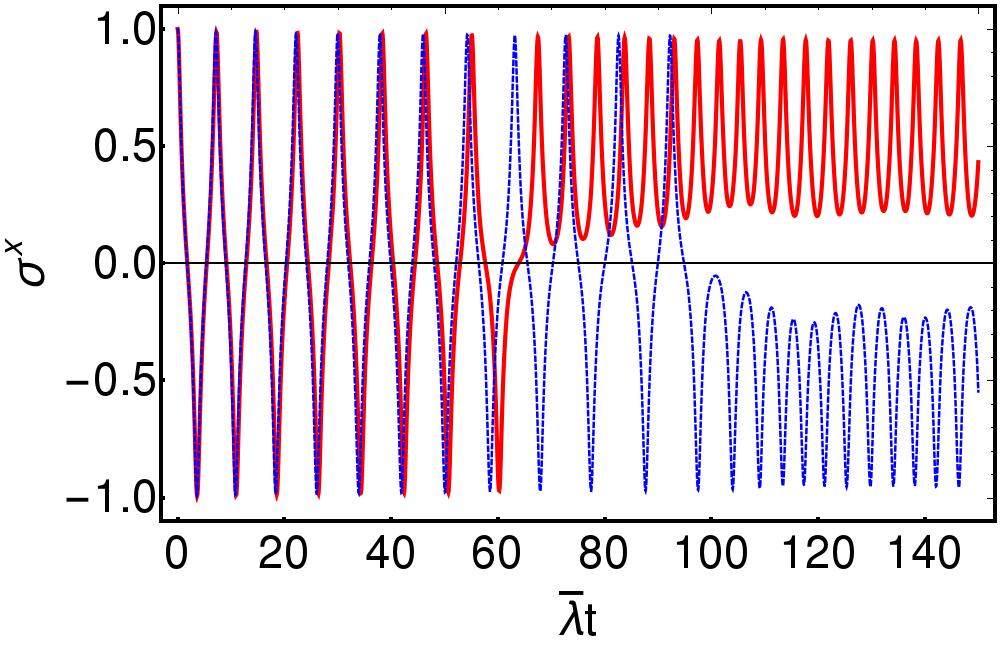} &
\includegraphics[width=7cm]{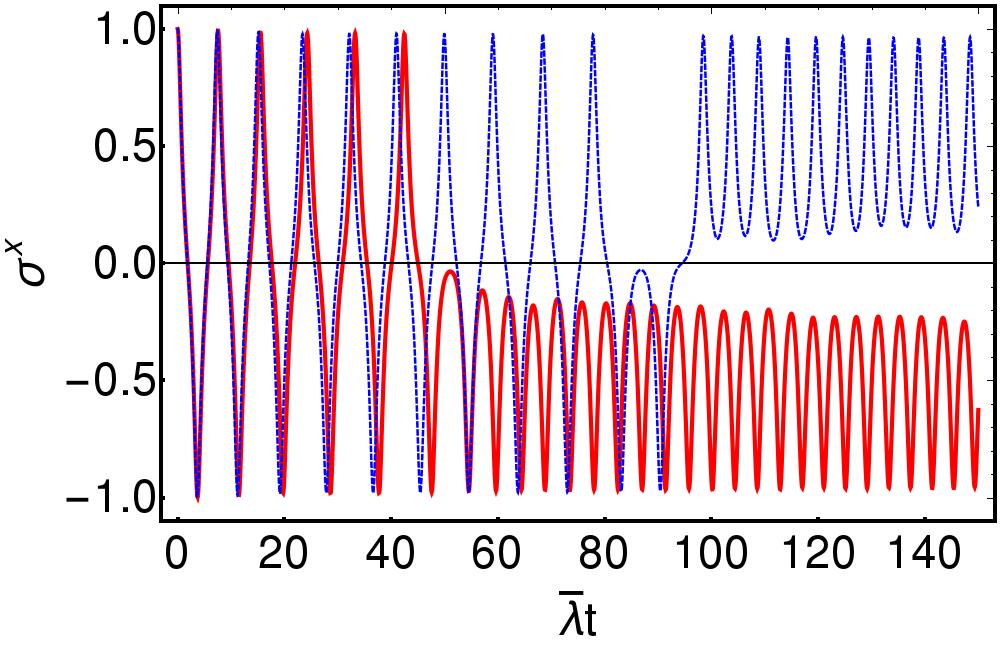} \\
\includegraphics[width=7cm]{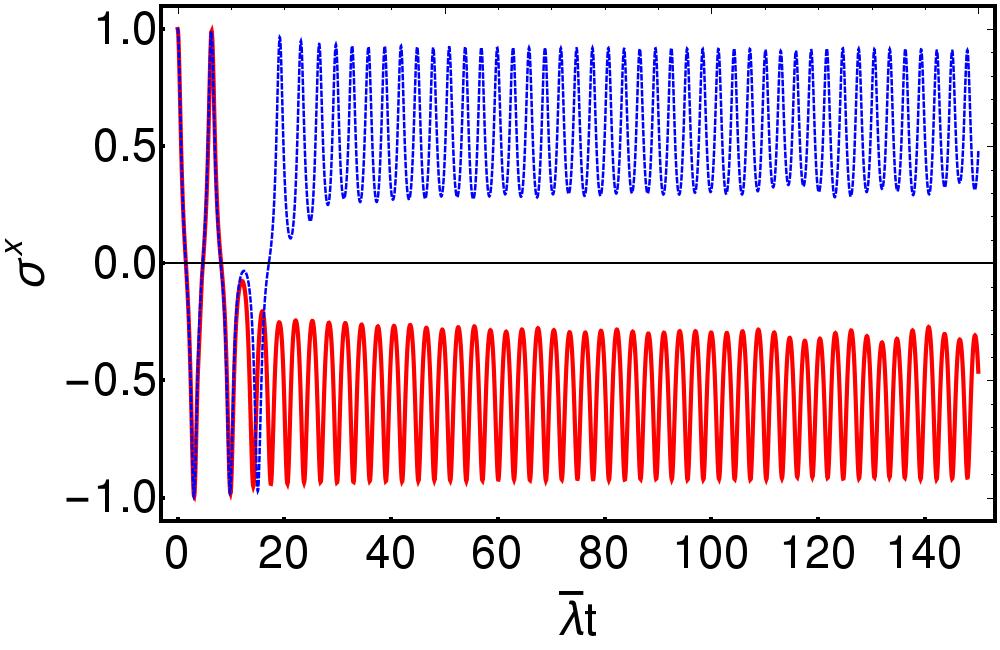} &
\includegraphics[width=7cm]{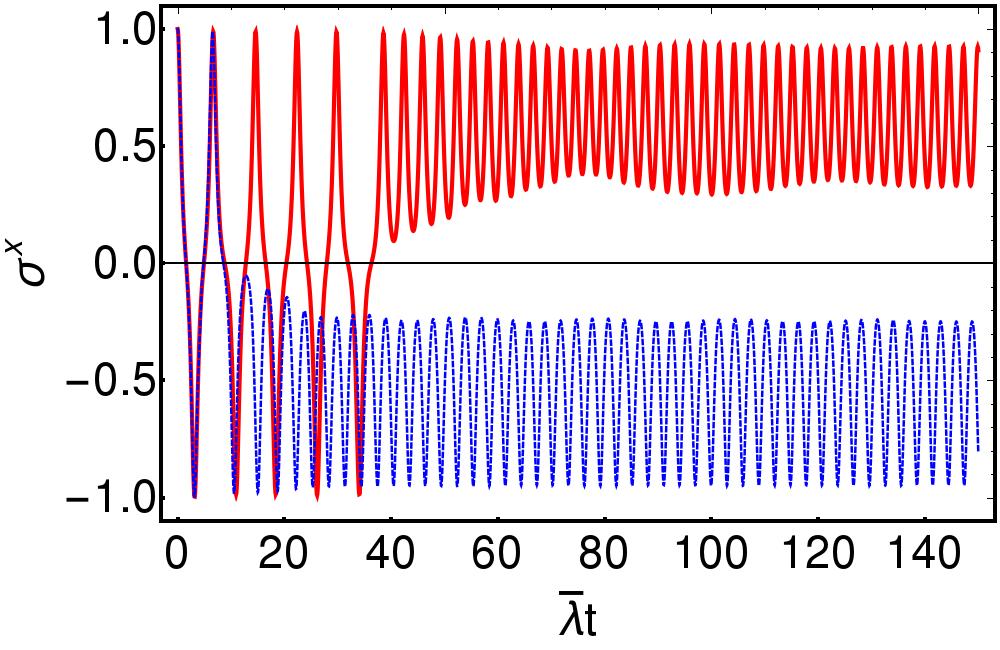}
\end{tabular}
\caption{
[Color online]
Evolution of the order parameter $\sigma^x(t)$ after a quench from a pure ferromagnetic state ($g_0=0$) in four different generalizations of the Ising Hamiltonian~\eqref{eq:startingH}.
Top left: XY spin chain with an infinite-range and a nearest-neighbor interaction, defined by Eq.~\eqref{eq:XYZ} with $\alpha_y=0.25$, $\alpha_z=0$, $g/\bar{\lambda}=1.03$ (solid red line) and $1.032$ (dashed blue), with $J/\bar{\lambda}=0.4$. 
Top right: XYZ spin chain with an infinite-range and a  nearest-neighbor interaction, defined by Eq.~\eqref{eq:XYZ} with  $\alpha_y=0.25$, $\alpha_z=0.125$,  $g/\bar{\lambda}=0.9$ (solid red line) and $0.902$ (dashed blue), with $J/\bar{\lambda}=0.4$.
Bottom left: Ising spin chain with an infinite-range and a  next-to-nearest-neighbor interaction, defined by Eq.~\eqref{eq:LR} with $v(r)=\delta_{r,1}+0.5 \delta_{r,2}$, $g/\bar{\lambda}=1.03$ (solid red line) and $1.031$ (dashed blue), with $J/\bar{\lambda}=0.2$.
Bottom right: Ising spin chain with an infinite-range and a  power-law decaying interaction, defined by Eq.~\eqref{eq:LR} with $v(r)=1/r^{2}$, $g/\bar{\lambda}=1.03$ (solid red) and $1.031$ (dashed blue), with $J/\bar{\lambda}=0.2$.
In all simulations, $N=100$. These trajectories have been obtained by numerically integrating the evolution equations given by the time-dependent spin wave theory, analogous to Eqs.~\eqref{eq:vacuummotion}, \eqref{eq:swmotiondelta}, derived for the generalized spin chains above through the same procedure as that explained in details in Sec.~\ref{sec:tdswt} for the Ising model.
}
\label{fig:generalCDP}
\end{figure*}

We now show  that the chaotic dynamical  phase is not peculiar to the model in Eq.~\eqref{eq:startingH1dnn}, but is actually expected to emerge in a rather general class of ferromagnetic spin systems, characterized by competition between long- and short-range interactions. 

First of all, the occurrence of the chaotic dynamical phase discussed in the previous section does not depend on the particular initial state we have chosen, as long as it has sufficiently strong magnetic ordering. In particular, the direction of the initial magnetization $\tr\big[\rho(t=0) \vec{S} \, \big] \propto (\sin\theta_0,0,\cos\theta_0)$ with $\theta_0\ne0$ is immaterial, and the initial state $\rho(t=0)$ needs not be pure. This class encompasses all low-temperature equilibrium ordered states of $H(g<g_{\text{cr}})$.

The chaotic dynamical phase occurs for arbitrary quantum spin magnitude, since a larger value of $s$ 
%represents   just a multiplicative constant in the spin wave equations derived in Sec~\ref{sec:methodweakfluct}.% (there, we adopted $s=1/2$ for definiteness).
%In the limit $s\to\infty$ the effect of fluctuations is suppressed 
just amounts to rescaling the coupling strength $J$ in Eqs.~\eqref{eq:vacuummotion}, and therefore to decreasing the overall effect of the feedback from fluctuations on the evolution of the collective order parameter. 
In addition, in the limit $s\to\infty$, quantum fluctuations in the pre-quench ground state are suppressed as a consequence of the individual spins approaching their classical limit. Accordingly, the chaotic dynamical phase progressively disappears. % \alessio{not with thermal initial states: comment}
However, thermal fluctuations can play a role similar  to that of quantum fluctuations when initial states in equilibrium with a finite temperature $T>0$ are considered, leading to a non-vanishing feedback and thus to an extended chaotic dynamical phase even in the classical limit.

We also expect  that the phenomena discussed here for a spin chain with an interaction characterized by $\mathbb{Z}_2$ symmetry should emerge also for other discrete symmetry groups.
%replacing a spin chain characterized by a $\mathbb{Z}_2$ symmetry with other discrete symmetry groups will not affect the qualitative aspects of our picture. 
In the case of a ``clock'' %permutation 
symmetry $\mathbb{Z}_n$, % elements, 
for instance, the dynamical order parameter is expected to
%to jump among the possible $n$  minima through spin wave assisted hops, until eventually one of them is selected. \alessio{no, emphasis on trapping. like a roulette}
get eventually  trapped into one of the $n$ distinct symmetry-breaking sectors, resulting in a multicolor version of the picture of Fig.~\ref{fig1} with different colors corresponding to the $n$ possible sectors. In this case, the appropriate heuristic analogy would be that of a ``roulette'' rather than a coin.

%We have explicitly checked in detail that  a broad class of Hamiltonians can host the chaotic dynamical ferromagnetic phase discussed in the previous Section.
%
%First of all, 
Furthermore, changing the  short-range spin-spin interaction term $J$ from ferromagnetic to antiferromagnetic ($J<0$) does not alter the structure of the phase diagram in Fig.~\ref{fig1}. 
Indeed, the time-dependent spin wave theory evolution equations \eqref{eq:vacuummotion} do not change when $J\mapsto -J$, provided the substitution of the summation variable $k\mapsto\pi -k$ is performed\footnote{Note, however, that completely different phenomena are expected in the presence of antiferromagnetic \emph{long-range} interactions, i.e., when $\lambda<0$: see, e.g., Ref.~\onlinecite{GiulianiLieb}.}.

 We now turn our attention to generalizations of the Ising Hamiltonian~\eqref{eq:startingH}. The top panel of Fig.~\ref{fig:generalCDP} shows the evolution of the order parameter $\sigma^x(t)$ for the XY (top left panel) and XYZ (top right) versions of the LMG model with a nearest-neighbor anisotropic perturbation, defined by
 %In particular, the first two models are defined by
 % we report  the existence of a chaotic phenomenon in the critical XY and XYZ versions of the Lipkin model perturbed by a short-range anisotropic coupling,
\begin{equation}
\begin{split}
&H_{XYZ}=-\frac{\lambda}{N}\sum_{i,j=1}^N \Big( \sigma^x_i \sigma^x_j + \alpha_y  \sigma^y_i \sigma^y_j + \alpha_z  \sigma^z_i \sigma^z_j \Big) \\ &- g \sum_{i=1}^N \sigma^z_i  - J\sum_{i=1}^N \Big( \sigma^x_i \sigma^x_{i+1} + \alpha_y  \sigma^y_i \sigma^y_{i+1} + \alpha_z  \sigma^z_i \sigma^z_{i+1} \Big).
\end{split}
\label{eq:XYZ}
\end{equation}
with $\alpha_z=0 $ (XY model) or non-vanishing values of $\alpha_{y,z}$ (XYZ model), while they reduce to the LMG model in Eq.~\eqref{eq:startingH1dnn} for $\alpha_y=\alpha_z=0$. Note that at the isotropic point $\alpha_y=1$ the discrete $\mathbb{Z}_2$ symmetry turns into a continuous $O(2)$ symmetry, as $\sigma^z$ is conserved. Consequently, the barrier separating the two ferromagnetic minima becomes increasingly shallow as this point is approached, which hinders the possibility for the collective order parameter to get trapped. Accordingly, the chaotic dynamical phase disappears in this limit.

 The bottom panel of Fig.~\ref{fig:generalCDP} shows the evolution of the order parameter $\sigma^x(t)$ for the LMG model with a next-to-nearest-neighbor (bottom left) or algebraically decaying (bottom right) perturbation, defined by
 %The last two models, instead, are encompassed by the Hamiltonian
%The LMG model perturbed by a short-range Ising type of perturbation~\eqref{eq:startingH1dnn} is recovered setting  $\alpha_y=\alpha_z=0$.
%The two bottom panels of Fig.~\ref{fig:generalCDP} present instead the chaotic phenomenon for the LMG model perturbed by second-nearest-neighbor spin-spin interactions or even power-law decaying ones, as  in the following Hamiltonian
\begin{equation}
\begin{split}
H_{LR}=&-\frac{\lambda}{N}\sum_{i,j=1}^N \sigma^x_i \sigma^x_j - g \sum_{i=1}^N \sigma^z_i  - J\sum_{i,r}^N v(r) \sigma^x_i \sigma^x_{i+r} ,
\end{split}
\label{eq:LR}
\end{equation}
where $v(r)$ decays to zero upon increasing the distance $r$ between the two interacting spins. For finite-range perturbations $v(r)$ has a compact support, while for power-law decaying interactions one has $v(r)\propto r^{-\alpha}$ with $\alpha>0$. 
The qualitative similarity of all the panels in Fig.~\ref{fig:generalCDP} with the evolution displayed in Fig.~\ref{fig2} demonstrates that the chaotic behavior observed in the latter case is actually a generic phenomenon which emerges also in the generalized models discussed above. In particular, the evolution of a certain initial state under the effect of two close post-quench Hamiltonians (red and blue curves) results into two markedly different asymptotic states. Although Fig.~\ref{fig:generalCDP} refers to specific choices of the various parameters involved, we verified that this strong sensitivity of the dynamics to the values of the parameters of the post-quench Hamiltonian persists in a neighborhood of the points considered.
We finally observe that the spatial dimensionality of the short-range perturbation does not play an important role, as well.
%We have also explicitly checked that the initial non-equilibrium state does not alter the possible onset of the chaotic dynamical phase. 
%\alessio{comment on other generalizations, initial states, thermal, etc}

In summary, we have shown that the emergence of a chaotic dynamical phase is an ubiquitous phenomenon that  requires essentially two sole physical ingredients, namely the spontaneous breaking of a discrete symmetry and a mean-field  model perturbed by an interaction term with a non-trivial spatial dependence, which introduces  fluctuations. % in the form of spin-waves.

\subsection{Correlation function of the local order parameter}
\label{sec:corr}

According to the picture presented above, the pre-thermal dynamics of the system can be understood in terms of the motion of a classical, macroscopic degree of freedom (the collective spin $\vec{\sigma}$) coupled to a weakly interacting many-body system (the ``bath'' of spin waves), which, in turn, %periodically 
is driven by the former: see Eqs.~\eqref{eq:timeindepH} and~\eqref{eq:eomtimeindep}. 
This driving mechanism is  %\emph{self-consistently} 
determined by the persistent precession of the collective spin and can be highlighted by studying 
 the time- and space-dependent equal-time correlation functions $\big\langle \sigma^x_j(t) \sigma^x_{j+r}(t) \big\rangle$, of the local order parameter $\langle \sigma^x_j(t) \rangle$.
Taking into account Eq.~\eqref{eq:approxH-P} at the leading order in the low-density expansion of Sec.~\ref{sec:methodweakfluct}, the connected correlation function $C^{xx}(r,t)$ can be expressed as
\begin{widetext}
\beq
\label{eq:correlationsxx}
\begin{split}
C^{xx}(r,t) \equiv  
 &\big\langle \sigma^x_j(t) \sigma^x_{j+r}(t) \big\rangle -  \big\langle \sigma^x_j(t) \big\rangle \big\langle \sigma^x_{j+r}(t) \big\rangle    \\ = &
      \left( \hat{X} \cdot \hat{x} \right)^2 \frac{1}{s} \big\langle q_j(t) q_{j+r}(t)\big\rangle  
    +\left( \hat{Y} \cdot \hat{x} \right)^2 \frac{1}{s} \big\langle p_j(t) p_{j+r}(t)\big\rangle + %\\ & +
    2  \left( \hat{X} \cdot \hat{x} \right) \left( \hat{Y} \cdot \hat{x} \right) \frac{1}{s} \bigg\langle\frac{ q_j(t) p_{j+r}(t)+p_j(t) q_{j+r}(t)}{2}\bigg\rangle 
    \\ = &
     \cos^2\theta(t) \cos^2\phi(t) \, \frac{1}{Ns} \sum_{k\ne0} \cos(kr) \, \Delta^{qq}_k(t)   + 
      \sin^2\phi(t) \, \frac{1}{Ns} \sum_{k\ne0} \cos(kr) \, \Delta^{pp}_k(t) \\ &
     - 2 \cos\theta(t) \cos\phi(t) \sin\phi(t) \, \frac{1}{Ns} \sum_{k\ne0} \cos(kr) \, \Delta^{qp}_k(t) .
\end{split}
\eeq
\end{widetext}
Analogous expressions can be readily obtained for $C^{\alpha\beta}(r,t)$, with $\alpha,\beta=x,y,z$.
In this section, for definiteness, we focus on the perturbed LMG model of Eq.~\eqref{eq:startingH1dnn}.

\subsubsection{Modulated light-cone effect}

%\begin{widetext}

\begin{figure*}[t!]
\centering
\begin{tabular}{cc}
\includegraphics[scale=0.16]{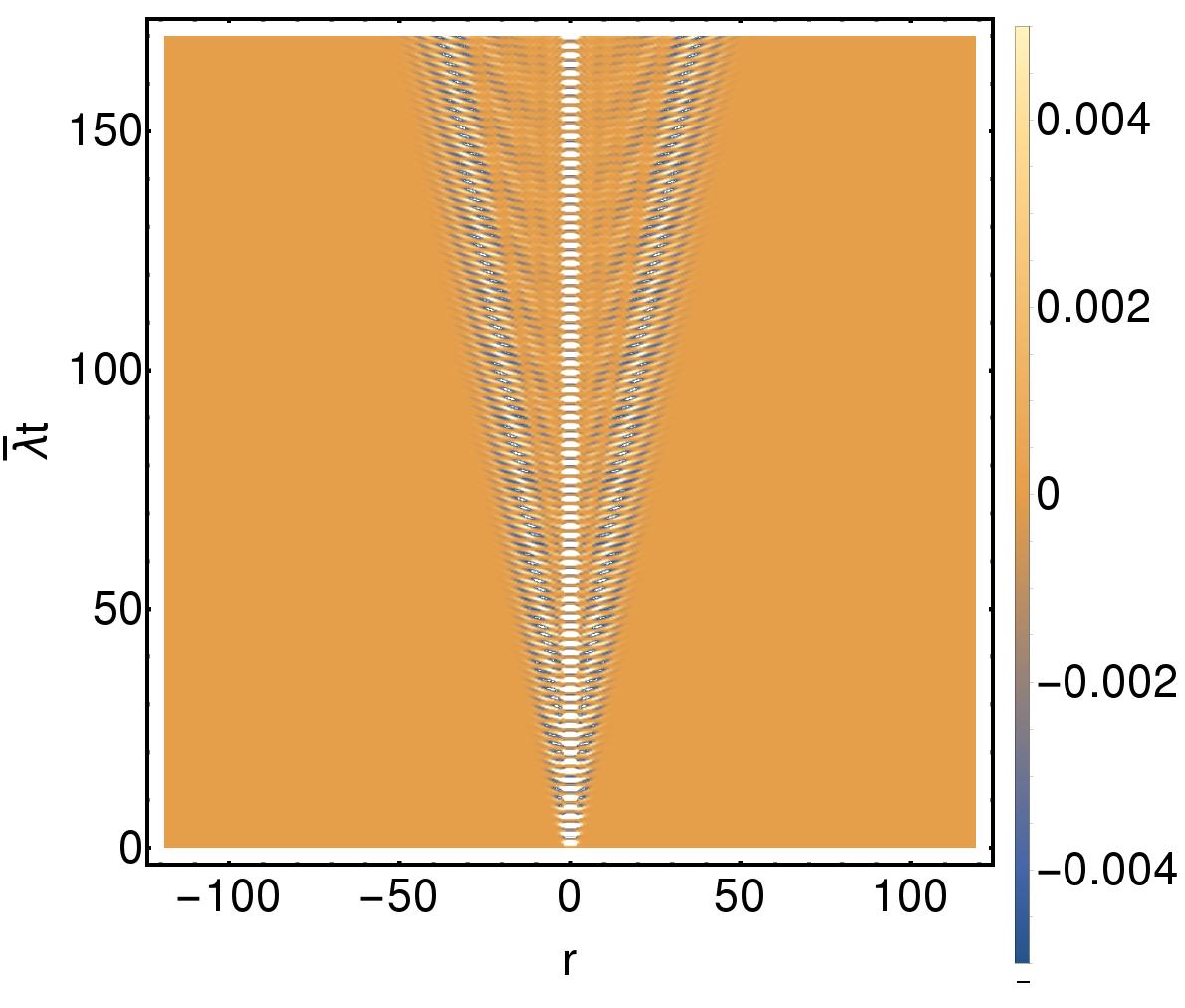} & \includegraphics[scale=0.16]{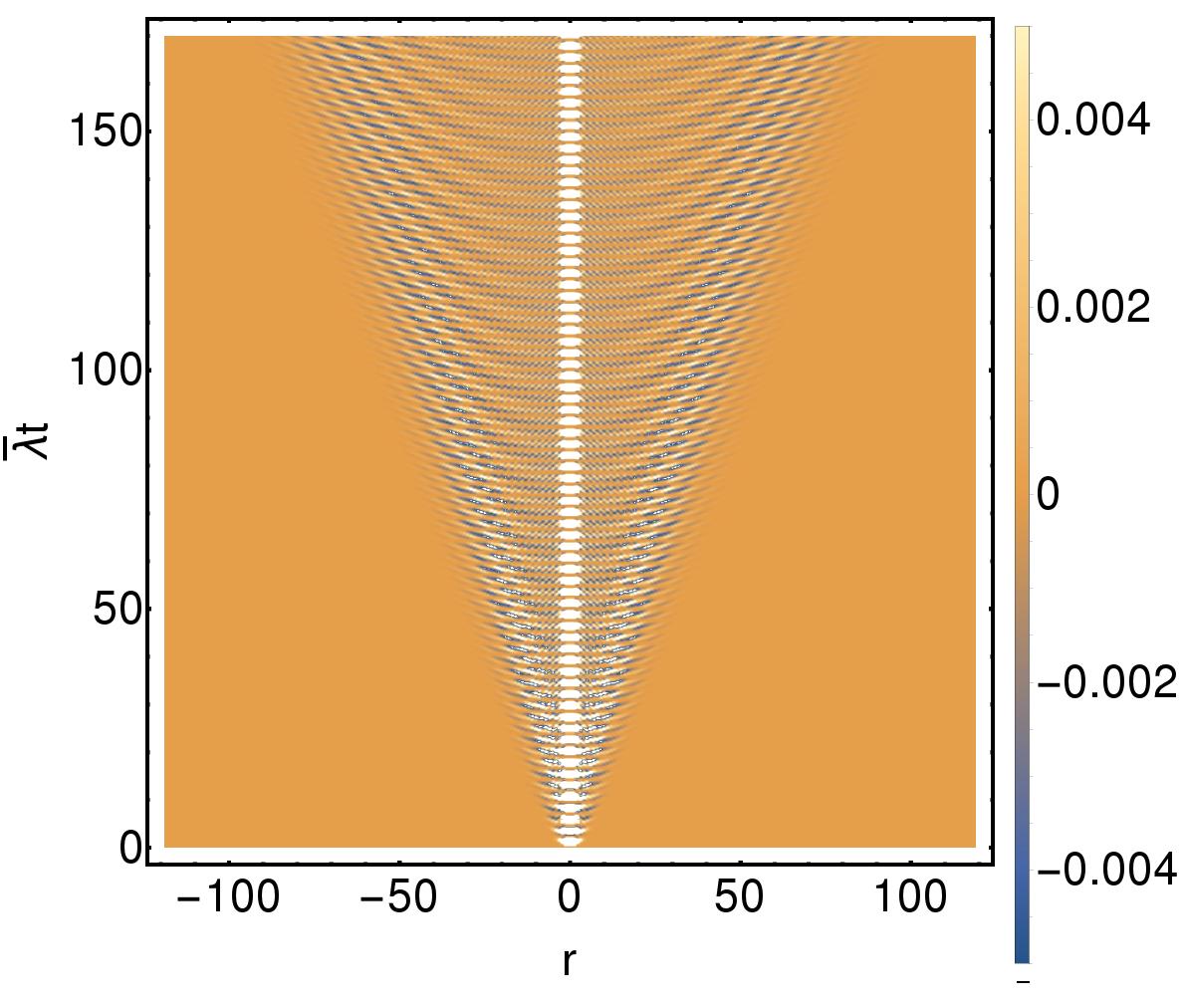} \\
\includegraphics[scale=0.16]{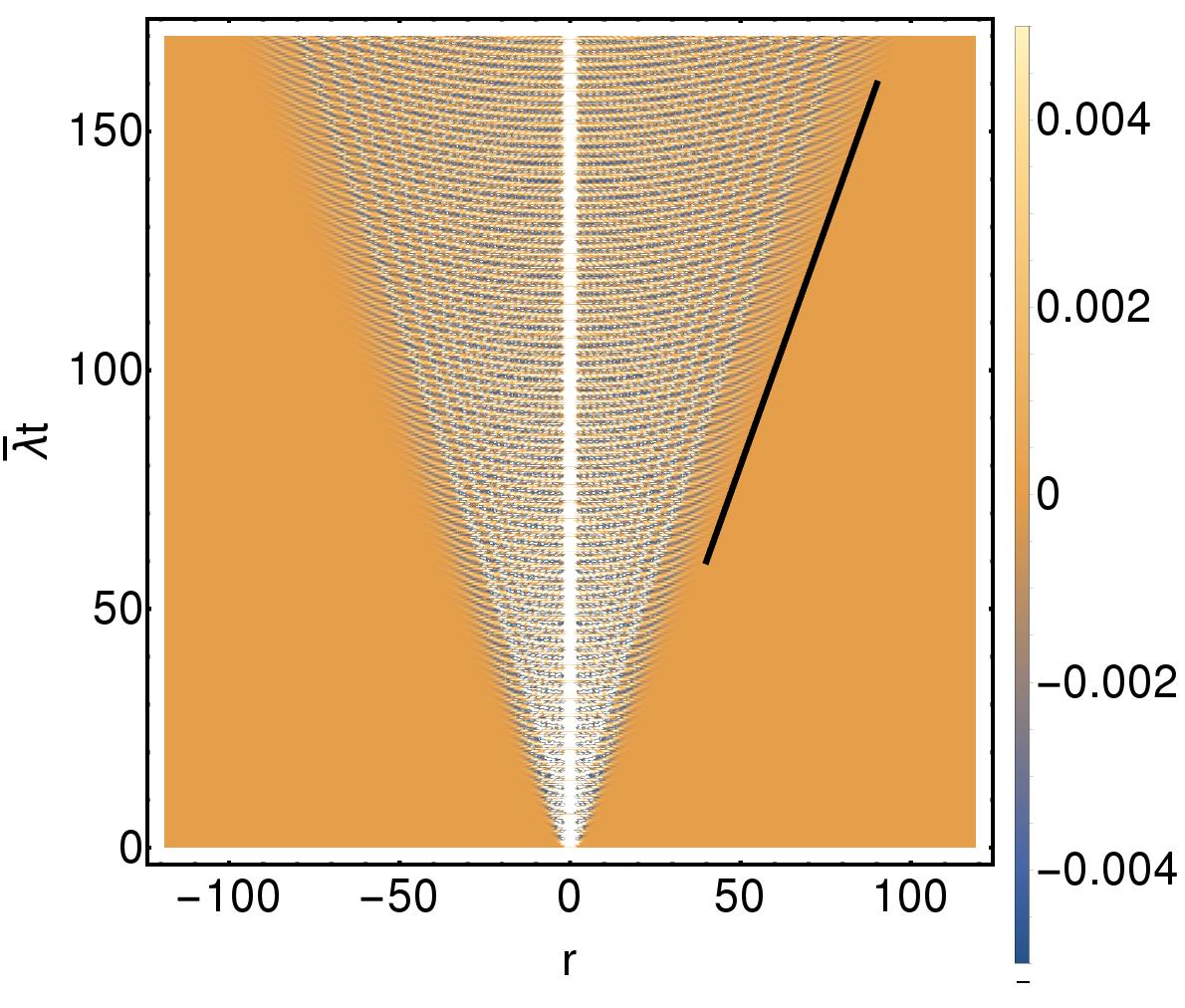} & \includegraphics[scale=0.16]{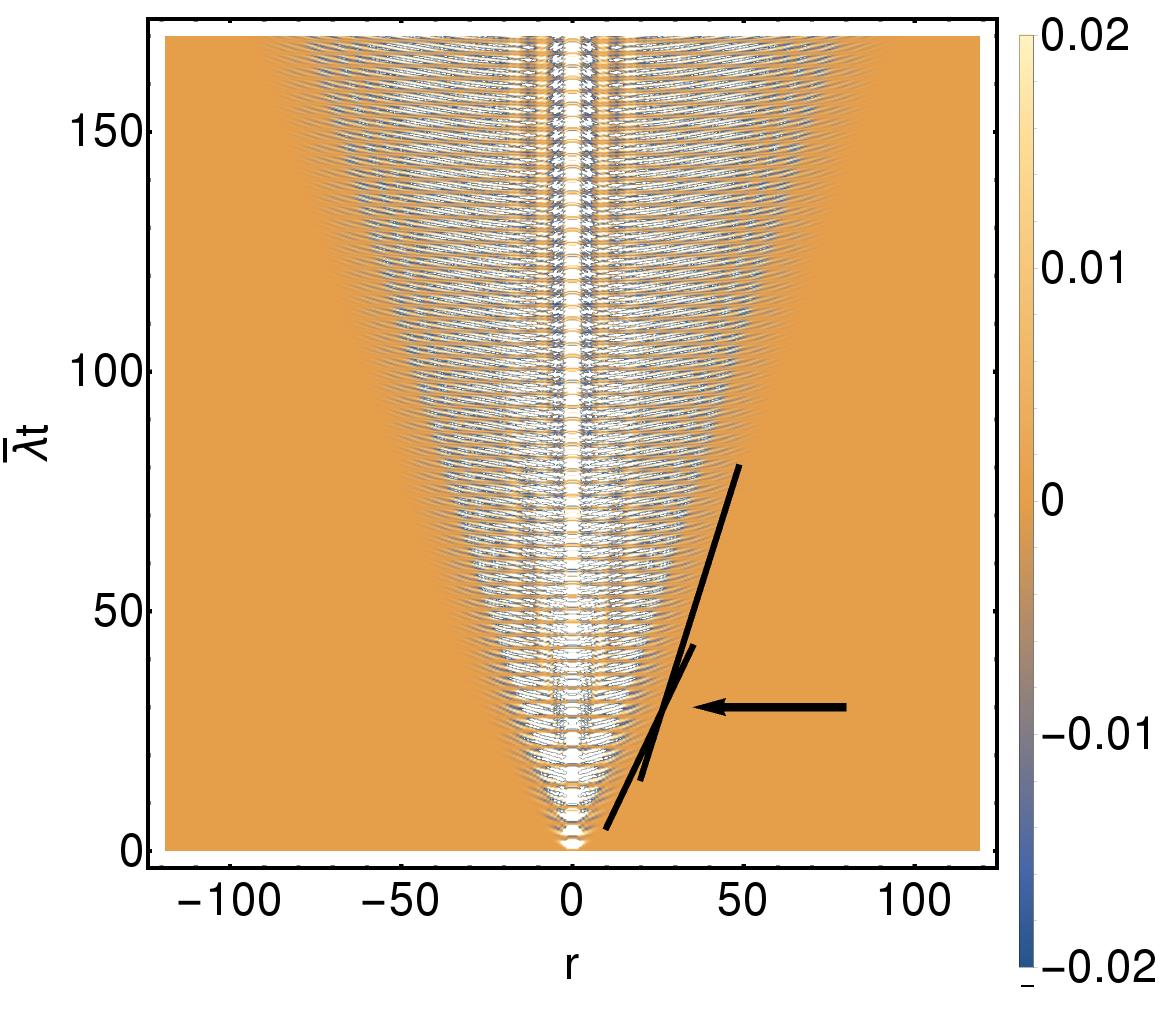} 
\end{tabular}
\caption{
[Color online]
%\bzcom{It would be nice to have the analytic lines on one side of the cone.} 
Space-time density plots of the dynamical correlation function $C^{xx}(r,t)$ [see Eq.~\eqref{eq:correlationsxx}] after a quench of the magnetic field $g$ from a fully polarized ferromagnetic state ($g_0=0$) to $g/\bar{\lambda}=0.7,0.9,1.025,3$, in clockwise order from top left. In all plots, $J/\bar{\lambda}=0.25$, $N=240$. For a small quench occurring deep in the ferromagnetic phase (top left), the overall amplitude of the correlation function is weak (few excitations are produced) and the light-cone is narrow due to an almost constant spin waves dispersion relation. The amplitude and the width become larger as the dynamical critical region is approached (top right). In the chaotic dynamical phase (bottom right), a ``knee'' is visible, marked by the black arrow, witnessing a change of the maximal velocity of propagation due to the trapping of the orbit, after a paramagnetic transient, into a ferromagnetic sector (notice the change of scale, highlighting a larger amplitude of the correlations). Finally, deep in the paramagnetic phase (bottom left), the maximal velocity approaches the value analytically predicted in Eq.~\eqref{eq:vmaxdeeppara}, indicated by the black line. An approximately periodic modulation of the amplitude of $C^{xx}(r,t)$ is visible in all cases, which reflects the approximately periodically driven nature of the spin waves, induced  by the precession of the collective spin.
}
\label{fig:lightcones}
\end{figure*}
%\end{widetext}

\begin{figure}[t]
\centering
%\begin{tabular}{cc}
\includegraphics[scale=0.3]{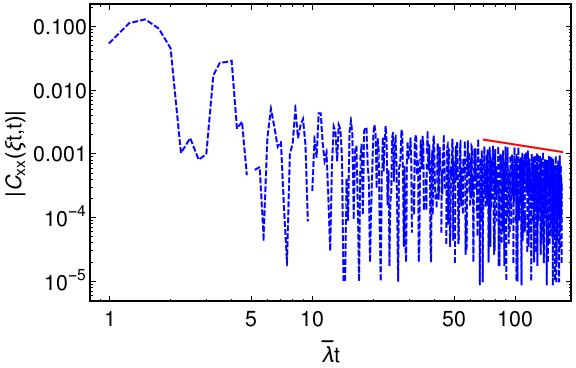} \\ %& 
\includegraphics[scale=0.3]{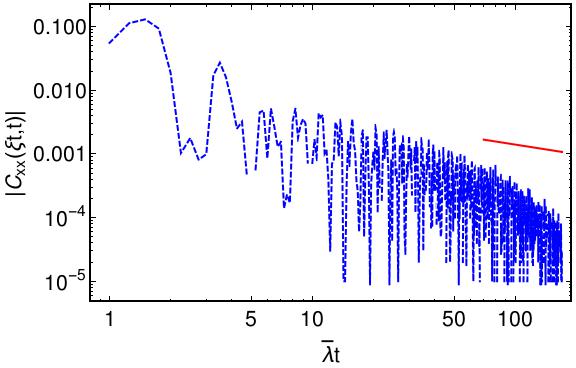}
%\end{tabular}
\caption{
[Color online]
Long-time behavior of the correlation function $C(\xi t,t)$ along two close spacetime rays with fixed $\xi=0.5\bar{\lambda}$ (left) and $0.55\bar{\lambda}$ (right) in a log-log scale, after a quench of the magnetic field from a pure ferromagnetic state ($g_0=0$) to $g/\bar{\lambda}=0.9$ with $J/\bar{\lambda}=0.25$, $N=240$, corresponding to the data of the top right panel of Fig.~\ref{fig:lightcones}. Apart from an (approximately) periodic modulation, $C(\xi t,t)$ decays, in the large spacetime limit, as a power-law (left) or as an exponential (right) as a function of time. The red line highlights the $t^{-1/2}$ decay suggested by the argument in the text, see Eq.~\eqref{eq:lightconescaling}. Data are consistent with a maximal velocity of propagation of the effective free quasiparticles between $0.25 \bar{\lambda}$ and $0.275\bar{\lambda}$ in this specific numerical instance.
}
\label{fig:spacetimescaling}
\end{figure}

Density plots of the equal-time correlation functions $C^{xx}(r,t)$ are shown in Fig.~\ref{fig:lightcones}. They are obtained by integrating the equations of motion \eqref{eq:vacuummotion} and \eqref{eq:swmotiondelta} with the same initial conditions as  in the previous sections  and by substituting their solution $\Big( \theta(t),\phi(t), \{ \Delta^{qq}_k(t),\Delta^{qp}_k(t),\Delta^{pp}_k(t) \} \Big)$ into Eq.~\eqref{eq:correlationsxx}, with $s=1/2$.

A light-cone effect\cite{Calabrese2006} is present for all values of the parameters $g$ and $J$, which is characterized by an exponentially fast decay in time of the correlation function for $\lvert r \rvert > 2 v_{\text{max}} t$ with a certain $v_{\text{max}}$, see Ref.~\onlinecite{LR}. In fact,  the infinite-range Hamiltonian generates a collective coherent precession of all the spins with no spatial structure, due to the full permutational symmetry of the spins. %, but does not generate a non-trivial spatial structure of correlations or entanglement. \alessio{explain logarithmic growth} 
However, a non-trivial spatial dependence of the dynamical correlations arises in the presence of the additional short-range interaction term, which results in a light-cone.
A closer inspection of the figures reveals that a (seemingly) periodic modulation is superimposed to the amplitude of the correlations. %The reason is that the instantaneous spin wave dispersion relation is  varying approximately periodically in time, due to the approximately periodic motion of the angles $\theta(t)$ and $\phi(t)$ representing the instantaneous configuration of the collective spin (see Eqs.~\eqref{}). 
The origin of this phenomenon can be explained in the following terms. Within the low-density expansion considered here, the quadratic bosonic Hamiltonian~\eqref{eq:effectiveH} governing the evolution of the spin waves has coefficients which depend parametrically on the angles $\theta(t),\phi(t)$. The latter evolve approximately periodically in time (cf. Fig.~\ref{fig:robustness}), resulting in an instantaneous dispersion relation of the spin waves with an approximately periodic time-dependence.
The ``stroboscopic'' dynamics of the spin waves at integer multiples of the ``period'' of the collective precession can thus be argued to relax to a periodic or stroboscopic  generalized Gibbs ensemble \cite{russosilva, Lazar}, which is known to occur in  quantum many-body systems subject to an \emph{external} periodic driving. 
We emphasize, however, that here the  periodic drive is given by the dynamics of the system itself, i.e., it is self-generated by the autonomous Hamiltonian   dynamics, without  external actions: the collective motion of the classical spin $\vec{\sigma}(t)$ generates an effective ``external'' drive for the spin waves, see Eq.~\eqref{eq:timeindepH}.

Let us now investigate the behavior of %the maximal velocity $v_{\text{max}}$ of propagation, i.e., of 
the slope of the light-cone edge as a function of the system's parameters. This quantity can  be derived in terms of the maximal velocity $v_{\text{max}}$ of propagation of the quasi-particles, which can be computed as the maximal slope of their effective dispersion relation $ \omega^{(\text{eff})}_k$ (see below). In the two limiting cases, $g\to0$ and $g\to\infty$, this  velocity  $v_{\text{max}}$ can be easily determined analytically. 

For $g\ll \bar{\lambda}$, the classical spin performs small oscillations near the initial ferromagnetic configuration, hence $\theta(t)\approx \pi/2$ and $\phi(t)\approx0$ (or $\pi$) for all  times. The dispersion relation in such a near-equilibrium condition has already been determined in Eq.~\eqref{eq:dispreldeepferro}: in fact, the corresponding dispersion relation is asymptotically flat, $ \omega^{(\text{eff})}_k \to 2 \bar{\lambda}$, as $g\to0$, and hence  $v_{\text{max}}$ approaches zero in this limit. This is confirmed by  numerical computations, as shown in the two top panels of Fig.~\ref{fig:lightcones} where the light-cone width shrinks as $g$ decreases. % (although hidden by  the low amplitude of the correlation function, due to the small amount of spin waves produced by a shallow quench).

In the opposite  limit $g\gg \bar{\lambda}$, the collective spin approximately rotates uniformly along the equator, $\theta(t)\approx \pi/2$, $\phi(t)\approx 2gt$, at frequency $2g$. The effective (Floquet) Hamiltonian~\cite{BukovReview} of the spin waves  is simply given by the time-averaged Hamiltonian to lowest order in the driving period $\pi/g$. Thus, by averaging in time the coefficients of Eq.~\eqref{eq:timeindepH}, we find the effective dispersion relation
\beq
\omega^{(\text{eff})}_k = 4 \sqrt{\bar{\lambda}(\bar{\lambda}- J \cos k)},
\eeq
and therefore, for small $J/\bar{\lambda}$, the maximal velocity of propagation is given by
\beq
\label{eq:vmaxdeeppara}
v_{\text{max}} = \max_k \bigg\lvert \frac{\partial \omega^{(\text{eff})}_k}{\partial k} \bigg\rvert \thicksim J.
\eeq
Comparing this prediction with the slope $2 v_{\text{max}} $ of the light-cone of correlations, we find fairly good agreement with the data shown in the bottom left panel of  Fig.~ \ref{fig:lightcones}.

A more precise quantitative determination of the light-cone edge from numerics  requires some care. In order to % illustrate a possible strategy for doing 
address this,
consider  a quantum system composed by free quasi-particles with dispersion relation $\omega_k$, and  assume parity symmetry, i.e.,  $\omega_k=\omega_{-k}$. An equal-time, two-point correlation function can be generically expressed as (see, e.g., Ref.~\onlinecite{CEF1})
\beq
C(r,t) = \int_{-\pi}^{\pi} \frac{dk}{2\pi} f(k) \, e^{ikr-i2\omega_k t},
\eeq
where the function $f$ depends on the model and on the quench. In the scaling limit of large $r$ and $t$ with fixed $r/t\equiv \xi$, $t\to\infty$, this correlation function shows a different asymptotic behavior along  rays within or outside the causal region delimited by the light cone $\lvert \xi \rvert \le 2 v_{\text{max}}$. Indeed, one finds
\begin{widetext}
\beq
\label{eq:lightconescaling}
C(\xi t,t) \underset{t\to\infty}{\thicksim} 
\begin{cases}
\sum_{k^*} f(k^*(\xi)) \frac{\exp \Big[ i \big(k^*(\xi) \xi - 2\omega_{k^*(\xi)} \big) t \Big]}{\sqrt{2\pi \omega''_{k^*(\xi)} t}}, & \text{for } \lvert\xi\rvert < 2 v_{\text{max}}, \\
A \exp \Big(- \delta(\xi) t \Big),   &  \text{for }   \lvert\xi\rvert > 2 v_{\text{max}} ,
\end{cases}
\eeq
\end{widetext}
where $k^*(\xi)$ is a solution to the equation $2 \partial \omega_k /\partial k (k) = \xi $, which exists only if $\lvert \xi \rvert <2 v_{\text{max}}$, and  the sum runs over the set of such solutions. %In fact, the quantity $\xi^*$ turns out to coincide with $v_{\text{max}} =\max_k  \big\lvert\frac{\partial \omega_k}{\partial k}\big\rvert$. 
Accordingly, upon increasing the time $t$, the correlation function decays to zero as $t^{-1/2}$ along rays \emph{within} the light-cone, whereas it decreases exponentially along rays \emph{outside} the light-cone (the latter is a general fact valid for all systems with  short-range interactions, as follows from the Lieb-Robinson bound~\cite{LR}). The proper way of extracting  $v_{\text{max}}$, and thus of defining the light-cone edge from the numerical data, is therefore by inspecting the decay of the correlation function along space-time rays and thereby discriminating power-law  from exponential decay: the critical ray which separates the two behaviors \emph{is} the light-cone edge, and its slope unambiguously determines the maximal velocity of propagation of excitations within the system. %The  light-cone effect is therefore due to ballistic quasi-particles traveling across the system and spreading correlations. 
%\begin{widetext}
Figure \ref{fig:spacetimescaling} shows that the two scaling behaviors in Eq.~\eqref{eq:lightconescaling} are indeed found in the the numerical data. This agrees with the picture of self-consistently periodically driven spin waves.

%\end{widetext}
\subsubsection{Dynamical correlations in the chaotic dynamical phase}

The chaotic dynamical ferromagnetic phase C in Fig.~\ref{fig1} leaves detectable signatures on the dynamics of the local order parameter correlation functions. 

The  self-consistent internal driving provided by the collective spin dynamics changes when the transient paramagnetic behavior turns into an evolution occurring eventually within one of the ferromagnetic sectors, as happens, e.g., in Fig \ref{fig2}. 
From the point of view of the spin waves, this can be seen as a change of their effective (Floquet) Hamiltonian, which, accordingly, results in a change of the associated ``speed of light''. Although the values of $v_{\text{max}}$ before and after this change from dynamical paramagnet to dynamical ferromagnet, are not very different, a variation of slope in the light-cone is visible in some of the numerical computations, e.g., those reported in the bottom right panel of Fig.~ \ref{fig:lightcones}.
They correspond  to the macroscopic, qualitative change in the internal driving provided by $\vec{\sigma}(t)$. This phenomenon is  {a   consequence of the existence of a chaotic dynamical ferromagnetic phase, and it can be seen as a further, characteristic hallmark of its peculiar nature}.

%\begin{widetext}

\subsection{Ramp dynamics}
\label{sec:ramp}

We now extend the analysis of the previous Section and of Ref.~\onlinecite{LeroseShort} for the system described by the Hamiltonian~\eqref{eq:startingH} to a time-dependent  ramp $g(t)$ of the transverse field,  describing the crossover  from the sudden quench (infinitely quick ramp)  to the adiabatic evolution (infinitely slow ramp). We consider a linear time-dependence
\beq
\label{eq:ramp}
g(t)=
\begin{cases}
g_0, & \text{for } t<0; \\
g_0 + (g-g_0)\frac{t}{\tau}, & \text{for } 0\le  t \le \tau; \\
g, & \text{for } t>\tau. \\
\end{cases}
\eeq
The parameter $\tau$ controls the total duration of the ramp. The system is initialized in the ground state of $H(g_0)$ at $t_0<0$ and then evolves with the time-dependent Hamiltonian $H(g(t)))$ at later times.
When $\bar{\lambda} \tau \ll 1$, the results approach those obtained for the quantum quench dynamics of the previous section, see Fig.~\ref{fig1}. 
Upon increasing $\tau$, we expect two effects. \textit{(i)} First, the final state for $t\ge\tau$, will be progressively closer to the adiabatic one (i.e., the ground state of the final Hamiltonian, since the system is initialized in a zero-temperature ground state); this implies that the dynamical critical point, separating non-equilibrium trajectories within one ferromagnetic well from the dynamically paramagnetic ones encircling both wells, will move towards the equilibrium critical point, which is asymptotically reached in the adiabatic limit $\tau\to\infty$. 
This phenomenon occurs also in the absence of fluctuations, i.e., within the LMG model. %\bzcom{Is it possible to find $\lambda_c=\lambda_c(\tau^{-1})$?}
\textit{(ii)} Second, in the presence of  fluctuations, an increasingly slower  protocol will deposit in the system a progressively smaller amount of energy  in the form of spin wave  excitations with $k\ne0$. 
By inspecting the Hamiltonian~\eqref{eq:timeindepH} or  the equations of motion, one notices that the  driving $g(t)$ directly affects only the dynamics of the collective spin. 
This macroscopic precession, in turn, causes the production of pairs of spin waves  with opposite quasi-momenta: %, ``emitted'' by the large classical spin in the course of its macroscopic precession. 
%the driving acts on the collective spin, $\sigma^x(t)$, which  ``drags'' the bath of spin-waves, which gets in turn excited. 
% 
Near the dynamical transition,  the self-generated bath of spin waves  dissipates the energy of the classical spin, causing its trapping into either of the ferromagnetic wells. Accordingly, the smaller the amount of spin waves, the smaller the region of the parameter space within which the trapping phenomenon can occur is, and we therefore expect that the chaotic dynamical ferromagnetic phase C will shrink  as $\tau$ increases.

\begin{figure*}[t]
\centering
\begin{tabular}{ccc}
\includegraphics[scale=0.13]{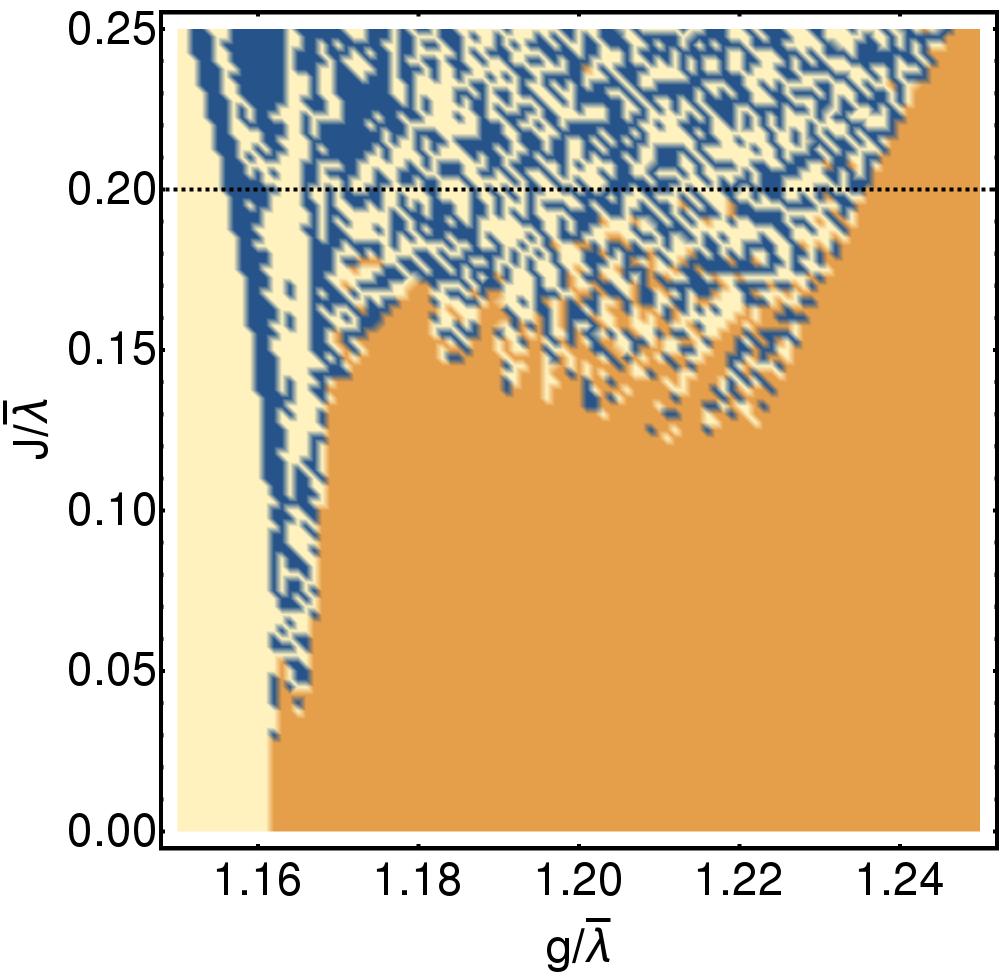} & \includegraphics[scale=0.13]{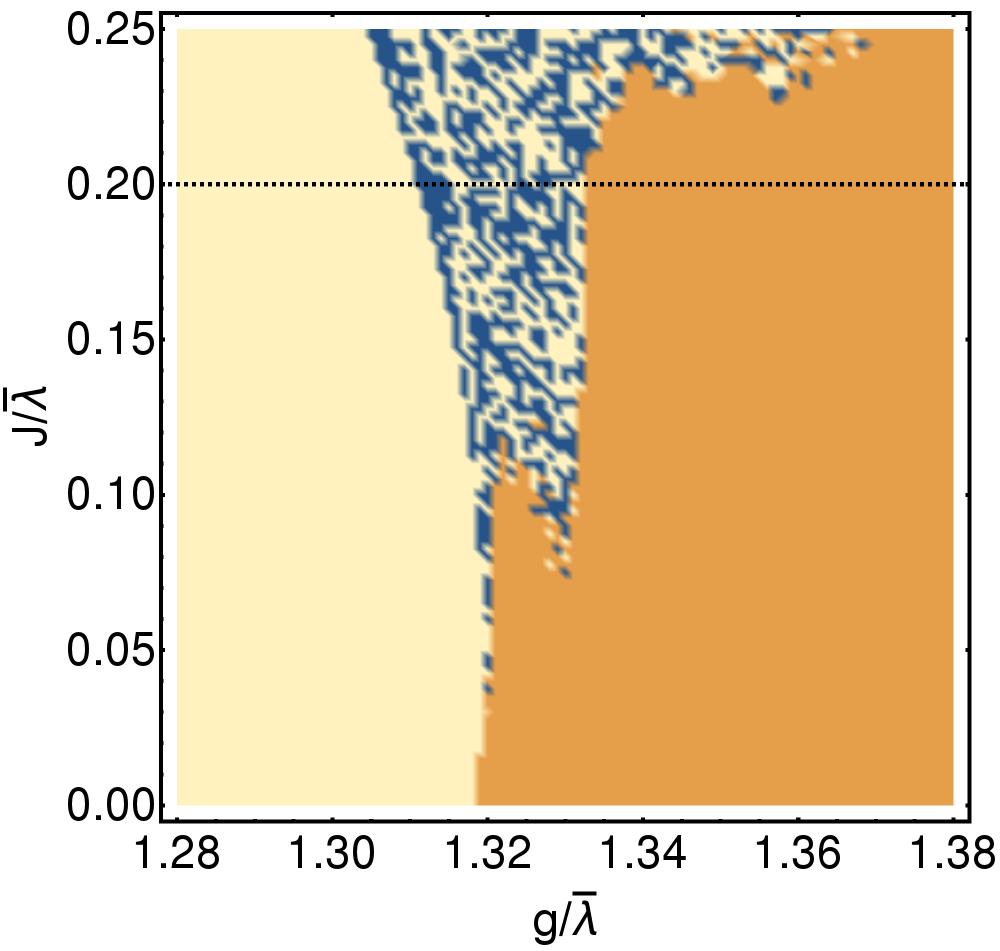} & \includegraphics[scale=0.13]{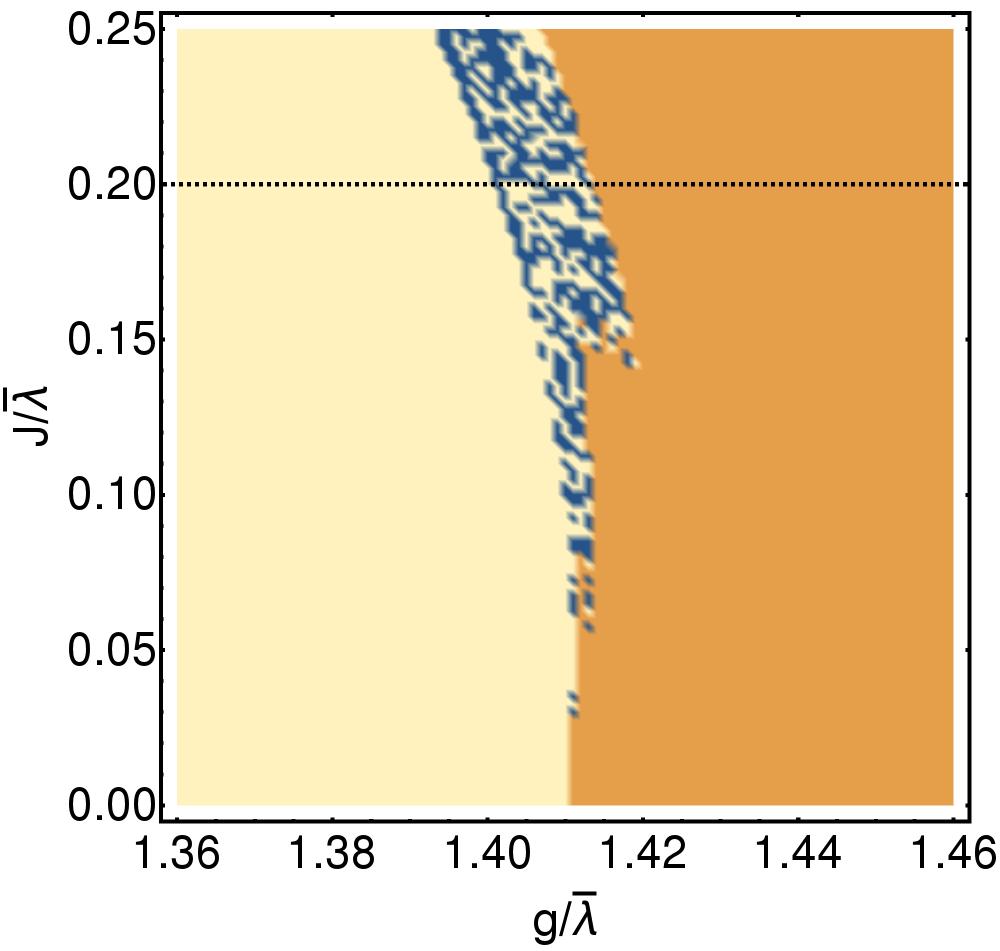} 
\end{tabular}
\caption{%\bzcom{Also here we could add the absolute value phase diagram. Just below the ones that are already there.}
[Color online]
Dynamical phase diagrams for linear ramps [see Eq.~\eqref{eq:ramp}] of the transverse magnetic field for the model described by Eq.~\eqref{eq:startingH} starting from $g_0=0$ (fully polarized ferromagnetic state), in the plane of the dimensionless final magnetic field $g/\bar{\lambda}$ and short-range interaction strength $J/\bar{\lambda}$, analogous to Fig.~\ref{fig1}. Here $N=100$. %and $\bar{\lambda}\equiv\lambda+J=1$, as in the previous figures. 
The color of each point %$(g/\bar{\lambda},J/\bar{\lambda})$ of 
in the diagrams indicates the asymptotic sign of the time-averaged order parameter, with the same graphical conventions as in Fig.~\ref{fig1}. The dimensionless duration $\bar{\lambda}\tau$ of the ramp is $0.7$ (left), $1.00$ (middle), $1.15$ (right). As the driving becomes slower, the mean-field dynamical critical point for $J\to0$ shifts from the sudden quench value $g_{\text{dyn}}/\lambda=1$ towards that in the adiabatic limit, i.e., the equilibrium critical point $g_{\text{cr}}/\lambda=2$ % -  (J/\bar{\lambda})^2 + \mathcal{O}(J/\bar{\lambda})^3 \approx 1.96$%=2$
 which is witnessed by the progressive shift rightwards along the horizontal axis of the border between the yellow and orange regions in the plot. Simultaneously, the chaotic dynamical ferromagnetic phase {shrinks}, due to the progressively smaller  amount of non-equilibrium excitations produced by the increasingly slower ramp.}
%Indeed, the \emph{equilibrium} amount of excited spin-waves is of order $\mathcal{O}(J^2)$, as computed in equation \eqref{eq:epsilongaussian}, so it is one order of magnitude smaller.
\label{fig:rampvsJ}
\end{figure*}

\begin{figure}[t]
\centering
\begin{tabular}{cc}
\includegraphics[scale=0.11]{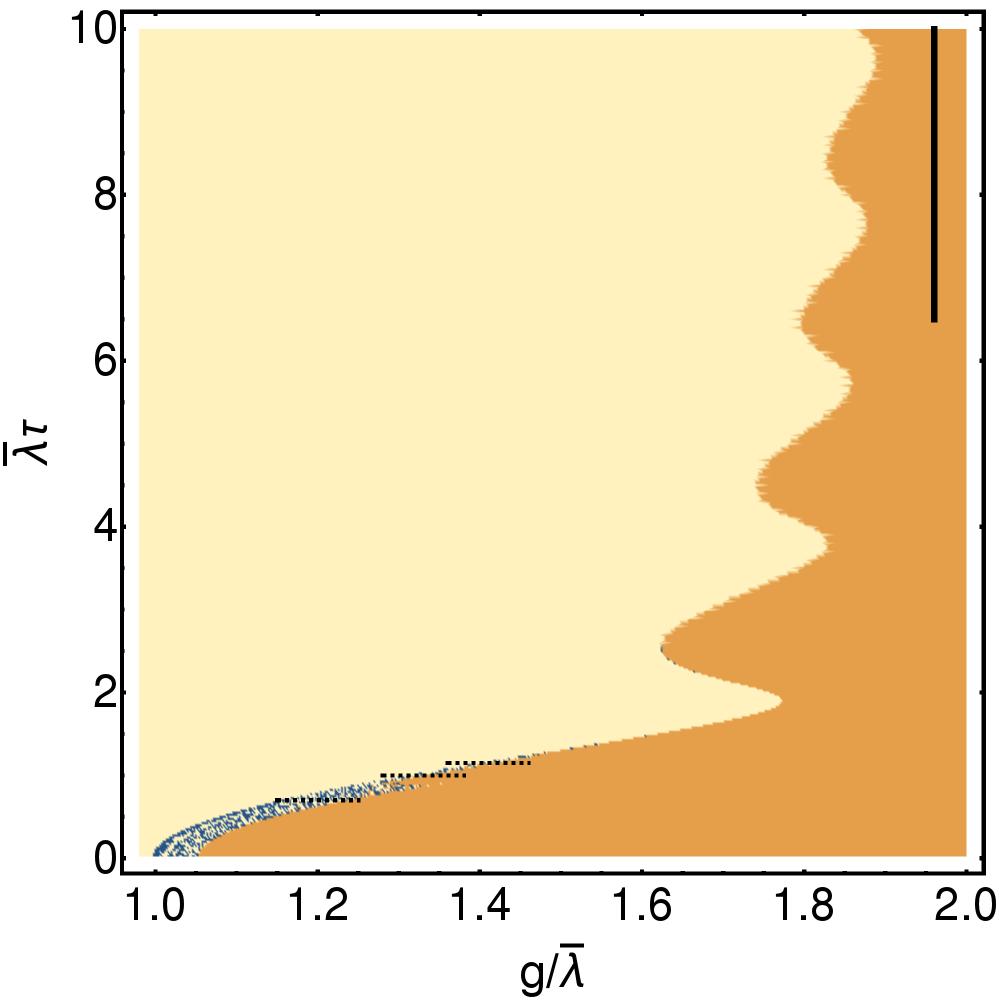} & 
\includegraphics[scale=0.11]{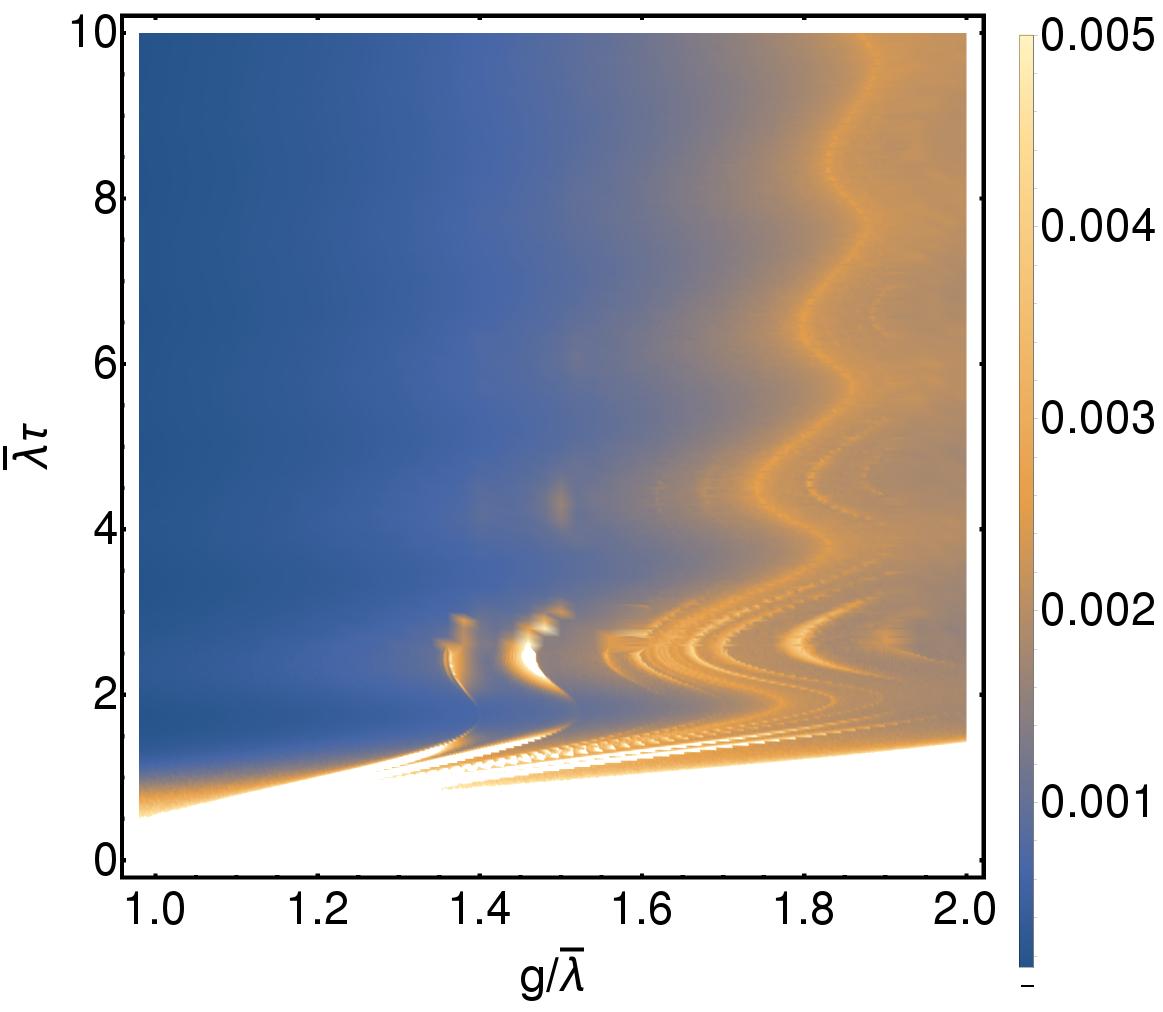}
\end{tabular}
\caption{
[Color online]
Left panel: Dynamical phase diagram for linear ramps of the magnetic field $g$ of the same model as in Fig.~\ref{fig:rampvsJ}, %starting from $g_0=0$ (pure ferromagnetic state), 
in the plane of the dimensionless final magnetic field $g/\bar{\lambda}$ and of the dimensionless ramp duration $\bar{\lambda}\tau$, but with fixed $J/\bar{\lambda}=0.2$.  %and $\bar{\lambda}\equiv\lambda+J=1$ as usual. 
The color of each point %$(g/\bar{\lambda},\bar{\lambda} \tau)$ 
of the diagram 
%represents the asymptotic sign of the order parameter, or, equivalently, the sign of its time-average. 
is assigned as in Figs.~\ref{fig:rampvsJ} or \ref{fig1}, and the diagram corresponds to taking a horizontal cut of those in Fig.~\ref{fig:rampvsJ} at fixed $J/\bar{\lambda}=0.2$ and varying $\tau$ continuously. As the ramp becomes slower, we notice two features: first, the  two boundaries of the chaotic phase shift from the sudden quench position around $g_{\text{dyn}}/\bar{\lambda}= 1$ towards the equilibrium critical point $g_{\text{cr}}/\bar{\lambda} =2 -  (5/8)(J/\bar{\lambda})^2 + \mathcal{O}(J/\bar{\lambda})^3 \approx 1.975$ [see Eq.~\eqref{eq:quantumcorrectioncrit}] in the adiabatic limit, marked by the black vertical line.
Second, the chaotic dynamical phase shrinks and practically disappears as $\tau$ is increased. Both these features are clearly visible in the picture. The ``oscillatory'' dependence of the phase boundary on $\tau$ is already present at the mean-field level.
Right panel: long-time average of the density $\epsilon(t)$ of spin wave excitations generated in the non-equilibrium dynamics. %This quantity is larger in the sudden quench limit $\tau\to0$  and decreases %to the equilibrium value  as the adiabatic limit $\tau\to\infty$ is approached.
}
\label{fig:rampvstau}
\end{figure}

This picture is confirmed by the numerical integration of the equations of motion, as one can see from Figs.~\ref{fig:rampvsJ} and~\ref{fig:rampvstau}.  
In particular, Fig.~\ref{fig:rampvsJ} shows how the dynamical phase diagram in Fig.~\ref{fig1} changes upon increasing, from left to right, the duration of the ramp $\tau$ in $g(t)$ which takes it from the initial value $g_0$ to the final value $g$.
As expected, the chaotic region C in the parameter space shrinks with its two boundaries getting increasingly closer to each other, while region C as a whole moves towards the line at which the transition occurs in equilibrium, see Eq.~\eqref{eq:quantumcorrectioncrit}. %\alessio{draw the equilibrium boundary?}

In order to highlight this shift and the fate of the chaotic phase, the left panel of Fig.~\ref{fig:rampvstau} shows a cut of the phase diagrams in Fig.~\ref{fig:rampvsJ} corresponding to a fixed value $J/\bar{\lambda}=0.2$ along the horizontal axis, and how the corresponding phases as a function of $g/\bar{\lambda}$ change as $\bar{\lambda} \tau$ increases well beyond the values considered in Fig.~\ref{fig:rampvsJ}. In order to assess the reliability of the spin wave approximation on which our analysis rely, the right panel of Fig.~\ref{fig:rampvstau} shows with colorcode how fast the long-time averaged spin wave density $\epsilon(t)$ [see Eq.~\eqref{eq:epsilondef}] decreases upon increasing the ramp duration and as a function of $g/\bar{\lambda}$ for the same conditions as in the left panel. 

%\alessio{comment here on Kibble-Zurek?}

%

%
%\end{widetext}

\section{Strong interactions}
\label{sec:bojan}

In order to check the robustness of the observed  phenomena in the presence of a nearest-neighbor interaction strength $J$ increased  beyond the perturbative regime considered in the previous sections, we  simulate numerically the evolution of the system by using the time-dependent variational principle developed in Refs.~\onlinecite{HLO14,LOV15}. This formulation requires a matrix product operator (MPO) representation of the Hamiltonian. Since the interaction strength of the infinite-range part of the Hamiltonian~\eqref{eq:startingH} scales with the system size $N$ it is not possible to rewrite the thermodynamic limit of the Hamiltonian in the MPO form and hence to simulate the time evolution directly in the thermodynamic limit. Accordingly, we performed finite-size simulations on long chains, up to $N=400$.

The reformulation of the Hamiltonian in the MPO form is done in two steps. First, we write an exact homogeneous MPO with a large bond dimension $D_{\rm MPO}=N+1$ and then use standard methods in order to find a compact inhomogeneous MPO representation with a bond dimension up to $D_{\rm MPO}=17$ and an error below $10^{-10}$. In all simulations  we used a time step of 0.02 (in units of $\lambda$), a matrix product state bond dimension up to $D=600$, and the second order single-site integrator proposed in Refs.~\onlinecite{HLO14,LOV15}.

\subsection{The trajectories and the phase diagram at large nearest-neighbor interactions}
\label{sec:strongcouple}

\begin{figure}[!htb]
  \includegraphics[width=8cm]{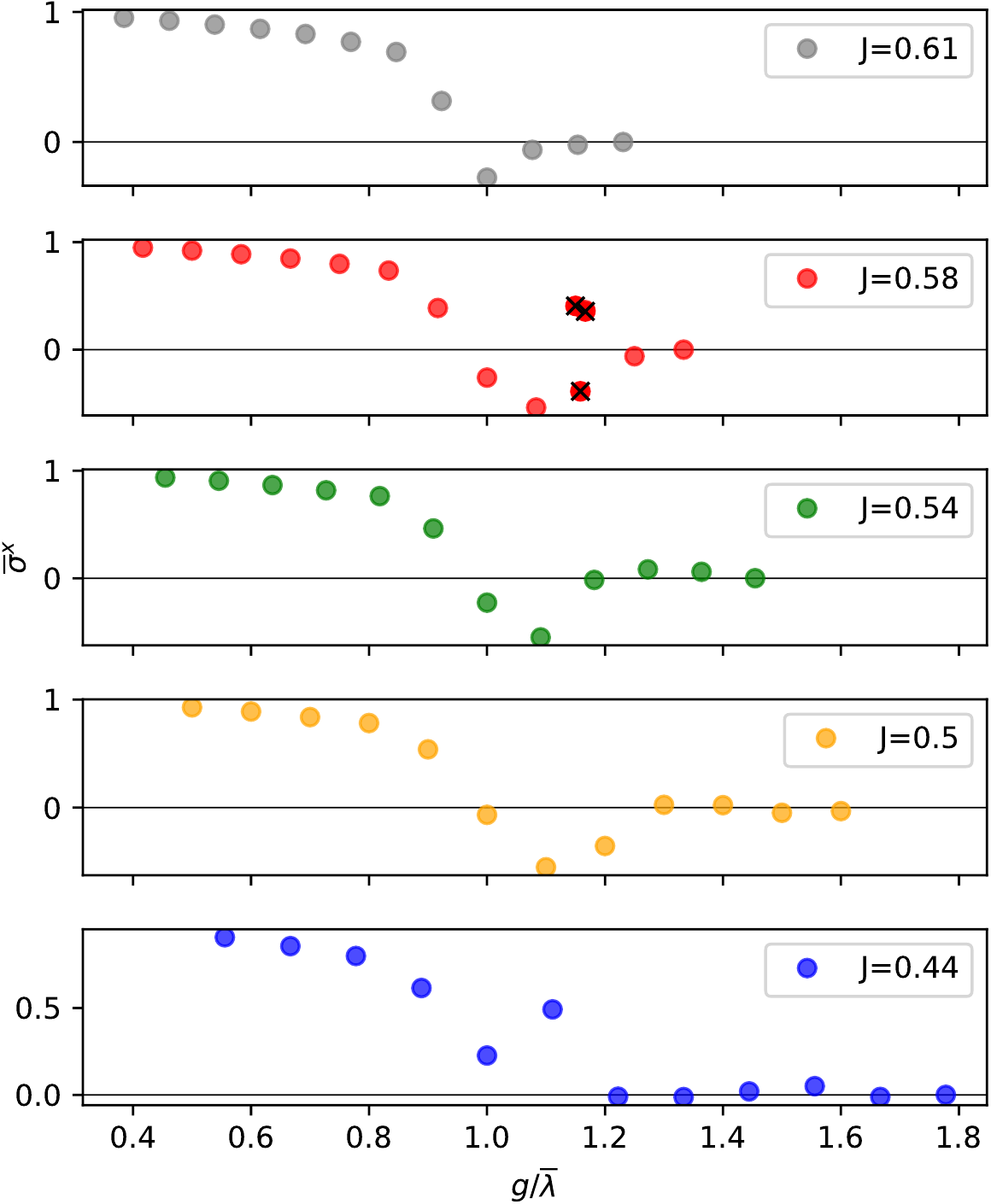}  
  \caption{
  [Color online]
Time-averaged order parameter $\overline{\sigma^x}$ of the model of Eq.~\eqref{eq:startingH1dnn} as a function of the post-quench values $g$ of the transverse field, for various values of the coupling $J$ around $0.5$, with $\bar{\lambda}=\lambda + J = 1$. The black crosses ($J=0.58$) correspond to a finer grid of values of $g$, with $\delta g = 0.008$ are shown in order to display the high sensitivity of the chaotic phase to post-quench parameters. These data show that the dynamically ferromagnetic and chaotic region persist also at large nearest-neighbor interactions. The data is calculated for system size $N=200$.}
  \label{fig:phase_diag}
\end{figure}

We first verify that the sensitivity of the evolution and of the final state to the values of the quench parameters observed in the perturbative regime of small spin wave density $\epsilon$ carries over to a larger nearest-neighbor interaction strength $J$. By extensive numerical simulations we show that indeed both phenomena persist as summarized in Fig.~\ref{fig:phase_diag}, where we show the dependence of the long-time time-averaged value of the order parameter $\overline{\sigma^x}$ as a function of the transverse field $g$ for various values of $J$ around $0.5 \bar{\lambda}$, i.e., $J\approx \lambda$. In particular, depending on the value of $J$, the dynamical ferromagnetic phase with $\overline{\sigma^x}>0$ at $g/ \bar{\lambda} \lesssim 0.9$ turns into a dynamical paramagnetic phase with $\overline{\sigma^x}=0$ at $g /  \bar{\lambda} \gtrsim 1.3$ via an intermediate region in which the ferromagnetic ordering is reversed as compared to the initial one, i.e., $\overline{\sigma^x}<0$. This is reminiscent of the ``stripes'' of reversed magnetization in Fig.~\ref{fig1} in the leftmost part of region C. We therefore expect a ``chaotic'' region to be present in between.

\begin{figure}[t!]
  \centering
  \includegraphics[width=8cm]{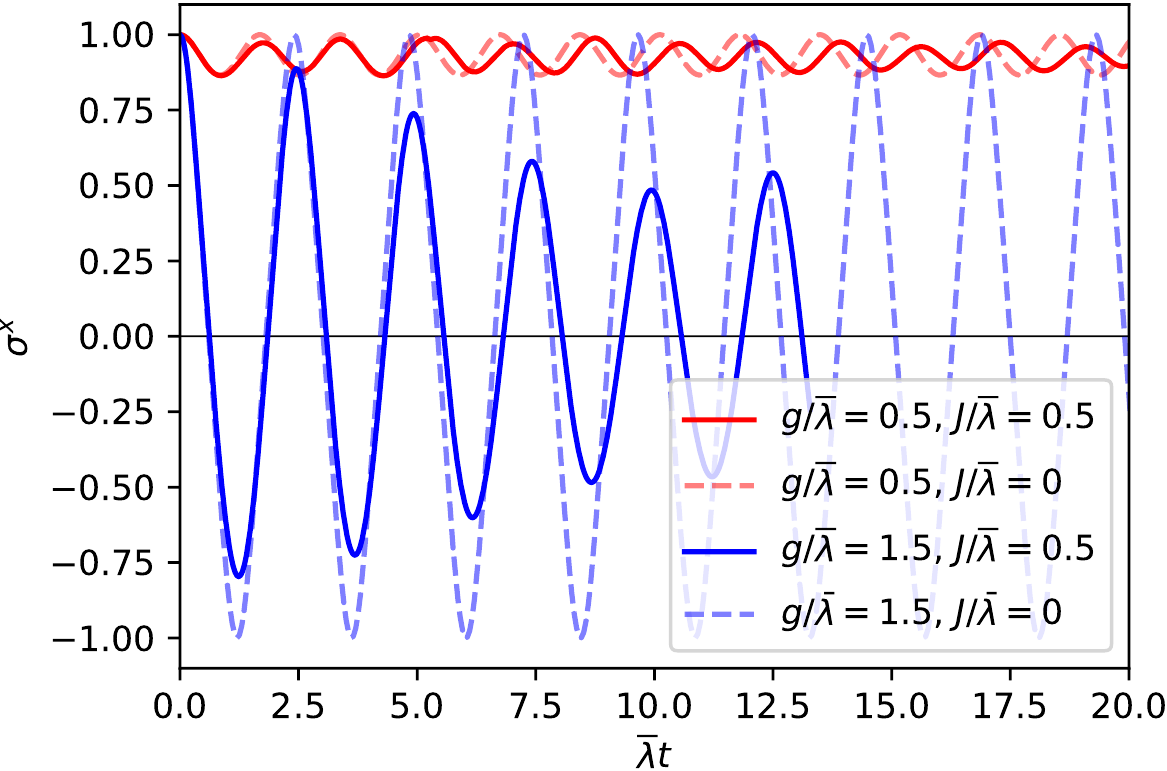}
  \caption{
 [Color online]
  Comparison between the evolution of the order parameter $\sigma^x$ at large $J$ (solid lines) with those of the mean-field model with $J=0$ (dashed lines) far from the critical region and for the same model as in Fig.~\ref{fig:phase_diag}. We observe that the evolution corresponding to both the ferromagnetic and paramagnetic phases are not altered qualitatively by the effects of quantum fluctuations. The decay of the oscillations amplitude upon increasing time is a finite-size effect.  In these simulations $N=400$ and $D=300$.
  }
  \label{fig:far_from_critical_traj}
\end{figure}

Far from the mean-field dynamical critical point $g_{\text{dyn}}=\bar{\lambda}$, the time evolution of the order parameter remains qualitatively similar to the mean-field case. As shown in Fig.~\ref{fig:far_from_critical_traj}, for $g/\bar{\lambda}=0.5$ and $g/\bar{\lambda}=1.5$ the ferromagnetic (red solid line) and paramagnetic (blue solid) trajectories are only slightly shifted with respect to the mean-field evolution ($J=0$, dashed red and blue lines) despite the large interaction strength $J=0.5 \bar{\lambda} = \lambda$.

\begin{figure}[htb!]
  \centering
  \includegraphics[width=8cm]{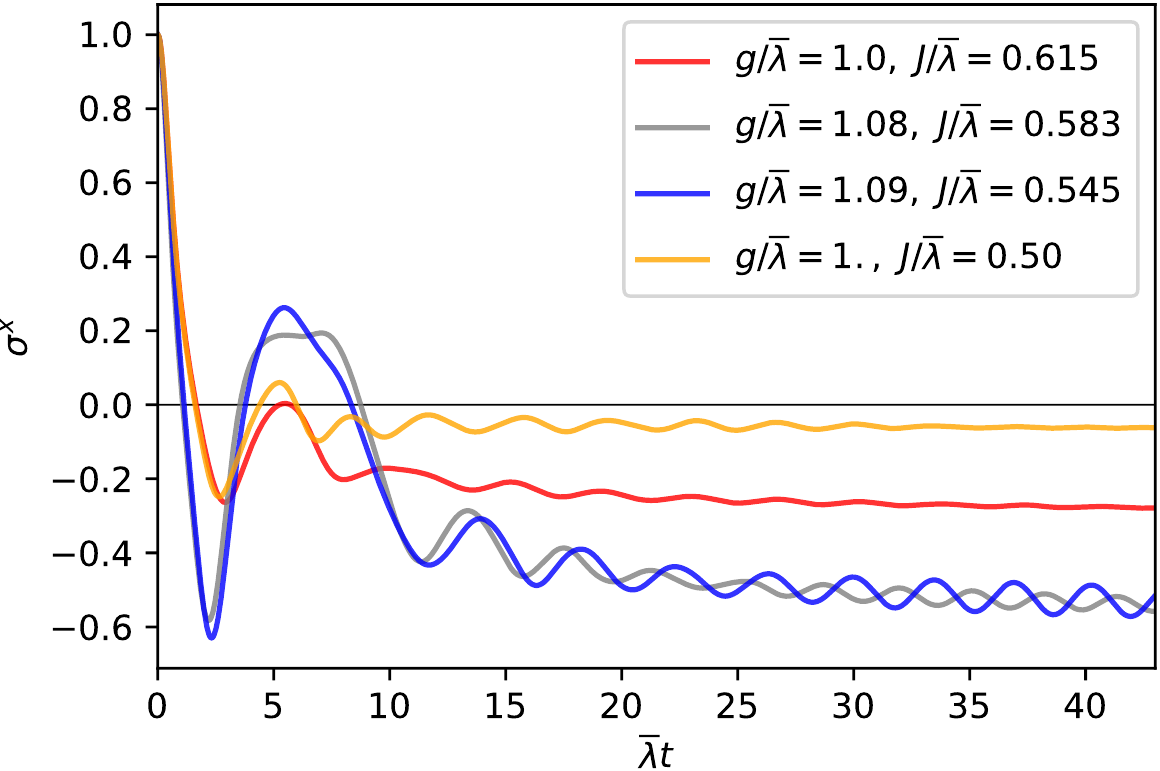}
  \caption{
  [Color online]
    Stability of the flipped ferromagnetic region. We show several trajectories within a wide range of different quench parameters as a part of the same region with a flipped final magnetization. This demonstrates stability of the flipped ferromagnetic region at large nearest-neighbor interactions. Simulations were performed with $N=200$, $D=200$.
  }
  \label{fig:flipped_ferro_traj}
\end{figure}

\begin{figure}[!htb]
  \includegraphics[width=8cm]{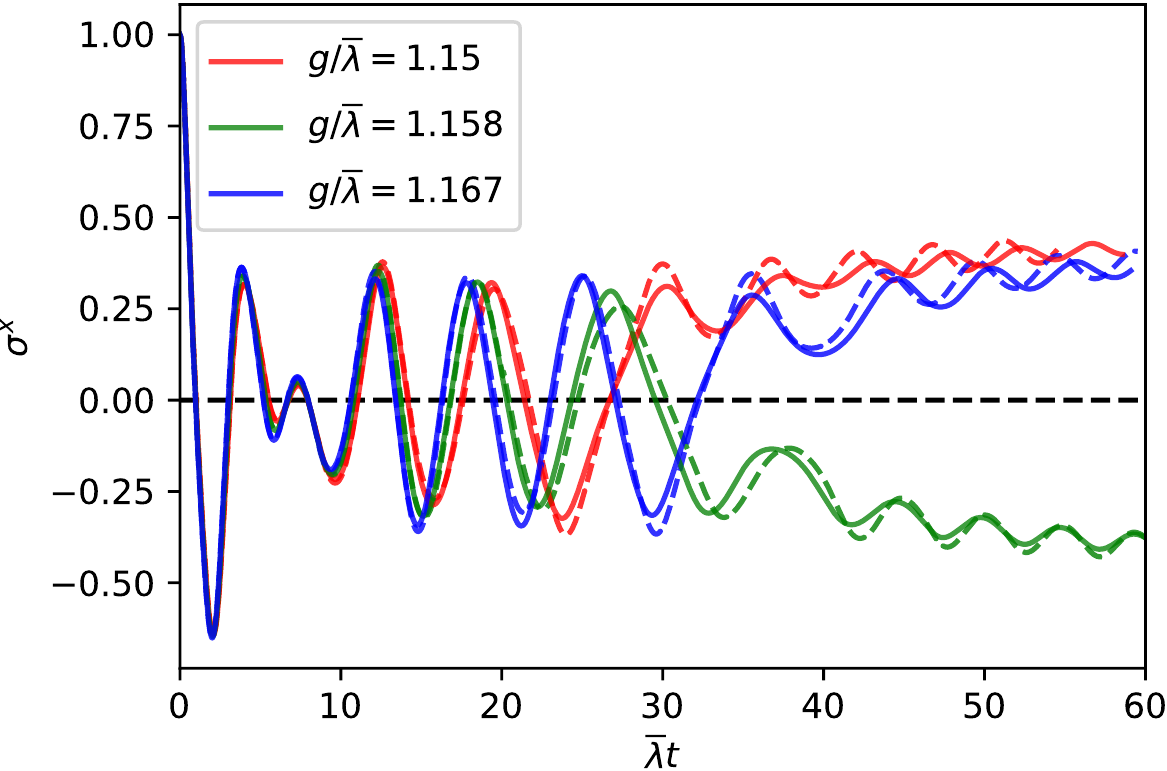}  
  \caption{
  [Color online]
  Evolution of the order parameter $\sigma^x(t)$ for a quench occurring close to the mean-field dynamical critical point within the chaotic phase. These curves at fixed $J/\bar{\lambda}=0.583$ show a sensitive dependence on the value of $g$, and they may oscillate for a long time before settling eventually in a sector with definite positive or negative order parameter. 
  By changing the quench parameter $g/\bar{\lambda}$ only slightly (approximately by $~0.08$) we observe a large change in the final magnetization which jumps from the positive to the negative sector and finally back to the positive sector. The curves in this plot correspond to the data points indicated by black crosses in Fig.~\ref{fig:phase_diag}. Simulations were performed with $N=200$, and with $D=600$ (full lines) and $D=500$ (dashed lines).}
  \label{fig:chaotic_flip}
\end{figure}

Upon getting closer to the mean-field dynamical critical point $g=\bar{\lambda}$ with $J\ne0$ one observes, instead, significant qualitative changes in the time evolution of $\sigma^x(t)$ compared to the case $J=0$, as shown in Figs.~\ref{fig:flipped_ferro_traj} and~\ref{fig:chaotic_flip}. 
In particular, in the region of parameters highlighted in Fig.~\ref{fig:phase_diag} one observes that the eventual sign of $\sigma^x(t)$ is reversed compared to the initial value and that it is attained possibly after a number of sign changes, as in Fig.~\ref{fig:flipped_ferro_traj}. This final sign reversal appears to be stable in longer simulations. The dynamics, however, become more complex when $g$ approaches the dynamical paramagnetic phase: as expected, the associated instability significantly affects the resulting evolution of $\sigma^x(t)$ which is displayed in Fig.~\ref{fig:chaotic_flip} and which is characterized by a sensitive dependence of the long-time magnetization on the quench parameter (magnetic field). Correspondingly, the time evolution of the order parameter looks irregular before it settles in one of the two sectors with a definite sign of $\sigma^x$. For some trajectories visible in both Figs.~\ref{fig:flipped_ferro_traj} and~\ref{fig:chaotic_flip}, the order parameter oscillates between them before it eventually reaches the final magnetization sector. In this case, the period of these oscillations progressively increases before the ``trapping'' occurs. %when $g/\bar{\lambda}$ approaches $1$. 
Similarly to the case of the propagation of correlations at weak interactions discussed in Sec.~\ref{sec:corr}, this change of the oscillation frequency corresponds to a transition from a dynamically paramagnetic to a dynamically ferromagnetic regime. Trajectories of Fig.~\ref{fig:chaotic_flip} are marked in the phase diagram shown in Fig.~\ref{fig:phase_diag} by black crosses.

In summary, the qualitative picture of the 
phases observed at small interactions persists also at large $J$. We emphasize the fact that for the values of $J$ used in the simulations reported in this Section, the accuracy of the spin-wave approach is poor and no quantitative agreement between the two methods has to be expected. In turn, at smaller values of $J$ and for the largest system sizes $N$ reached in these simulations, the time scale over which the collective magnetization gets trapped into a ferromagnetic sector is larger than the Ehrenfest time scale $T_{\text{Eh}} \lesssim \mathcal{O}(\sqrt{N})$ over which the motion of the collective magnetization is approximately classical (see Secs. \ref{sec:finitesizeLMG} and \ref{sec:cat}). This fact makes it difficult to observe the chaotic dynamical phase in this regime with MPS-TDVP. For this reason, the two methods used in this paper effectively explore complementary regimes of the dynamics of the system and they cannot be quantitatively compared.

\subsection{Correspondence between the zeros of the order parameter and the cusps of the return probability}
\label{sec:loschmidt}

In addition to the dynamical phase transition in the qualitative features of the long-time behavior of the order parameter discussed so far, another type of dynamical criticality related to non-analytic behavior of the return probability to the ground state manifold has been proposed in Refs.~\onlinecite{Heyl2013,Heyl2014dynamical}, as discussed in the Introduction. In Ref.~\onlinecite{Zunkovic2016} an intimate connection between them has been observed by studying a transverse field Ising model with variable-range interactions similar to that discussed in this work. We provide here numerical evidence that this correspondence remains valid also when the mean-field infinite-range Hamiltonian is perturbed by nearest-neighbor interactions studied in this work. In particular we demonstrate that the cusps in the time evolution of the  return probability $P(t)$ to the ground state manifold are simultaneous with the zeros of the time-dependent order parameter $\sigma^x(t)$, since $P(t)=P_1(t)+P_2(t)$, where %. These cusps actually appear at times at which the probability 
$P_1(t)$ is the probability to return to the initial state, %equals the probability 
and $P_2(t)$ is the probability to end up in the state with opposite longitudinal magnetization, cusp singularities are expected whenever $P_1(t)=P_2(t)$. Notice that both $P_1$ and $P_2$ become dramatically small as a function of time as a result of the excitation of spin waves.

\begin{figure}[!htb]
  \includegraphics[width=8cm]{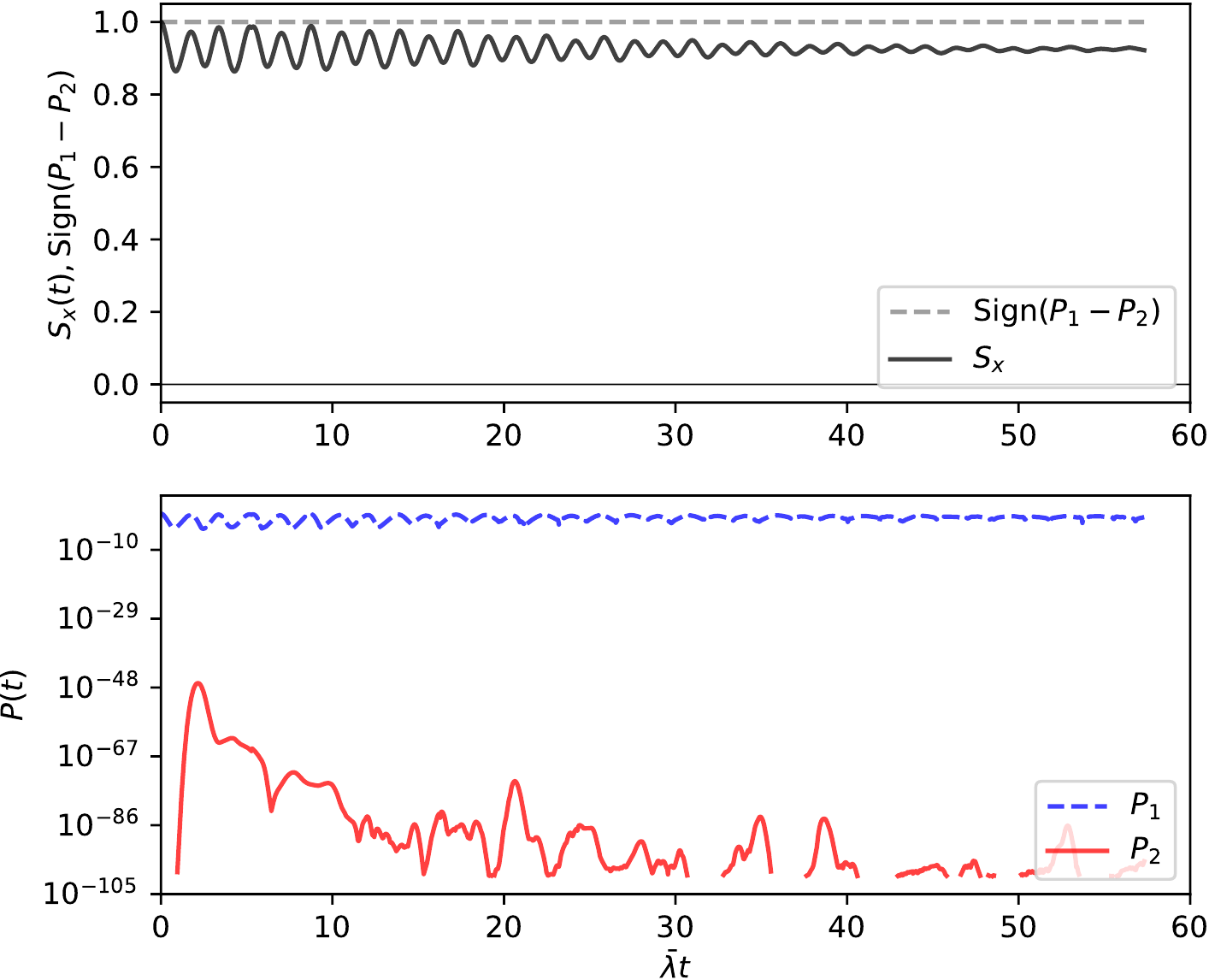}   \includegraphics[width=8cm]{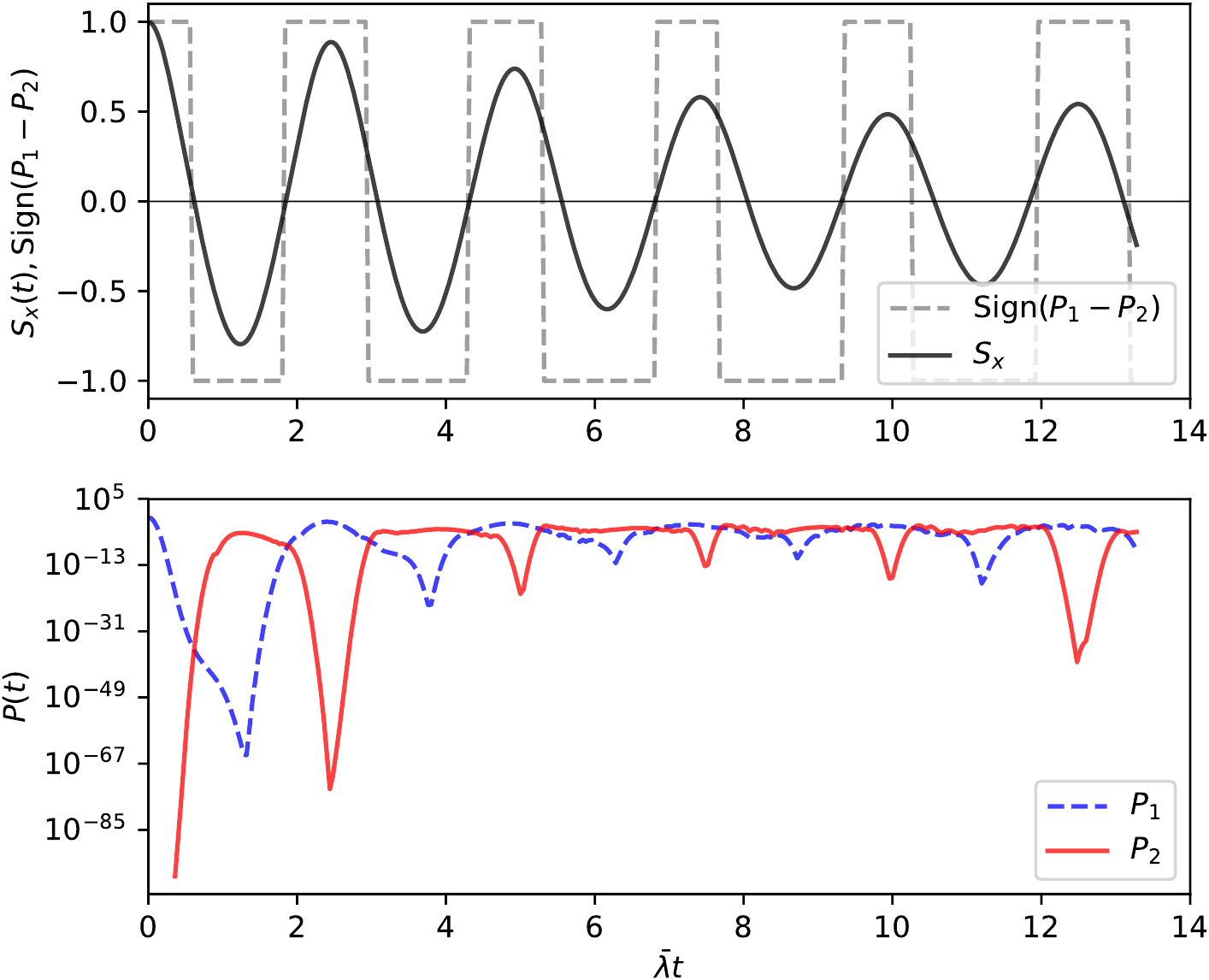}
  \caption{
  [Color online]
  %DPT correspondence in the ferromagnetic region (left) and in the paramagnetic region (right). 
  Relationship between the vanishing of the order parameter $\sigma^x$ and the change of sign of the difference between the probability $P_1$ to return to the initial state and the probability $P_2$ to reach the state with the opposite magnetization.
  Upper panels: comparison between the order parameter (orange) and the ${\rm sign}$ of $(P_1-P_2)$ (blue). Lower  panels:  evolution of the return probabilities $P_1$ (blue) and $P_2$ (red). In the ferromagnetic region (left) we observe that the order parameter remains close to one, which corresponds to a large difference in the probabilities $P_1$ and $P_2$. The return probability to the initial state, $P_1$, remains at all times much larger than the return probability  to a state with the opposite magnetization, $P_2$. On the other hand, in the paramagnetic region (right panels) the oder parameter periodically changes the sign. These changes correspond well with the cusps in the return probability which appear at points where $P_1-P_2$ changes its sign. Parameters: $D=300$, $N=400$, $J/\bar{\lambda}=0.5$, $g/\bar{\lambda}=0.5$\,(top), 1.5\,(bottom).}
  \label{fig:dpt_noncrit}
\end{figure}

Deep in the ferromagnetic region, shown in Fig.~\ref{fig:dpt_noncrit},  the order parameter $\sigma^x(t)$ remains positive, and, in fact, we observe that the probability $P_1$ to return to the initial state is always much larger than the probability $P_2$ to reach the state with an opposite magnetization. This region thus corresponds to a non-vanishing order parameter and the absence of cusps in the return probability. 

\begin{figure}[!htb]
  \includegraphics[width=8cm]{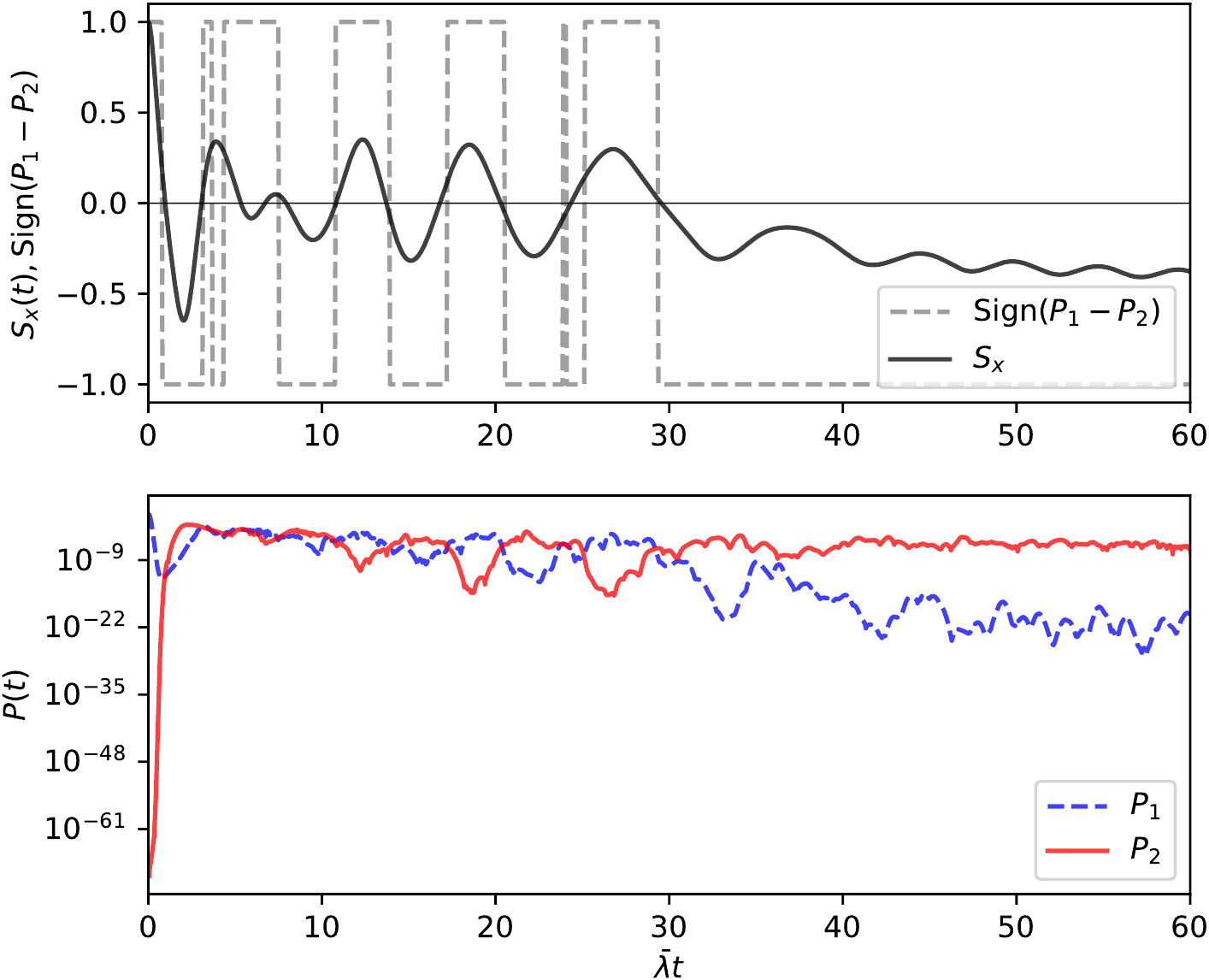} \includegraphics[width=8cm]{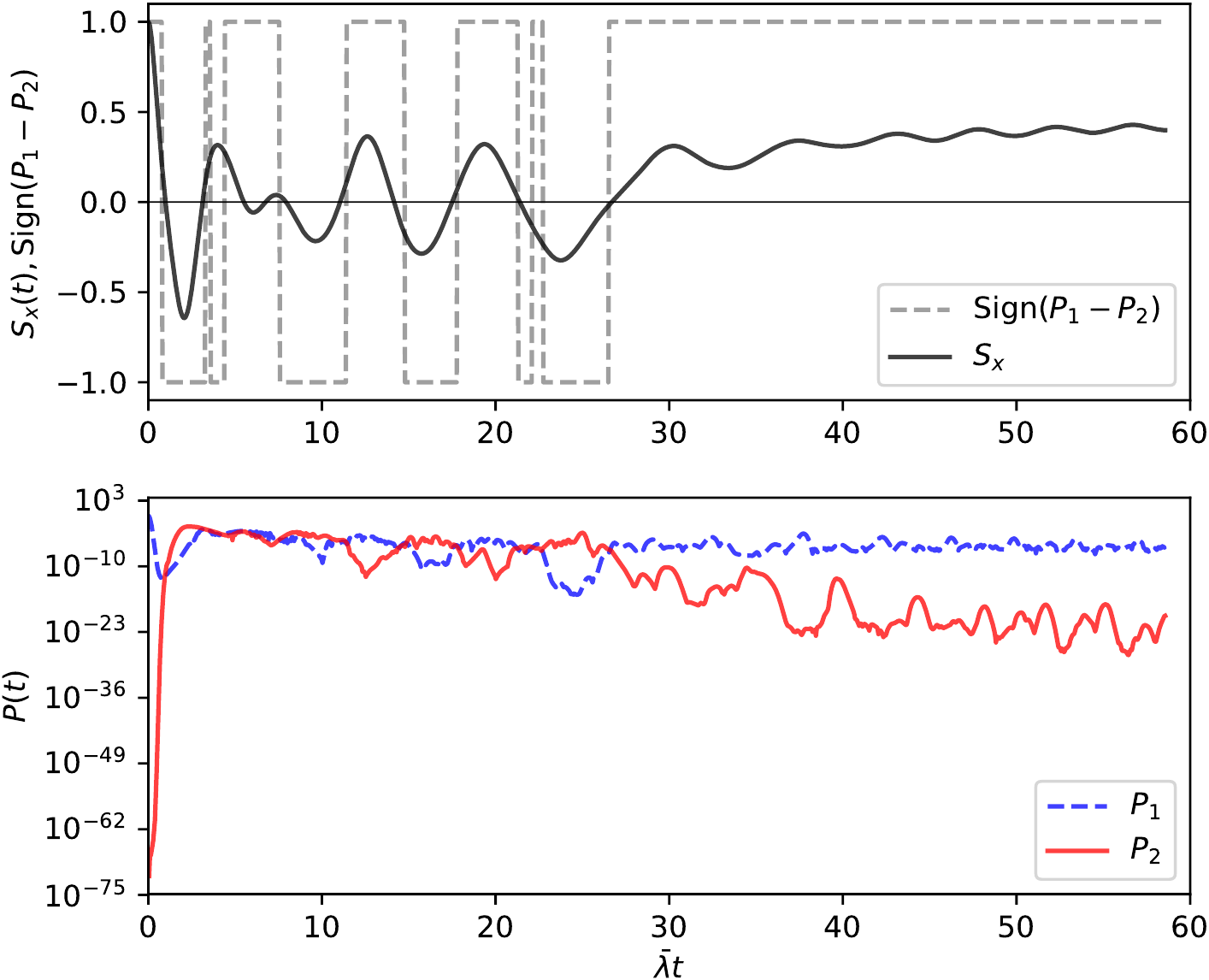}
  \caption{
  [Color online]
    Same plots as in Fig.~\ref{fig:dpt_noncrit} with different values of the parameters corresponding to the flipped and chaotic ferromagnetic regions. Initially the zeros of the order parameter correspond precisely to the cusps in the return probability, i.e., to $P_1=P_2$. At intermediate times the quality of this correspondence decreases,  due to finite-size effects (and is improved by increasing the system size). (left) At late times the return probability $P_2$ to the state with the opposite magnetization, becomes larger than the return probability $P_1$ to the initial state. This corresponds well to the flipped time-dependent order parameter at late times. (right) By changing the parameters only slightly we the final magnetization changes its sign, but the correspondence between the zeros of the order parameter and the cusps in the return probability remains valid. Parameters: $D=600$, $N=200$, $J/\bar{\lambda}=0.583$, $g/\bar{\lambda}=1.15$\,(top), 1.158\,(bottom).}
  \label{fig:dpt_crit}
\end{figure}

By increasing the magnetic field $g$ we enter a region where the order parameter $\sigma^x$ vanishes at certain times but later remains finite for a long time. Also in this region we observe that the zeros of the order parameter are close to the cusps in the return probability as shown in Fig.~\ref{fig:dpt_crit}.
Similarly the reversal of the final magnetization of one of the trajectories (left panels) corresponds to a larger probability $P_2$, which in this case becomes much larger than the probability to return to the initial state $P_1$.
The correspondence remains valid also in the paramagnetic region, as shown in Fig.~\ref{fig:dpt_noncrit}. 

Finally, we remark that the convergence properties with respect to the bond dimension are better in the ferromagnetic and paramagnetic regions than in the chaotic region. Far from the critical point the simulations for the system size $N=400$ converged already with bond dimension $D=300$. On the other hand, in the chaotic region we needed bond dimensions around $D=600$. Therefore, we could perform simulations only up to the system size $N=200$. In addition, the probabilities $P_1$ and $P_2$ are closer in the chaotic region, where the system initially oscillates between the positive and negative magnetization sector and only later remains in one or the other. Hence, the finite-size effects are noticeably larger in this region. We, however checked that the zeros of the order parameter and the cusps in the return probability move closer as the system size is increased. Despite relatively large finite-size effects in the chaotic region we observe a clear tendency that $P_1>P_2$ when $\sigma^x>0$ and $P_1<P_2$ when $\sigma^x<0$. In summary, this analysis shows that the correspondence between the zeros in the time-evolution of the order parameter and the occurrence of cusps in the return probability remains valid also in the presence of strong nearest-neighbor interactions.

In order to reduce spurious sign changes of $P_1-P_2$ due to finite-size effects in Figs.~\ref{fig:dpt_crit} and~\ref{fig:dpt_noncrit}, we performed a moving average of $P_1, P_2$ over a small time window $0.2\overline{\lambda}$.

% In summary, this analysis shows  that the correspondence between the zeros in the time-evolution of the order parameter and the occurrence of cusps in that of the return probability remains valid also in the presence of strong nearest-neighbor interactions. Although we observe some deviations at intermediate times, our numerical data suggests that they are due to finite-size effects. The return probabilities are global quantities, and thus a larger system size and bond dimension are necessary for them to converge.

\section{Finite-size effects%: Schroedinger cats
}
\label{sec:cat}

\begin{figure*}[tb]
  \includegraphics[width=0.8 \textwidth]{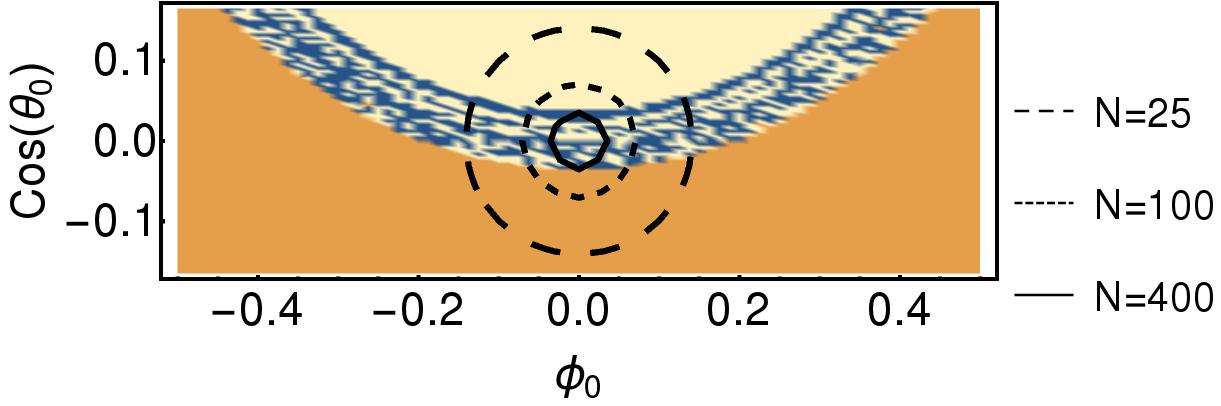}
  \caption{
  [Color online]
 Non-equilibrium phase diagram  for quenches with fixed post-quench parameters $g/\bar{\lambda}=1.03$, $J/\bar{\lambda}=0.25$ as a function of the direction on the Bloch sphere of the  pre-quench fully polarized spin-coherent initial state, parameterized by the canonically conjugated phase space coordinates $\phi_0$ and $\cos\theta_0$ (cf. Sec.~\ref{sec:timeindependent}). As in Fig.~\ref{fig1}, this plot is obtained via numerically integrating the evolution equations of the time-dependent spin wave theory in the thermodynamic limit, with the same graphical conventions thereof.
 The lowest-order finite-size correction consists in replacing a classical, uncertainty-free initial condition, specified by $(\phi_0,\cos\theta_0)$, with a Gaussian wavepacket in phase space centered around it, with linear extension $\sqrt{\hbar_{\text{eff}}}=1/\sqrt{Ns}$, which takes into account the quantum uncertainty at the lowest order in the semiclassical expansion, as discussed in Sec.~\ref{sec:finitesizeLMG}. The circles superimposed to the diagram indicate the width of these Gaussian distributions centered around $(0,0)$ for various values of $N$, corresponding to quenches from a ground state of the pre-quench Hamiltonian with $g_0=0$ considered in all simulations reported in this work. We see that for $N\lesssim 10^2$ 
 %the chaotic dynamical phase gets completely blurred by 
 the corresponding wavepacket encompasses initial conditions eventually belonging to all possible phases of the model. Accordingly, one expects the chaotic dynamical phase to be blurred by these
 quantum fluctuations when $N$ is sufficiently small. 
 This effect is more severe when $N$ is in the range $\lesssim 16$ accessible to full exact diagonalization of the Hamiltonian~\eqref{eq:startingH1dnn}, %initial states always comprise contributions from the three phases, 
 which makes it hard to observe signatures of the chaotic dynamical phase via this exact method. The latter is observed in MPS-TDVP simulations with $N$ in the range $10^2 \div 10^3$ and stronger perturbation $J/\bar{\lambda}\approx 0.5$, as reported in Sec.~\ref{sec:bojan}, in correspondence of which the extension of the  regions with a uniform sign of the asymptotic magnetization becomes sufficiently large compared to the coarse-graining scale $1/\sqrt{N}$.
 }
  \label{fig:initialdata}
\end{figure*}

In this last section, we discuss the relevance of the finite-size effects in the chaotic dynamical phase. For the sake of definiteness, we will focus on the model in Eq.~\eqref{eq:startingH1dnn}.

% role of N in the spin wave analysis
The first observation in order is that the spin wave technique developed in Sec.~\ref{sec:methodweakfluct} is rigorously valid in the thermodynamic limit, in which, as thoroughly discussed in Sec.~\ref{sec:finitesizeLMG},  the collective spin can be treated as a classical degree of freedom. 
For the LMG model with $\tilde{J}_{k\ne0}=0$, the spin-wave expansion~\eqref{eq:Hsw} of the Hamiltonian %provides a direct way 
allows one to compute the modifications to the classical evolution equations by accounting for the feedback from the quantum fluctuations of the $k=0$ mode in terms of $\Delta^{qq}_0$, $\Delta^{qp}_0$, $\Delta^{pp}_0$, which are suppressed as $N^{-1}$ [see Eq.~\eqref{eq:Hsw}]. % The resulting corrections to the trajectory of $\theta$ and $\phi$ are suppressed as $N^{-1}$. 
The presence of integrability-breaking perturbations $\tilde{J}_{k\ne0}\ne0$, as discussed in Sec.~\ref{sec:methodweakfluct}, activates the quantum feedback from all the spin waves modes with $k\ne0$, such as $\frac{1}{Ns} \sum_{k\ne0} \tilde{J}_k  \Delta^{qq}_k (t)$ and similar terms in Eq.~\eqref{eq:vacuummotion}. In contrast to the feedback from the zero-mode fluctuations, these latter terms have a finite limit as $N\to\infty$. For this reason, they have been properly taken into account in Sec.~\ref{sec:methodweakfluct} and thereafter, while the feedback from the quantum fluctuations of the $k=0$ mode has been neglected throughout this paper. 

In view of the above argument, all the results based on the time-dependent spin wave theory assume that the thermodynamic limit is taken at fixed time. The neglected finite-size effects typically set in at the (divergent) Ehrenfest time scale $T_{\text{Eh}} \sim \mathcal{O}(\sqrt{N})$ discussed in Sec.~\ref{sec:finitesizeLMG}. Thus, in Eqs.~\eqref{eq:vacuummotion} and~\eqref{eq:swmotiondelta} , as well as  in all simulations reported in Sec.~\ref{sec:results}, the parameter $N$ plays the role of a discretization of the Brillouin zone integrals such as $\frac{1}{Ns} \sum_{k\ne0} \tilde{J}_k  \Delta^{qq}_k (t) \sim \int_{-\pi}^{\pi} \frac{dk}{2\pi s} \tilde{J}_k  \Delta^{qq}_k (t)$, rather than properly accounting for actual finite-size effects.

% how large should N be?
The spin wave analysis carried out in the previous sections predicts the occurrence of a chaotic dynamical phase in perturbed mean-field models in the thermodynamic limit $N\to\infty$, which corresponds to a transient paramagnetic evolution followed by localization of the collective spin within one of the two ferromagnetic wells. It is then important to estimate how large $N$  should be in practice for this phenomenon to occur. In order to do this, we resort both to semiclassical arguments (cf. Sec.~\ref{sec:finitesizeLMG}) as well as to MPS-TDVP simulations (cf. Sec.~\ref{sec:bojan}).
%As we have shown in Sec.~\ref{sec:finitesizeLMG}, 
As explained in Sec.~\ref{sec:finitesizeLMG}, within the lowest-order semiclassical expansion or TWA \cite{TWA}, the quantum corrections to the classical motion amount to the replacement of the classical trajectory with the classical (Liouville) evolution of a Gaussian wavepacket in phase space centered around an initial condition $\theta_0,\phi_0$. 
The width of this distribution is given by the quantum uncertainty of the transverse components  $\tilde{\sigma}_{k=0}^X/N$, $\tilde{\sigma}_{k=0}^Y/N$ of the rescaled collective spin, which amount to $ \sqrt{\tilde{q}_0^2} /\sqrt{Ns}$ and $\sqrt{\tilde{p}_0^2} / \sqrt{Ns}$, respectively [see Eq.~\eqref{eq:approxH-Pfourier}], and hence are proportional to $1/\sqrt{N}$. 
The result of this approximation is visualized in Fig.~\ref{fig:initialdata}, where the asymptotic magnetization is shown for quenches to $g/\bar{\lambda}=1.03$, $J/\bar{\lambda}=0.25$ as a function of the initial condition $(\phi_0,\cos\theta_0)$, with the same color conventions as in Fig.~\ref{fig1}: The dynamical order parameter for a system of finite size $N$ corresponds to replacing the asymptotic magnetization of each point in this non-equilibrium phase diagram with its average over a Gaussian distribution centered at $(\phi_0=0,\cos\theta_0=0)$ with width of order $1/\sqrt{N}$, pictorially represented by the circles in Fig.~\ref{fig:initialdata}, corresponding to various values of $N$. This width can be viewed as a phase space coarse-graining scale in the presence of a system with finite size $N$.
%\footnote{More precisely, this should be done in a non-equilibrium phase diagram as a function of the initial conditions $\theta(t=0)$ and $\phi(t=0)$. However, the latter is qualitatively similar to Fig.~\ref{fig1}.}
We see that within the chaotic dynamical phase, for small $N$, the wavepacket encompasses several initial conditions $\theta_0,\phi_0$ whose evolutions end up into distinct ferromagnetic sectors. 
Accordingly, after a transient, the actual many-body wavefunction  is expected to realize a \emph{quantum} superposition of two wavepackets localized in the two distinct ferromagnetic sectors, i.e., a so-called \emph{cat state}.  %\footnote{Note that these expected violations of the cluster decomposition principle\cite{Weinberg} are not forbidden in the systems under consideration, because strict locality of the interactions is a primary assumption which fails here.} 
Hence, for sufficiently small $N$, this quantum superposition is expected to blur the critical region C in the diagram.
A classical-like behavior characterized by a non-vanishing average magnetization is expected to be seen only when the size of the initial wavepacket becomes smaller than the distance between the phase space boundaries of regions with a definite sign of the asymptotic magnetization, which happens for sufficiently large $N$.
Estimates based on the spin wave approximation in Sec.~\ref{sec:results} and in particular on the size of the largest spots of region C in Fig.~\ref{fig1} and in analogous diagrams for a range of parameters suggest that the \emph{minimal} system size $N$ required in order to observe localization of the wavepacket within a single sector should lie in the range $ 10^2 \div 10^3$.   This agrees with the observed convergence of the MPS-TDVP simulations within this region upon raising the bond dimension, as shown in Sec.~\ref{sec:bojan} and in Ref.~\onlinecite{LeroseShort}. 
% 

% what happens at smaller N?
For smaller $N$, the remnant of the chaotic dynamical phase is expected to be the formation of cat states, which can in principle be detected by inspecting the evolution of the full statistics of the order parameter, rather than only of its the average.
This study can actually be carried out via exact diagonalization of the Hamiltonian for a system size $N$ up to $16$. In this regime, however, finite-size effects are predominant (cf. Fig.~\ref{fig:ConvergenceToClassical}). This can be easily understood within the semiclassical picture illustrated in Fig.~\ref{fig:initialdata}, by observing that the order of magnitude $\mathcal{O}(1/\sqrt{N})$ of the width of a spin-coherent wavepacket  is comparable with the global width $\mathcal{O}(1)$ of the whole phase space when $N=16$, whereby the evolution results in a complicated superposition and interference of ferromagnetic, paramagnetic and ``chaotic'' classical trajectories. For this reason we do not report the relative results here, and we leave a detailed investigation of this issue via finer numerical techniques to future studies. %\alessio{Why not with TDVP? Motivate?}
%Indeed, one can see a paramagnetic transient followed by a fluctuating distribution seemingly characterized by two peaks in the two ferromagnetic sector, as shown in Fig.~\ref{fig:cat}. 

We emphasize, however, that by tuning continuously the parameters  across a phase boundary one should observe cat states for arbitrarily large $N$: in fact, the evolution governed by a finite matrix has to depend smoothly on the parameters, and the time-evolved wavefunction cannot undergo ``discontinuous'' transitions at finite $N$, as is well-known from general theory.
However, by the above semiclassical arguments, such cat states are expected to be confined within thin ``layers'' around the phase boundaries, whose width should shrink upon increasing $N$.

We conclude with two remarks on the crossover between quantum-mechanical behavior and its classical limit in the phenomena presented in this work as well as in Ref.~\onlinecite{LeroseShort}.
First, we note that the scale $1/\sqrt{N}$, which is associated with the extension of the Wigner function of spin-coherent states, clearly represents the characteristic distance in phase space beyond which two spin-coherent states become effectively orthogonal in Hilbert space. In fact, two  initial spin-coherent configurations separated by a smaller distance in phase space cannot localize in two distinct sectors after a certain time $t$, because they have non-vanishing initial overlap and hence cannot become orthogonal at any time due to unitarity of quantum time-evolution. Accordingly, the ``collective chaotic behavior'' unveiled in this work can arise in this quantum system only when the phase space coarse-graining scale $1/\sqrt{N}$ provides a sufficiently fine resolution for resolving the different classical outcomes. 
There is no basic contradiction in the emergence of a classical-like chaos in a quantum system, as long as two nearby initial conditions, whose classical evolution undergoes exponentially fast separation, correspond to orthogonal initial vectors in Hilbert space. In this work we have shown that this can occur in systems with competition of long- and short-range interactions driven close to a dynamical phase transition, provided the system size $N$ is sufficiently large.

The second, related, remark is that, although the emergence of a classical-like collective chaotic behavior involves a subtle interplay between the thermodynamic limit and the time evolution, in practice there is no sharp distinction between cat states and classical sensitivity of the asymptotic magnetization with respect to the parameters of the system. In fact, in both cases, experimental measurements of the collective magnetization will result in a distribution characterized by two peaks, and understanding whether the origin of such a macroscopic superposition is quantum-coherent (as in a cat state) or classical-incoherent (as would result from unavoidable experimental errors) would actually be unfeasible, due to fast decoherence of the cat state (as for the original Schr{\oe}dinger's cat!).

%mention exact solution? mention improved ED?

\section{Conclusions}
\label{sec:conclusions}
%
%\alessio{add all refs to sections and to letter}
This paper has been devoted to the analysis of the  impact of non-equilibrium fluctuations on mean-field %the quench dynamics of quantum spin chains supporting a 
dynamical phase transitions.
We have considered as unperturbed Hamiltonian a spin model with all-to-all couplings~\eqref{eq:MFH}, whose dynamics in the thermodynamic limit are equivalent to the classical evolution of the collective spin orientation
%which becomes exactly solvable in thermodynamic limit via its mean-field solution 
(see Sec.~\ref{sec:MF}). 
%
%As a consequence, its . 
%
This  has allowed us analyse the effect of any integrability-breaking interaction   in terms of a systematic spin-wave expansion %defined in a reference frame co-moving with the  classical motion of 
%which describe deviations with respect to the instantaneous direction of the average collective spin 
(see Sec.~\ref{sec:methodweakfluct}).
Through this time-dependent spin wave theory %we have investigated the onset of novel dynamical phases induced by non-equilibrium quantum fluctuations and 
we have found a general phenomenon concerning perturbations of dynamical critical points:
%
%Near a dynamical critical point point, %of a non-equilibrium phase diagram, 
%spin-wave 
fluctuations dominate the dynamics and act as a self-generated quantum friction, %drive chaotically the order parameter in and out of its ferromagnetic sectors, until the dynamics of spin waves dephase, and 
which makes the order parameter eventually remain trapped in  one of them in a pseudo-random fashion. 
We refer to this phase as  chaotic, since the asymptotic sign of the order parameter depends with extreme sensitivity on the initial conditions  and on the Hamiltonian parameters
% ruling the non-equilibrium dynamics: even tiny differences in the various couplings controlling the evolution can result in the system ending into a certain ferromagnetic sector, rather than in any of the other ones available, compatibly with the conservation of energy 
(see Secs.~\ref{sec:quenchchaos} and~\ref{sec:generalchaos}).
The existence of this peculiar dynamical behavior has been benchmarked with numerical methods based on a time-dependent variational principle developed on the matrix product state manifold, and shown to persist even for stronger integrability-breaking couplings (see Sec.~\ref{sec:bojan}).
We have also studied the signatures of this novel dynamical phase on the space-time dependent correlation functions (see Sec.~\ref{sec:corr}), as well as demonstrated that for sufficiently slow ramps of the transverse magnetic field the chaotic phase gradually fades away %because  the energy of the spin waves decreases as the ramp becomes increasingly adiabatic 
(see Sec.~\ref{sec:ramp}).

%In this paper we have considered  spin models. 
A straightforward and interesting extension of our analysis would consist in considering the sudden quench of integrability-breaking terms in the Dicke model, describing the interaction of several two-level atoms (spins) with a collective cavity photonic mode. 
The Dicke model possesses a rich dynamical phase diagram resulting from a quantum quench of the light-matter coupling~\cite{SciollaBiroliMF,Brandes,Keel10, Keel12}. 
%\revcomment{CUT? Contrary to the fully-connected  Ising ferromagnet  considered in this paper, the order parameter of the superradiant phase transition of the Dicke model is the expectation value of the photonic creation (or, equivalently, annihilation) operator.} %(though the former is mappable into the latter via adiabatic elimination).
%
The potential onset of a similar chaotic dynamical phase, monitored by photonic observables directly accessible in cavity quantum-electrodynamics experiments, could represent % an important step forward the establishment 
a welcome experimental verification of the phenomena discussed in this work. %\alessio{comments on exp with trapped ions?}

Finally, it would be interesting to inspect the effect on mean-field dynamical critical phases of a weak spatial disorder, which can be accounted for in the time-dependent spin wave treatment:
a natural, intriguing question would be to establish whether the competition of quantum fluctuations and classical spatial inhomogeneities would  enhance or suppress the novel non-equilibrium phase discussed in this paper.

%\alessio{add comments on: 1) experiments with ion traps, including higher dimensions and higher spins 2) Z3 symmetry 3) dissipation-enhanced chaotic behavior}

%
\emph{Acknowledgments.} We  thank E. Demler, F. Essler, Y. Gefen, A. Rosch for useful discussions. 
%
%A. L. acknowledges hospitality from the University of Cologne.
%
J. M. acknowledges support from the Alexander Von Humboldt foundation.
J. M. is supported by the European Union's Horizon 2020 research and innovation
programme under the Marie Sklodowska-Curie grant agreement No 745608 (QUAKE4PRELIMAT).
B. \v{Z}. is supported by the Advanced grant of European Research Council (ERC), No.
694544 – OMNES

\bibliography{biblio}

\end{document}